\documentclass{aa}  
\usepackage{graphicx}
\usepackage{txfonts}
\usepackage{color}
\newcommand{\mm}{\, \mu \mathrm{m}}

\newcommand{\kms}{\, \mathrm{km}\,\mathrm{s}^{-1}}

\newcommand{\pc}{\, \mathrm{pc}}

\begin{document}

\title{Star formation and gas flows in the centre\\ of the NUGA galaxy \object{NGC 1808} observed with SINFONI\thanks{Based on observations with ESO-VLT, STS-Cologne GTO proposal ID 094.B-0009(A) and ESO archival data, proposal nos 074.A-9011(A) and 075.B-0648(A) }}
\titlerunning{Star formation in NGC 1808}
\author{Gerold Busch \inst{1},
Andreas Eckart \inst{1,2},
M\'onica Valencia-S. \inst{1},
Nastaran Fazeli \inst{1},
Julia Scharw\"achter \inst{3},
Françoise Combes\inst{4}, \and
Santiago Garc\'ia-Burillo \inst{5}
}
\authorrunning{Gerold Busch et al.}

\institute{
I. Physikalisches Institut der Universit\"at zu K\"oln, Z\"ulpicher Str. 77, 50937 K\"oln, Germany\\
\email{busch@ph1.uni-koeln.de, eckart@ph1.uni-koeln.de}
\and
Max-Planck-Institut f\"ur Radioastronomie, Auf dem H\"ugel 69, 53121 Bonn, Germany
\and
Gemini Observatory, Northern Operations Center, 670 N. A'ohoku Place, Hilo, HI, 96720, USA
\and
LERMA, Observatoire de Paris, College de France, PSL, CNRS, Sorbonne Univ., UPMC, 75014 Paris, France
\and
Observatorio Astron\'omico Nacional (OAN) - Observatorio de Madrid, Alfonso XII 3, 28014 Madrid, Spain
}

%\date{Received ???; accepted ???}

\abstract{\object{NGC 1808} is a nearby barred spiral galaxy which hosts young stellar clusters in a patchy circumnuclear ring with a radius of $\sim 240\,\mathrm{pc}$. In order to study the gaseous and stellar kinematics and the star formation properties of the clusters, we perform seeing-limited $H+K$-band near-infrared integral-field spectroscopy with SINFONI of the inner $\sim 600\,\mathrm{pc}$. From the $M_\mathrm{BH}-\sigma_*$ relation, we find a black hole mass of a few $10^7\,M_\odot$. We estimate the age of the young stellar clusters in the circumnuclear ring to be $\lesssim 10\,\mathrm{Myr}$. No age gradient along the ring is visible. However, the starburst age is comparable to the travel time along the ring, indicating that the clusters almost completed a full orbit along the ring during their life time. In the central $\sim 600\,\mathrm{pc}$, we find a hot molecular gas mass of $\sim 730\,M_\odot$ which, with standard conversion factors, corresponds to a large cold molecular gas reservoir of several $10^8\,M_\odot$, in accordance with CO measurements from the literature. The gaseous and stellar kinematics show several deviations from pure disc motion, including a circumnuclear disc and signs of a nuclear bar potential. In addition, we confirm streaming motions on $\sim 200\,\mathrm{pc}$ scale that have recently been detected in CO(1-0) emission. Due to our enhanced angular resolution of $<1\arcsec$, we find further streaming motion within the inner arcsecond, that have not been detected until now. Despite the flow of gas towards the centre, no signs for significant AGN activity are found. This raises the questions what determines whether the infalling gas will fuel an AGN or star formation.}


\maketitle

\section{Introduction}

It is now well established that galaxies, at least those with a massive spheroidal component, harbour a supermassive black hole (SMBH) in their nucleus \citep[e.g.][]{1995ARA&A..33..581K,2013ARA&A..51..511K}. Active galactic nuclei (AGNs) are fuelled by accretion onto the central SMBH. Tight relations between the mass of the SMBH and properties of the host galaxy (mainly its spheroidal component) are interpreted as sign for a common evolution of SMBH and host galaxy \citep{2000ApJ...539L...9F,2000ApJ...539L..13G,2003ApJ...589L..21M,2004ApJ...604L..89H,2007ApJ...655...77G,2009ApJ...698..198G,2012ApJ...746..113G,2014ApJ...780...70L,2016ApJ...817...21S,2016ApJ...821...88S,2016ASSL..418..263G}. Therefore, a full understanding of how AGNs are fuelled and what prevents other galaxies from being fuelled (and thereby being ``quiescent'' instead of ``active'') is crucial to understand the evolution of galaxies from high redshift to the Local Universe.

While in high-luminosity AGNs the onset of nuclear activity is linked to large-scale (kpc) perturbations like bars and galaxy interactions which can trigger a gas inflow \citep[e.g.][]{1988ApJ...325...74S,2010MNRAS.407.1529H,2013MNRAS.429.2924H}, low-luminosity AGNs seem to be dominated by secular evolution \citep[e.g.][]{2008ApJS..175..356H,2010MNRAS.407.1529H,2011Natur.469..374K}. Possible feeding mechanisms are bars, secondary/nuclear bars, $m=1$ instabilites, warps, nuclear spirals, stellar winds \citep[][and references therein]{2005A&A...441.1011G,2012JPhCS.372a2050G}, and cover a large range of scales from host galaxy ($\gtrsim 1\,\mathrm{kpc}$) to nuclear scales ($\lesssim\,\mathrm{pc}$). A further difficulty is caused by the different time-scales of star formation and AGN activity, in particular the fact that even during a total AGN duty cycle of $10^8\,\mathrm{yr}$, the AGN might ``flicker'' on much lower time scales of $10^5\,\mathrm{yr}$ \citep[e.g.][]{2014ApJ...782....9H,2015MNRAS.451.2517S}.

The NUGA survey (NUclei of GAlaxies) was established to systematically investigate the issue of nuclear fuelling for nearby galaxies. NUGA started off as an IRAM key project \citep[PIs: Santiago Garc\'ia-Burillo and Fran\c{c}oise Combes; see][]{2003A&A...407..485G} on the northern hemisphere and is continued now on the southern hemisphere with the Atacama Large Millimeter/submillimeter Array (ALMA) as the millimetric investigations can prospectively be performed with a superior angular resolution and sensitivity \citep{2013A&A...558A.124C,2014A&A...565A..97C,2014A&A...567A.125G}. 
NUGA comprises a sample of 30 nearby AGNs covering all stages of nuclear activity (Seyferts - LINERs - starbursts). The combined datasets allow a first systematic study of gas kinematics covering scales from a few tens parsec to the outer few tens kiloparsec.

Complementary integral-field spectroscopy (IFS) data are taken in the near-infrared \citep[see e.g. the cases of NGC1433 and NGC1566;][]{2014A&A...567A.119S,2015A&A...583A.104S}. With the NIR data we can include information on the hot molecular and atomic gas and their excitation mechanisms \citep[e.g.][]{2007A&A...466..451Z,2013MNRAS.428.2389M,2014A&A...567A.119S}, as well as properties of the central engine \citep[in particular black hole mass and hidden broad line region, see case of \object{NGC 7172} in][]{2012A&A...544A.105S}. Furthermore IFS in the NIR is the ideal tool to study stellar populations and star formation in (dust-obscured) centres of galaxies \citep[e.g.][]{2008AJ....135..479B,2009MNRAS.393..783R,2009ApJ...698.1852B,2012A&A...544A.129V,2014MNRAS.438..329F,2015A&A...575A.128B,2015A&A...583A.104S}. IFS also allows finding kinematically decoupled regions like nuclear discs and spatially resolving inflows and outflows in many galaxies \citep[e.g.][]{2008MNRAS.385.1129R,2010MNRAS.402..819S,2011ApJ...739...69M,2013MNRAS.429.2315S,2014ApJ...792..101D,2015MNRAS.451.3587R,2015MNRAS.453.1727D}.

In this paper, we present near-infrared integral-field spectroscopy of the galaxy \object{NGC 1808} from the NUGA sample which is a (R)SAB(s)a barred spiral galaxy \citep{1991rc3..book.....D}. \object{NGC 1808} has been early reported to contain ``hotspots'' near the galaxy nucleus \citep{1958PASP...70..364M,1965PASP...77..287S,1975Ap&SS..33..173P}. These hotspots are mainly prominent in radio and/or near+mid-infrared emission and trace young stellar clusters and their associated supernova remnants \citep{1992MNRAS.259..293F,1994MNRAS.268..203C,1994ApJ...425...72K,1996A&A...313..771K,1996AJ....112..918T,2005A&A...438..803G,2008A&A...487..519G}. The near-infrared spots are located in a circumnuclear ring with radius $r\sim 280\,\mathrm{pc}$ \citep{2010MNRAS.402.2462C}. Some authors have reported that the galaxy contains a weak AGN \citep[following the classification of][]{1985A&A...145..425V}, however, others disagree \citep[e.g.][]{1992MNRAS.259..293F,1993AJ....105..486P,2001ApJ...557..626K,2015ApJS..217...12D}. We discuss the possible presence of an AGN in Sect.~\ref{sec:AGN}.

Prominent dust lanes on kiloparsec scales (pointing from the nucleus in north-east direction) that are visible in optical images and peculiar motions of \ion{H}{i} are interpreted as indications for an outflow of neutral and ionised gas \citep{1993AJ....105..486P,1993A&A...268...14K}. A possible tidal interaction with NGC 1792 has been discussed \citep{1990A&A...240..237D,1993A&A...268...14K}.

\object{NGC 1808} has recently been mapped in $^{12}\,\mathrm{CO}\,(J=1-0)$ by  \cite{2016ApJ...823...68S} at a resolution of $2\arcsec$ ($\sim 100\,\mathrm{pc}$). In the centre, they find a compact circumnuclear disc ($r<200\,\mathrm{pc}$) and a molecular gas ring with radius $r\sim 500\,\mathrm{pc}$. Analysing the gas kinematics, they find several components of non-circular motion: a spiral pattern in the inner disc ($r<400\,\mathrm{pc}$) and gas streaming motion on the inner side of one spiral arm, as well as a molecular gas outflow from the nuclear starburst region ($r<250\,\mathrm{pc}$) which is spatially coincident with one of the mentioned dust lanes.

This paper presents a first comprehensive near-infrared IFS study of the central kpc of \object{NGC 1808 }. In comparison to previous studies of the emission line distribution, our data has higher spatial resolution ($\lesssim 1\arcsec$) and much higher sensitivity so that we can trace the morphology in great detail. The combination of a high spectral coverage ($1.45 - 2.45\,\mu\mathrm{m}$ for the $H+K$-band grating) and high spectral resolution ($R=4000$ for the $K$-band grating) allows us to get detailled maps of stellar and gaseous kinematics and study diagnostic line ratios in a spatially resolved way. Only with instruments like SINFONI, we can study combined kinematical information and excitation mechanisms from a single, homogeneous data set.

\begin{figure}
\centering
\includegraphics[width=\columnwidth]{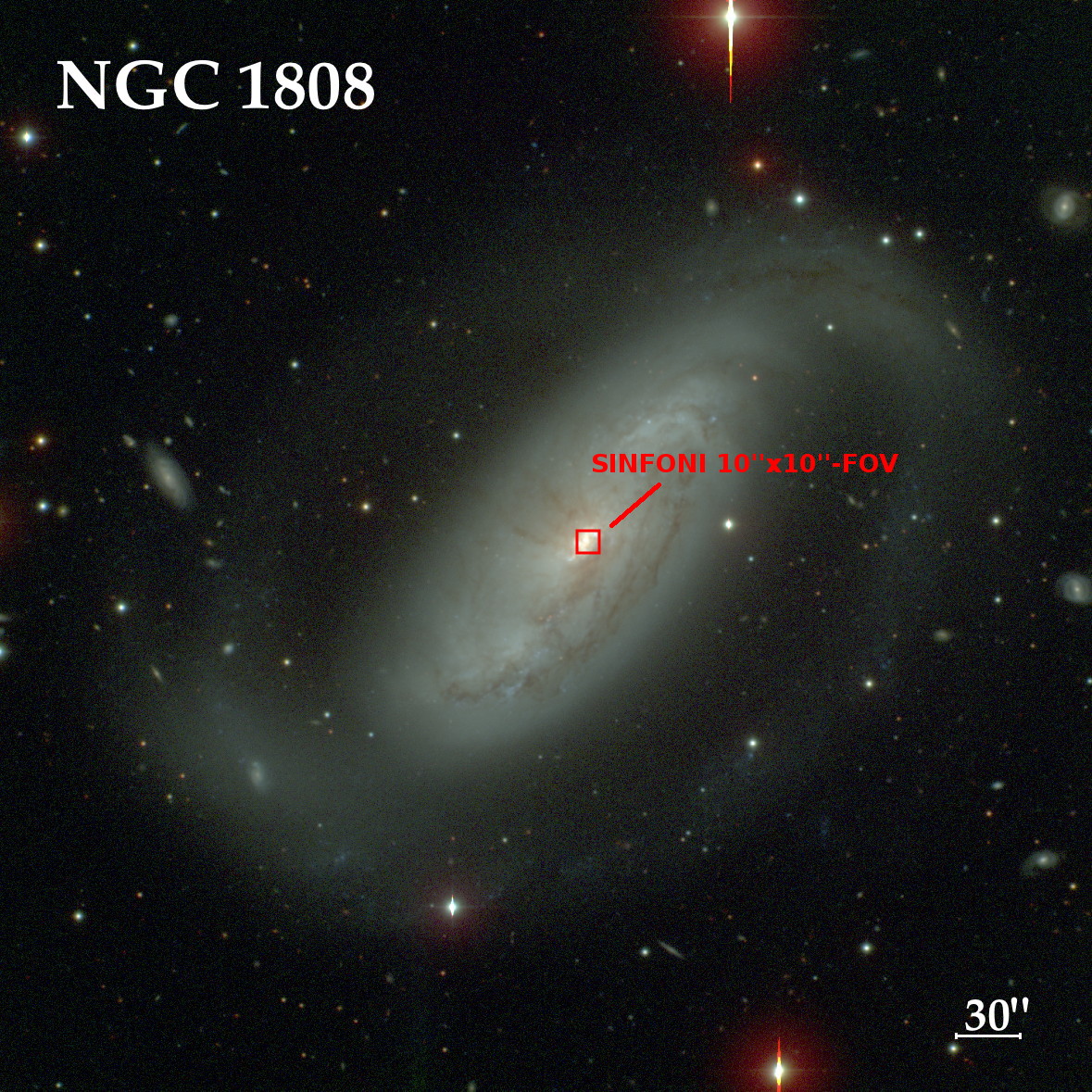}
\caption{Optical image of the barred spiral galaxy NGC 1808. The $10\arcsec \times 10\arcsec$ field-of-view of SINFONI is indicated. Image courtesy: Carnegie-Irvine Galaxy Survey \citep{2011ApJS..197...21H}.}
\label{fig:ngc1808_cgs}
\end{figure}

Figure \ref{fig:ngc1808_cgs} shows an optical image of NGC 1808, with the FOV of the SINFONI near-infrared integral-field spectroscopy data which is analysed in detail in this paper, indicated. Figure \ref{fig:hst} shows a HST false-colour image of the central region.

The paper is structured as follows: In Sect.~\ref{sec:observation} we present the observations used for this analysis. In Sect.~\ref{sec:results} we present the results of the analysis of the SINFONI data cubes (in particular emission line maps as well as stellar and gaseous kinematics), that we then discuss with regard to previous results from the literature in the following Sect.~\ref{sec:discussion}. Section \ref{sec:conclusions} gives a short summary and conclusions from this work.

In order to calculate spatial scales and luminosities, we adopt a luminosity distance of $D_L=(12.8\pm 1.2)\,\mathrm{Mpc}$ \citep{2009AJ....138..323T} which corresponds to a scale of $62\,\mathrm{pc}\,\mathrm{arcsec}^{-1}$.

\section{Observations and data reduction}
\label{sec:observation}

\subsection{SINFONI near-infrared integral-field spectroscopy}
The analysis of NGC 1808 in this paper is based on near-infrared integral-field spectroscopy data obtained with SINFONI \citep{2003SPIE.4841.1548E,2004Msngr.117...17B} at the Very Large Telescope (VLT) of the European Southern Observatory (ESO) in Chile. 

We observed the central region of NGC 1808 in seeing-limited mode on October 6th, 2014. The field-of-view (FOV) of single exposures in this mode is $8 \arcsec \times 8 \arcsec$ which results in a spatial sampling of 0.125 arcsec pixel$^{-1}$. We used a jitter pattern with offsets $\pm 1\arcsec$ to minimize the effect of bad pixels and increase the FOV of the combined cube to $10\arcsec \times 10 \arcsec$ (which corresponds to a linear scale of 620 pc). We spent 1500s ($10\times 150$s) on-source time using the $H+K$ grating with a spectral resolution of $R\approx 1500$. This grating has a large simultaneous wavelength coverage which is useful for analysis of emission line ratios. Furthermore, we spent 3000s ($20\times 150$s) on-source time using the $K$-band grating which results in a higher spectral resolution of $R\approx 4000$ which is ideal to analyse the stellar dynamics.

Furthermore, we use seeing-limited (FOV: $8\arcsec \times 8 \arcsec$) SINFONI $J$-band data from the ESO-archive (proposal nos. 074.A-9011(A) and 075.B-0648(A)) that have a total on-source exposure time of 1200s. This data are needed to determine the extinction by using the line ratio between the hydrogen emission lines Br$\gamma$ in the $K$-band and Pa$\beta$ in the $J$-band.

\begin{figure}
\centering
\includegraphics[width=\columnwidth]{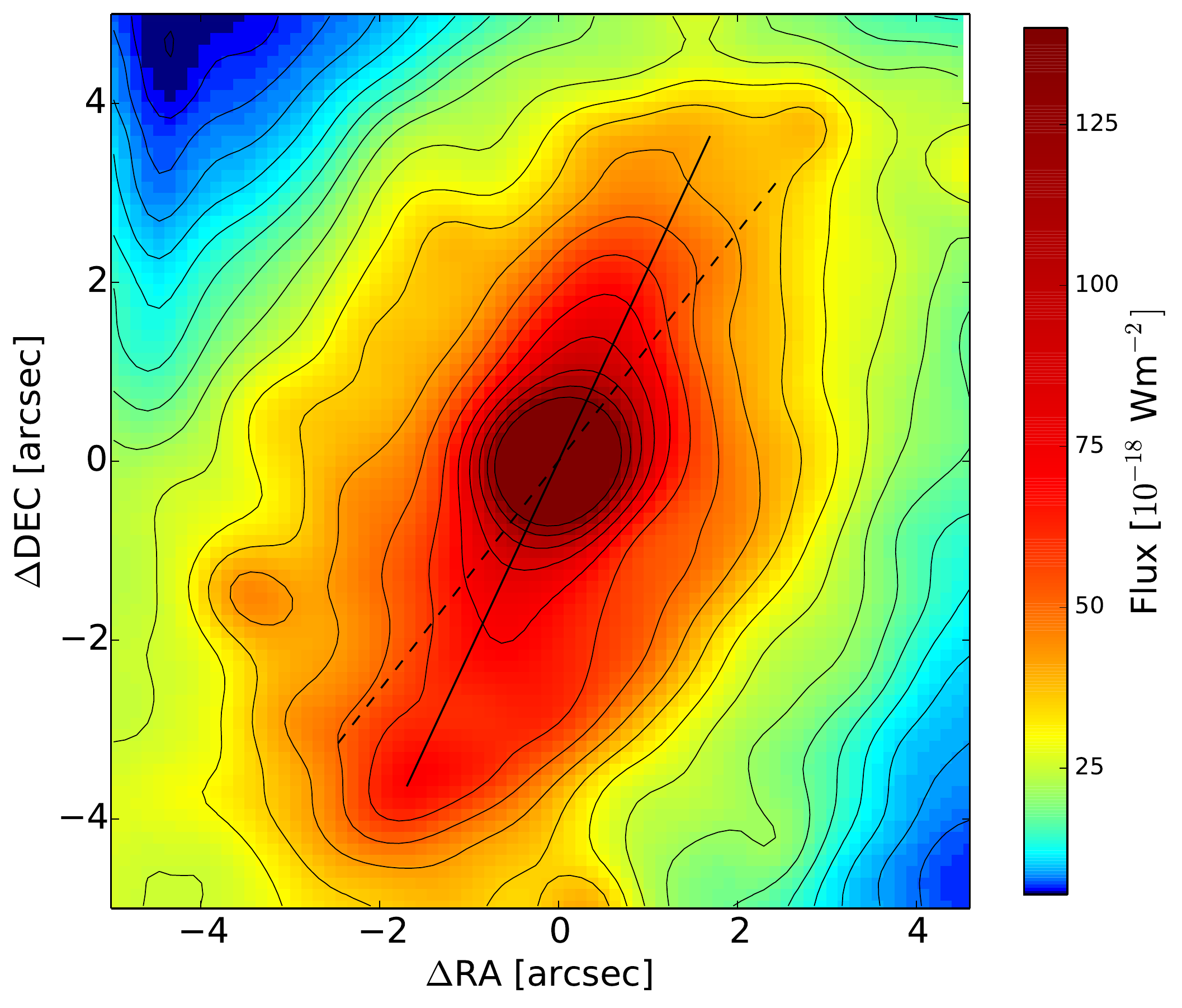}
\caption{$K$-band continuum image extracted from the SINFONI data cube. The solid line marks the PA of the nuclear bar ($\sim 115^\circ$, measured from north to east), while the dashed line marks the line-of-nodes of the stellar kinematics (Sect.~\ref{sec:stellkin_disc}).}
\label{fig:continuum}
\end{figure}

The pipeline which is delivered by ESO was used for data reduction up to single-exposure-cube reconstruction. For alignment, final coaddition, and telluric correction, we use our own \textsc{Python} and \textsc{Idl} routines. For a more detailed description of the reduction and calibration, we refer the reader to \citet{2015A&A...575A.128B} and \citet{2014A&A...567A.119S}. Figure \ref{fig:continuum} shows a $K$-band continuum image that was extracted from the SINFONI data cube.

\subsection{Hubble Space Telescope imaging data}
\begin{figure}
\centering
\includegraphics[width=\columnwidth]{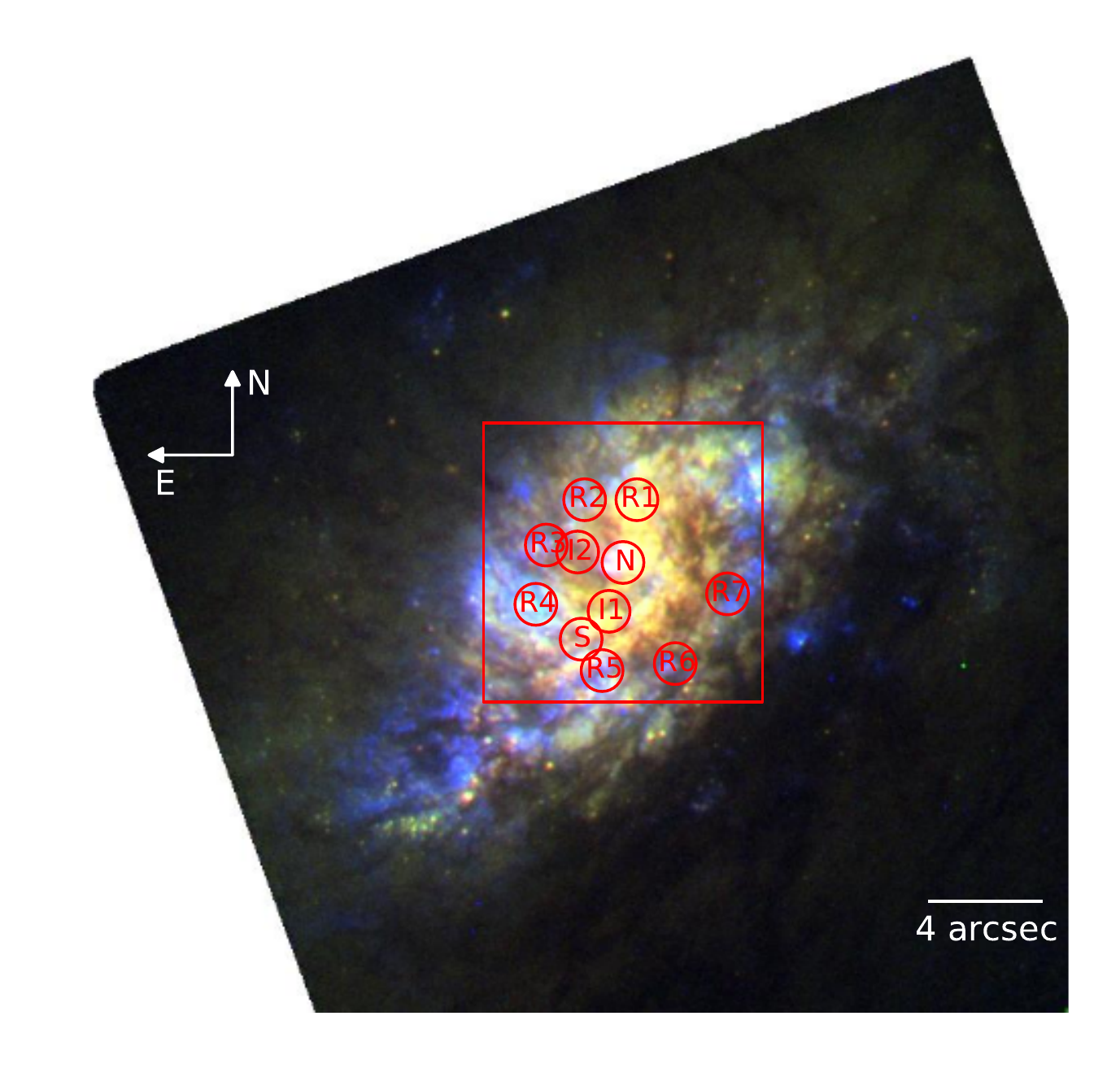}
\caption{HST false-colour image of the central region of NGC 1808 (red: F814W, green: F675W, blue: F658N). The apertures that are analysed in detail are indicated by red circles.}
\label{fig:hst}
\end{figure}

We retrieved an image of the central region of NGC 1808 from the Hubble Legacy Archive\footnote{http://hla.stsci.edu/}. The images were taken with the \emph{Wide Field and Planety Camera 2} of the \emph{Hubble Space Telescope} in the F814W, F675W, and F658N filter. The observations took place in August 1998 (proposal ID 6872, PI: James Flood). In Fig. \ref{fig:hst}, we show a RGB composition of these images (red: F814W, green: F675W, blue: F658N) and indicate the apertures of the spots that we analyse in detail.

\section{Results}
\label{sec:results}

\subsection{Emission-line flux distributions}
\label{sec:fluxdistr}

The NIR emission-lines trace the hot molecular and ionised atomic gas phase. This gas is predominantly associated with the sites of young star formation (\ion{H}{ii} regions, shocks, hot surfaces of molecular clouds) or non-thermal nuclear activity (nuclear ionisation, winds and shock regions). 

The data cubes show a large variety of emission lines, including several hydrogen recombination lines (that trace fully ionised regions), molecular hydrogen lines, and the shock tracer [\ion{Fe}{ii}] (that traces partially ionised regions), as well as stellar absorption lines.

We use the \textsc{Python} implementation of \textsc{Mpfitexpr}  which is based on the Levenberg-Marquardt algorithm \citep{2009ASPC..411..251M,jore} to generate maps of the emission-line flux distributions. The line fluxes were determined by fitting a Gaussian function to the line profile. The continuum was subtracted by fitting a linear function to the continuum emission in two spectral windows, left and right from the emission line. The emission-line maps are clipped to have a uncertainties less than 30\%.

\begin{figure*}
\centering
\includegraphics[width=0.33\linewidth]{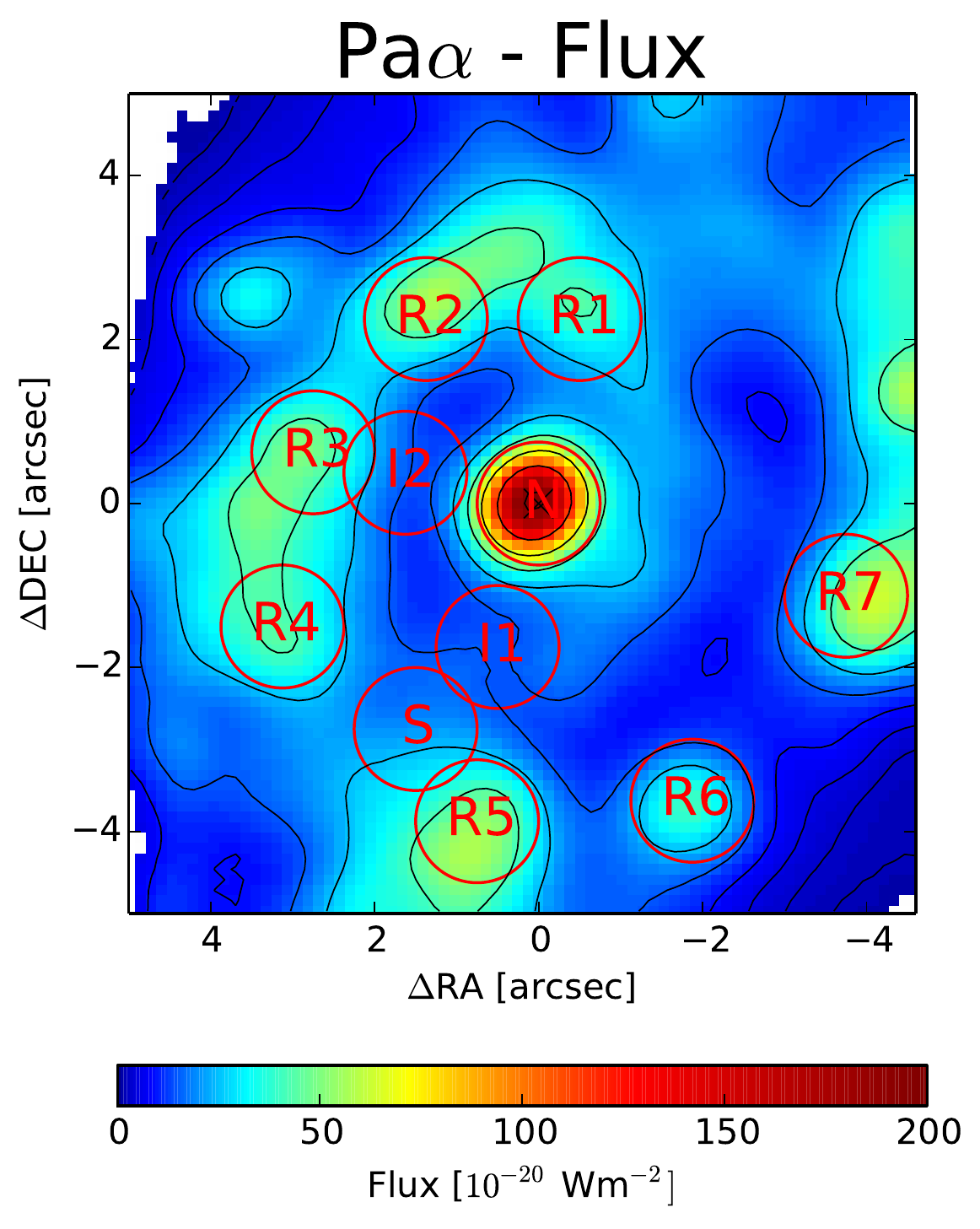}
\includegraphics[width=0.33\linewidth]{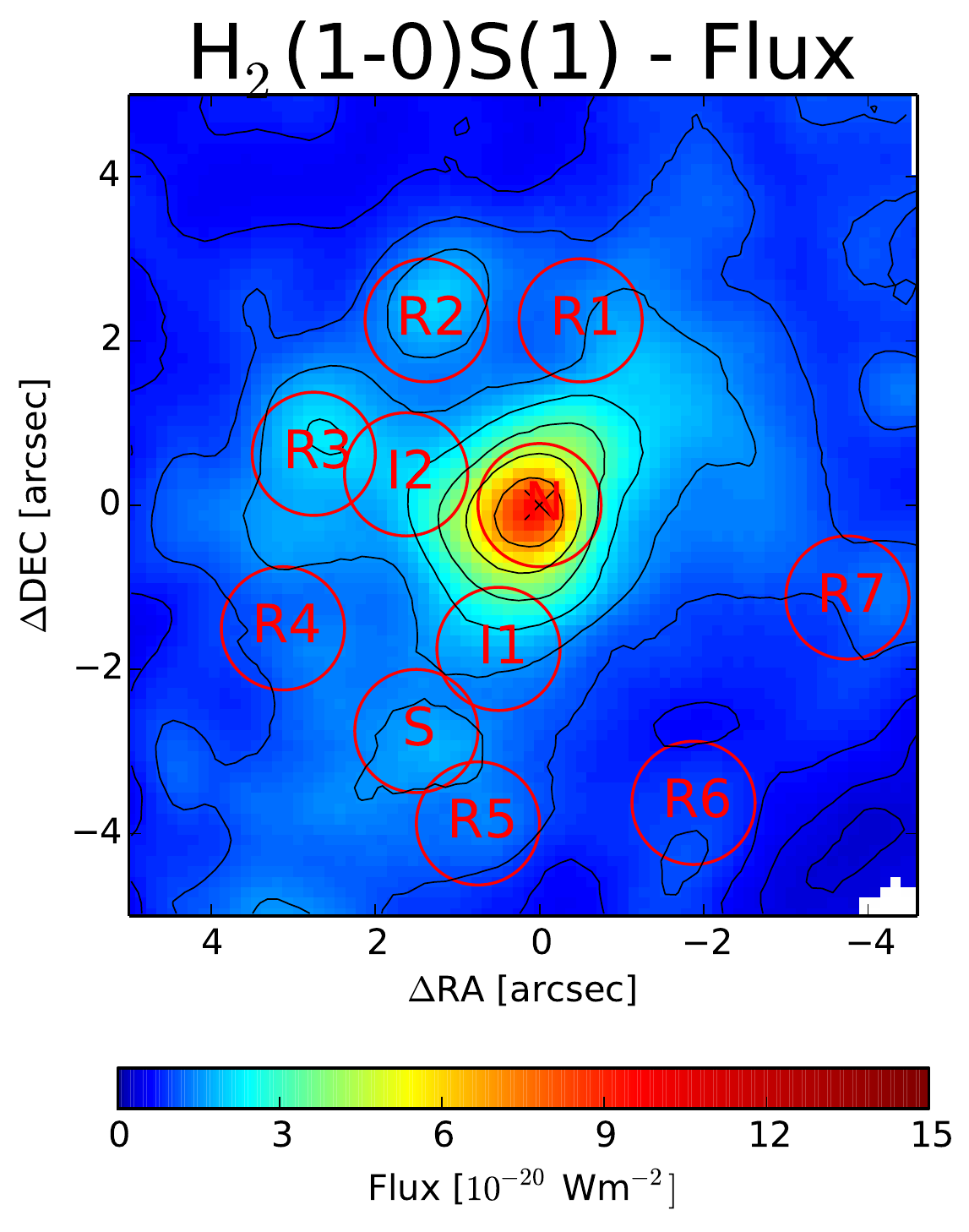}
\includegraphics[width=0.33\linewidth]{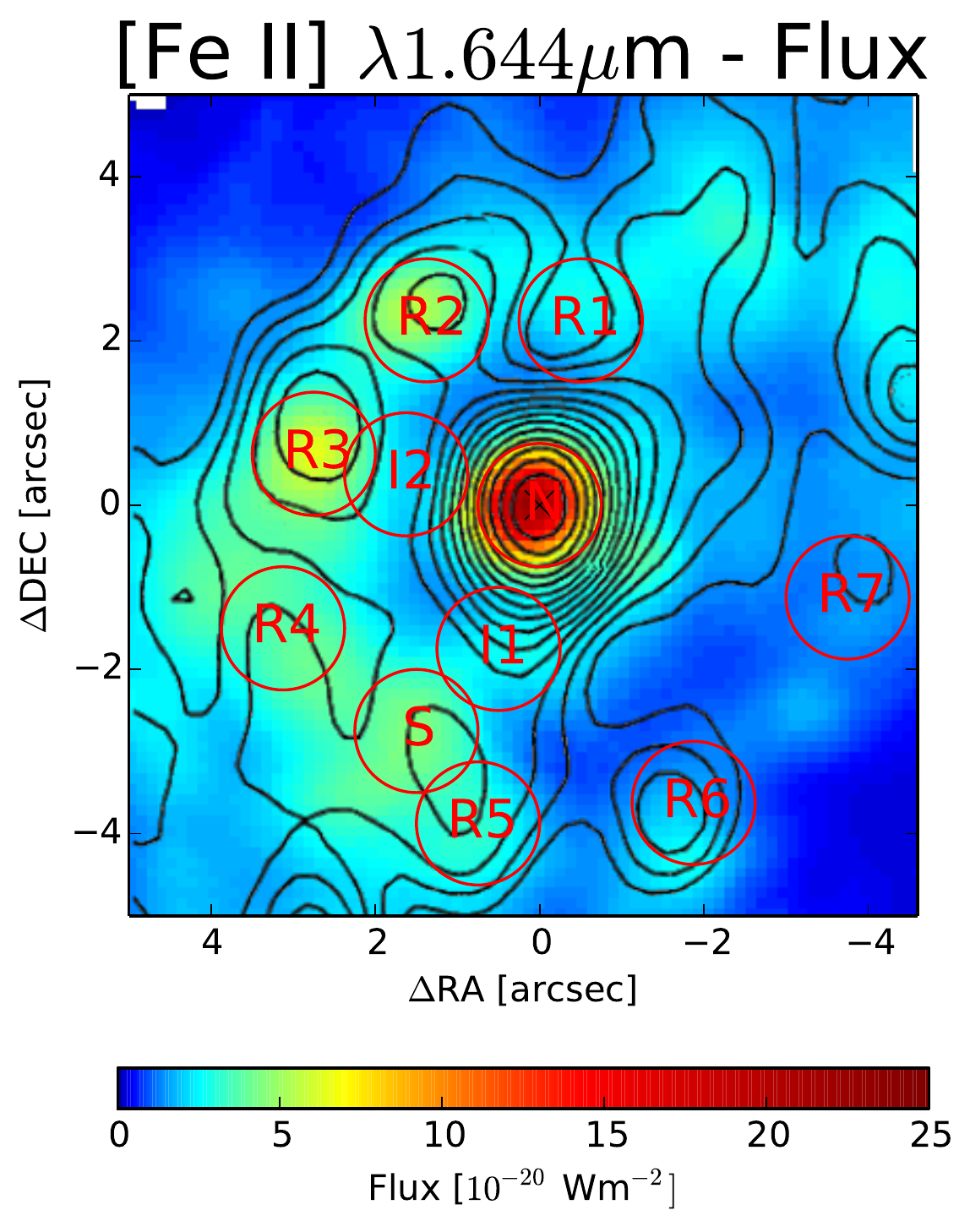}
\caption{Flux maps of the Pa$\alpha$, H$_2 \lambda2.12\mu$m, and [\ion{Fe}{ii}]$\lambda 1.64\mm$ emission lines. The [\ion{Fe}{ii}] map is overlaid with the 3.6cm radio-continuum image of \citet{1994MNRAS.268..203C} as printed in \citet{1996A&A...313..771K}. The molecular hydrogen shows a more centrally concentrated distribution that is different from the distribution of ionised hydrogen and the shock tracer [\ion{Fe}{ii}] which show emission in the centre but also in a circumnuclear ring with radius $4\arcsec \approx 240\pc$.}
\label{fig:deepmaps}
\end{figure*}

In Fig.~\ref{fig:deepmaps}, we show the maps of Pa$\alpha$, H$_2\lambda2.12\mm$, and the $H$-band emission-line [\ion{Fe}{ii}] at $1.644\mm$ which we analyse in the following.

All three mapped gas types (Pa$\alpha$, H$_2 \lambda 2.122\mm$, [\ion{Fe}{ii}]) have their absolute peak in the centre, coinciding with the peak of the continuum flux, which is marked by a cross. Apart from this they show quite different flux distributions:

Ionised gas as traced by the hydrogen recombination lines Pa$\alpha$ and Br$\gamma$ is located in a ring around the centre which has a radius of $\sim 4\arcsec$ ($\approx 240\,\mathrm{pc}$)\footnote{Note that this circumnuclear ring is not the same as the $500\,\mathrm{pc}$ molecular gas ring that \cite{2016ApJ...823...68S} find in their recent CO data. Our FOV is limited to the region that they call ``circumnuclear disk''.}. With our high spatial-resolution data, we confirm the asymmetry mentioned by \citet{1994ApJ...425...72K} and \citet{1996A&A...313..771K}: The patches to the south-east show higher flux levels compared to the west side. Furthermore, there seem to be several gaps in the ring, the largest in the north-west, that give the ring an overall patchy structure. We stress that also in the region between ring and nucleus, Pa$\alpha$ and Br$\gamma$ are detected but at a lower intensity than in the ring. A tail-like structure in the south-east is visible in the maps of \citet{1996A&A...313..771K} (that have a FOV of $20\arcsec \times 20\arcsec$ but lower angular resolution) and is probably part of gas spiral arms that can often be seen on kpc scales \citep[see e.g.][]{2015A&A...575A.128B}.

In the emission of hot molecular hydrogen (traced by the H$_2 \lambda 2.12\mm$ emission line) the circumnuclear ring, found in Pa$\alpha$ and Br$\gamma$, is less prominent. Some of the patches in the ring show an enhanced flux level also in H$_2$ but the contrast between ring and interring region is much lower in H$_2$ than in Pa$\alpha$. Furthermore, we find that the flux in the center is more extended than in Pa$\alpha$. To quantify this finding, we fit a Gaussian to the central peak. For H$_2$, the FWHM is $1\farcs3$, but only $0\farcs9$ for Pa$\alpha$. Furthermore, we note that we find strong H$_2$ flux between ring and nucleus in eastern direction (marked with ``I2'' in Fig.~\ref{fig:deepmaps}).

At first glance, the flux distribution of [\ion{Fe}{ii}] looks very similar to that of Pa$\alpha$ and Br$\gamma$, i.e. it also shows a prominent circumnuclear ring. However two differences are striking: First, some Pa$\alpha$ patches show much lower flux level in [\ion{Fe}{ii}] compared to the others. We will use this to estimate starburst ages. Second, the [\ion{Fe}{ii}] shows strong emission in the southeast side of the ring (marked with ``S'' in Fig.~\ref{fig:deepmaps}) that does not coincide with a Pa$\alpha$ and Br$\gamma$ peak. We will analyse this region in more detail later.

\begin{figure*}
\centering
\includegraphics[width=\linewidth]{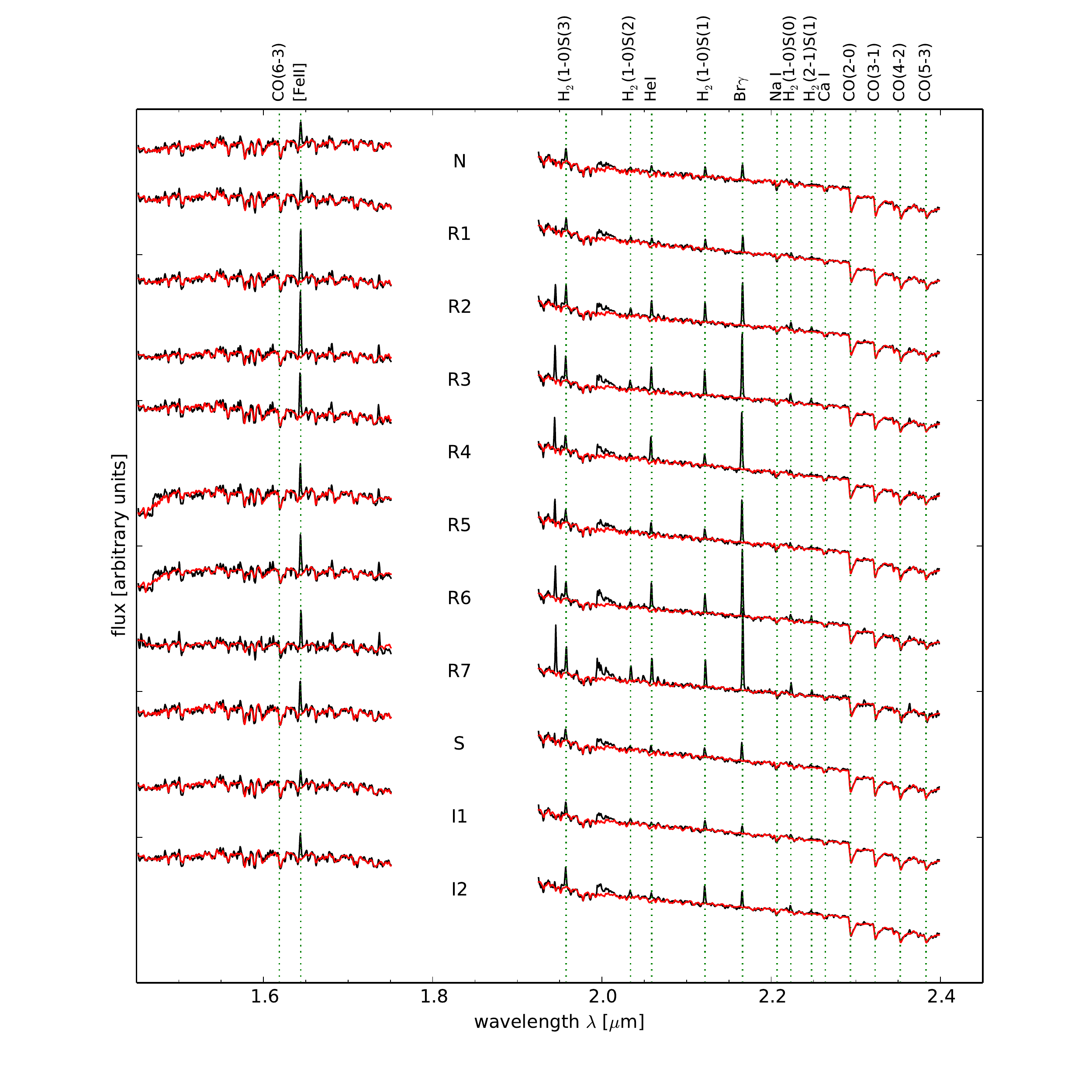}
\caption{Spectra of the nuclear regions plus seven apertures in the circumnuclear ring (R1-7), a possibly shocked region (S), and two in the inter-ring region (I1-2). The positions where the apertures are extracted, are indicated in Figs.~\ref{fig:hst} and \ref{fig:deepmaps}. Several emission and absorption lines are indicated with green vertical lines. The stellar continuum fit is shown in red.}
\label{fig:spectra}
\end{figure*}

In order to analyse the excitation properties, star formation, and more in the nucleus (N), ring (R1 to R7), and in between (I1, I2, and S), we chose eleven apertures that are marked in Figs.~\ref{fig:hst} and \ref{fig:deepmaps}. We extract spectra from circular apertures with radii $r=0\farcs75$. The spectra are shown in Fig.~\ref{fig:spectra}.

Before we measure the emission lines, we perform a subtraction of the stellar continuum with the \textsc{Python} implementation of the Penalized Pixel-Fitting method \citep[pPXF,][]{2004PASP..116..138C}. We use a set of synthetic model spectra by \citet{2007A&A...468..205L} with solar abundances, an effective temperature range $T_\mathrm{eff}=2900-5900\,\mathrm{K}$, gravities of $\log(g/\mathrm{cm}\,\mathrm{s}^{-2})=[-1.0,-0.5,0.0,+1.0]$, and masses $1\,M_\odot$ and $15\,M_\odot$. The modelled stellar continuum is shown in Fig.~\ref{fig:spectra} in red. We then subtract these models from the original spectra and use the residual for the emission line fits.

For the emission line fits, we use the \textsc{Python} implementation of \textsc{Mpfitexpr} which is based on the Levenberg-Marquardt algorithm \citep{2009ASPC..411..251M,jore}. All emission lines are fitted with Gaussian functions. In order to estimate the uncertainties of the parameters, we perform a Monte Carlo simulation with 100 iterations. In each iteration Gaussian noise is added to the input spectrum, with the width corresponding to the root-mean-square of the residual spectrum (input spectrum -- fit). We then take the mean of the 100 fit results as best fit and the standard deviations as uncertainties of the fit parameter \citep[see also][]{2016A&A...587A.138B}. The measured line fluxes are presented in Tables \ref{tab:emissionlines} and \ref{tab:h2lines}.

\begin{table*}
\centering
\caption{Emission lines fluxes in the selected apertures, and derived quantities: extinction and ionised gas mass. The emission line fluxes are corrected for extinction and given in units of $10^{-18}\,\mathrm{W}\,\mathrm{m}^{-2}$. All apertures have a radius of $0\farcs75$ which corresponds to a physical area of $6800\pc^2$.}
\label{tab:emissionlines}
\begin{tabular}{ccccccccc}
\hline \hline
 & Pa$\beta$ & [\ion{Fe}{ii}] & Pa$\alpha$ & Br$\delta$ & Br$\gamma$ & $A_V$ & \ion{H}{ii} mass \\
aperture & $\lambda 1.282\mm$ & $\lambda 1.644\mm$ & $\lambda 1.876\mm$ & $\lambda 1.945\mm$ & $\lambda 2.166\mm$ & [mag] & [$10^5\,M_\odot$] \\
\hline

N & $87.54 \pm 6.85$ & $35.41 \pm 4.00$ & $197.65 \pm 21.02$ & $5.89 \pm 0.95$ & $15.19 \pm 0.82$ & $3.3 \pm 0.3$ & $6.8$ \\
R1 & $18.26 \pm 1.33$ & $5.13 \pm 0.53$ & $50.24 \pm 1.92$ & $1.16 \pm 0.13$ & $2.98 \pm 0.13$ & $2.0 \pm 0.2$ & $1.7$ \\
R2 & $28.97 \pm 1.03$ & $8.70 \pm 0.37$ & $69.36 \pm 1.53$ & $2.96 \pm 0.10$ & $5.00 \pm 0.11$ & $2.7 \pm 0.1$ & $2.4$ \\
R3 & $40.95 \pm 1.00$ & $10.02 \pm 0.32$ & $66.81 \pm 1.39$ & $4.48 \pm 0.11$ & $7.05 \pm 0.11$ & $2.8 \pm 0.1$ & $2.3$ \\
R4 & $40.09 \pm 0.92$ & $6.59 \pm 0.39$ & $53.28 \pm 1.13$ & $3.84 \pm 0.13$ & $6.86 \pm 0.11$ & $1.5 \pm 0.1$ & $1.8$ \\
R5 & $17.45 \pm 0.63$ & $5.86 \pm 0.44$ & $67.31 \pm 1.51$ & $3.34 \pm 0.12$ & $5.82 \pm 0.13$ & $1.7 \pm 0.1$ & $2.3$ \\
R6 & $20.31 \pm 0.59$ & $4.35 \pm 0.25$ & $46.73 \pm 1.00$ & $2.70 \pm 0.05$ & $4.70 \pm 0.07$ & $2.9 \pm 0.1$ & $1.6$ \\
R7\tablefootmark{a} & --- & $1.80 \pm 0.10$ & $52.50 \pm 0.60$ & $1.94 \pm 0.06$ & $3.59 \pm 0.05$ & --- & $1.8$ \\
S & $19.43 \pm 1.37$ & $7.63 \pm 0.66$ & $29.96 \pm 1.36$ & $1.29 \pm 0.13$ & $3.32 \pm 0.15$ & $1.7 \pm 0.2$ & $1.0$ \\
I1 & $11.73 \pm 1.55$ & $6.18 \pm 0.76$ & $22.67 \pm 1.63$ & $0.53 \pm 0.22$ & $1.94 \pm 0.18$ & $2.5 \pm 0.4$ & $0.8$ \\
I2 & $15.00 \pm 1.02$ & $5.11 \pm 0.44$ & $25.53 \pm 1.17$ & $1.08 \pm 0.06$ & $2.39 \pm 0.10$ & $2.2 \pm 0.2$ & $0.9$ \\

\hline
\end{tabular}
\tablefoot{
\tablefoottext{a}{Values are not corrected for extinction because extinction could not be determined in this aperture.}
}
\end{table*}

\begin{table*}
\centering
\caption{Molecular hydrogen emission lines fluxes in the selected apertures, and derived quantities: hot and cold gas mass, and (cold) gas density. The emission line fluxes are corrected for extinction and given in units of $10^{-18}\,\mathrm{W}\,\mathrm{m}^{-2}$. All apertures have a radius of $0\farcs75$ which corresponds to a physical area of $6800\pc^2$.}
\label{tab:h2lines}
\begin{tabular}{ccccccccc}
\hline \hline
 & H$_2$(1-0)S(3) & H$_2$(1-0)S(2) & H$_2$(1-0)S(1) & H$_2$(1-0)S(0) & H$_2$(2-1)S(1) & hot H$_2$ mass & cold gas mass & $\log(\Sigma_\mathrm{gas})$ \\
aperture & $\lambda 1.958\mm$ & $\lambda 2.034\mm$ & $\lambda 2.122\mm$ & $\lambda 2.223\mm$ & $\lambda 2.248\mm$ & [$M_\odot$] & [$10^6\,M_\odot$] & [$M_\odot\,\mathrm{pc}^{-1}$]  \\
\hline

N & $15.27 \pm 2.06$ & $3.50 \pm 1.49$ & $9.42 \pm 1.19$ & $2.51 \pm 1.33$ & $1.42 \pm 0.84$ & $78$ & $24-125$ & $4.06$ \\
R1 & $3.01 \pm 0.22$ & $0.85 \pm 0.22$ & $1.86 \pm 0.15$ & $0.78 \pm 0.17$ & $0.62 \pm 0.22$ & $15$ & $5-25$ & $3.36$ \\
R2 & $3.00 \pm 0.17$ & $1.09 \pm 0.13$ & $2.39 \pm 0.09$ & $0.91 \pm 0.12$ & $0.50 \pm 0.12$ & $20$ & $6-32$ & $3.47$ \\
R3 & $3.14 \pm 0.12$ & $1.27 \pm 0.14$ & $2.81 \pm 0.10$ & $1.05 \pm 0.08$ & $0.61 \pm 0.12$ & $23$ & $7-37$ & $3.54$ \\
R4 & $2.07 \pm 0.15$ & $0.80 \pm 0.20$ & $1.45 \pm 0.11$ & $0.65 \pm 0.12$ & $0.26 \pm 0.09$ & $12$ & $4-19$ & $3.25$ \\
R5 & $2.58 \pm 0.26$ & $0.52 \pm 0.23$ & $1.56 \pm 0.17$ & $0.47 \pm 0.17$ & $0.45 \pm 0.19$ & $13$ & $4-21$ & $3.28$ \\
R6 & $1.64 \pm 0.10$ & $0.65 \pm 0.15$ & $1.39 \pm 0.06$ & $0.50 \pm 0.07$ & $0.35 \pm 0.07$ & $12$ & $3-18$ & $3.23$ \\
R7\tablefootmark{a} & $1.23 \pm 0.07$ & $0.77 \pm 0.07$ & $1.25 \pm 0.04$ & $0.57 \pm 0.03$ & $0.23 \pm 0.03$ & $10$ & $3-17$ & $3.19$ \\
S & $3.06 \pm 0.27$ & $0.92 \pm 0.25$ & $1.96 \pm 0.19$ & $0.76 \pm 0.19$ & $0.62 \pm 0.16$ & $16$ & $5-26$ & $3.38$ \\
I1 & $4.04 \pm 0.31$ & $1.46 \pm 0.23$ & $2.46 \pm 0.23$ & $1.04 \pm 0.21$ & $0.65 \pm 0.24$ & $20$ & $6-33$ & $3.48$ \\
I2 & $3.77 \pm 0.20$ & $1.36 \pm 0.20$ & $2.81 \pm 0.14$ & $0.98 \pm 0.09$ & $0.65 \pm 0.18$ & $23$ & $7-37$ & $3.54$ \\

\hline
\end{tabular}
\tablefoot{
\tablefoottext{a}{Values are not corrected for extinction because extinction could not be determined in this aperture.}
}
\end{table*}

\subsection{Emission-line ratios}

In the following section, we calculate and map emission-line ratios. Emission line ratios allow us to trace the extinction and the UV radiation field of the ISM as well as possible shocks.

\subsubsection{Extinction correction}

Especially in the case of AGNs and starburst galaxies, reddening induced by dust can significantly affect the measured emission line fluxes. Near-infrared observations have the advantage that this effect is much less prominent than in the optical (extinction is lower by approximately a factor of 10 in magnitude). Therefore observations, e.g. of the Galactic Centre or dust enshrouded galactic nuclei \citep[see e.g.][]{2012A&A...544A.105S}, are often only feasible when going from optical to longer near-infrared wavelengths. However, even in the near-infrared, extinction effects can still be significant and should therefore be corrected.

Emission-line fluxes can be corrected following the relation $F_\mathrm{intr} = F_\mathrm{obs} \times 10^{0.4\times A(\lambda)}$ between observed flux $F_\mathrm{obs}$ and intrinsic flux $F_\mathrm{intr}$. We use the extinction law by \citet{2000ApJ...533..682C} to get an expression for the extinction at wavelength $\lambda$ (in $\mu$m):
\begin{equation}
A(\lambda) = \frac{A_V}{R'_V} \times \left[ 2.659 \left( -1.857 + \frac{1.040}{\lambda} \right) + R'_V \right]
\end{equation}
with $R'_V=4.05$ and visual extinction $A_V$. 

The internal gas extinction can now be estimated by comparing the observed flux ratio of two hydrogen recombination lines (in our case we use Pa$\beta$ and Br$\gamma$) with the theoretical ratio between these two lines. For a case B recombination scenario with a typical electron density of $n_e=10^4\,\mathrm{cm}^{-3}$ and a temperature of $10^4\,\mathrm{K}$ this ratio is Pa$\beta$/Br$\gamma$=5.89 \citep{2006agna.book.....O}. The visual extinction is then given by 
\begin{equation}
A_V = 11.52 \times \log\left(\frac{5.89}{F_{\mathrm{obs,Pa}\beta}/F_{\mathrm{obs,Br}\gamma}} \right).
\end{equation}

Since the $J$-band data cube does not include the region around aperture R7, we cannot determine the extinction in this aperture. For all other apertures, we list the visual extinction $A_V$ in Table \ref{tab:emissionlines}. The derived values of $A_V$ range from around 1.5 to 3.3 and are therefore a bit lower than but still consistent with the typical values from previous studies \citep{1994ApJ...425...72K,1996A&A...313..771K,2012A&A...540A.116R} that range from 2.5 to 5. 

\begin{figure*}
\centering
\includegraphics[height=0.30\linewidth]{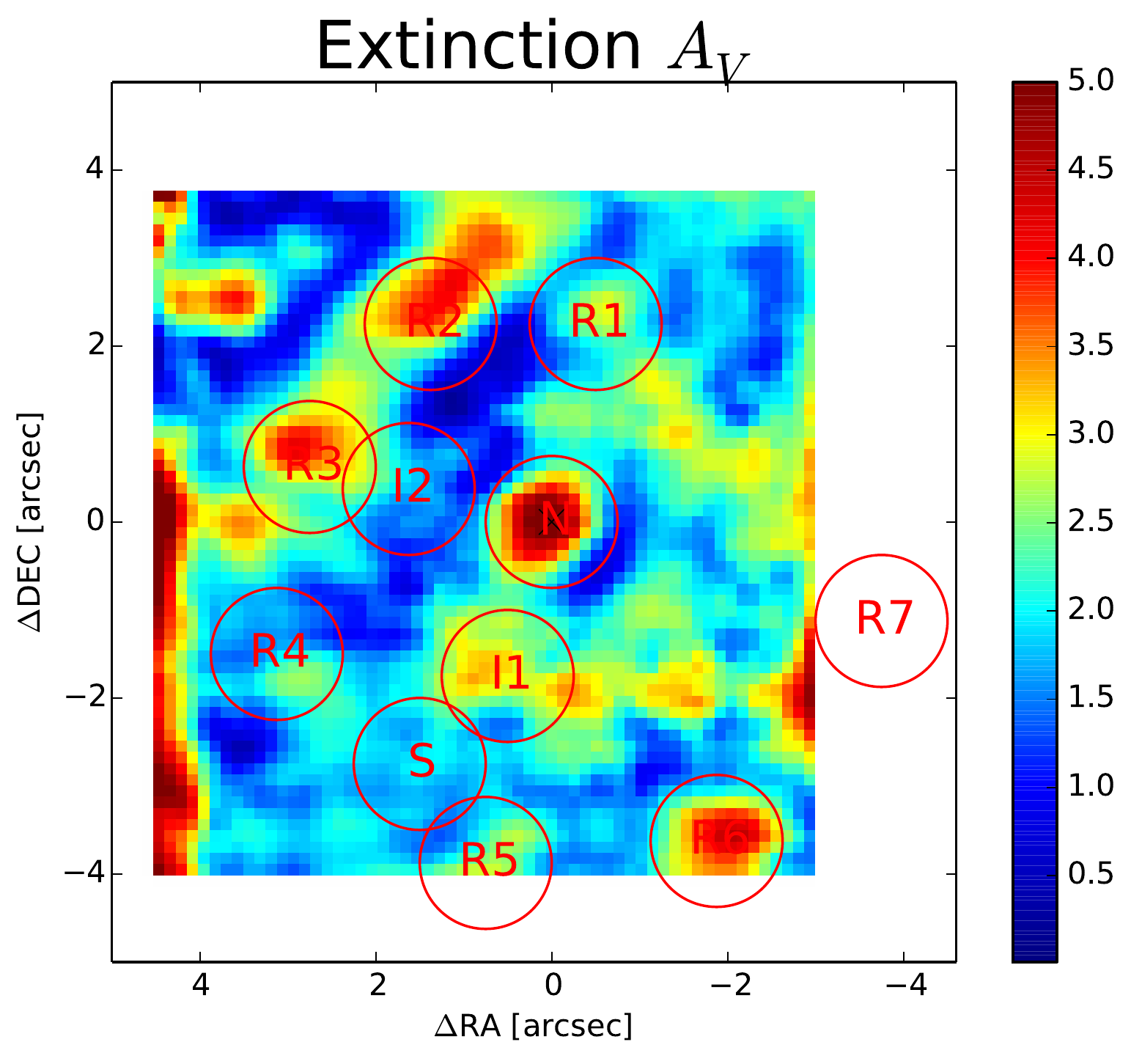}
\includegraphics[height=0.30\linewidth]{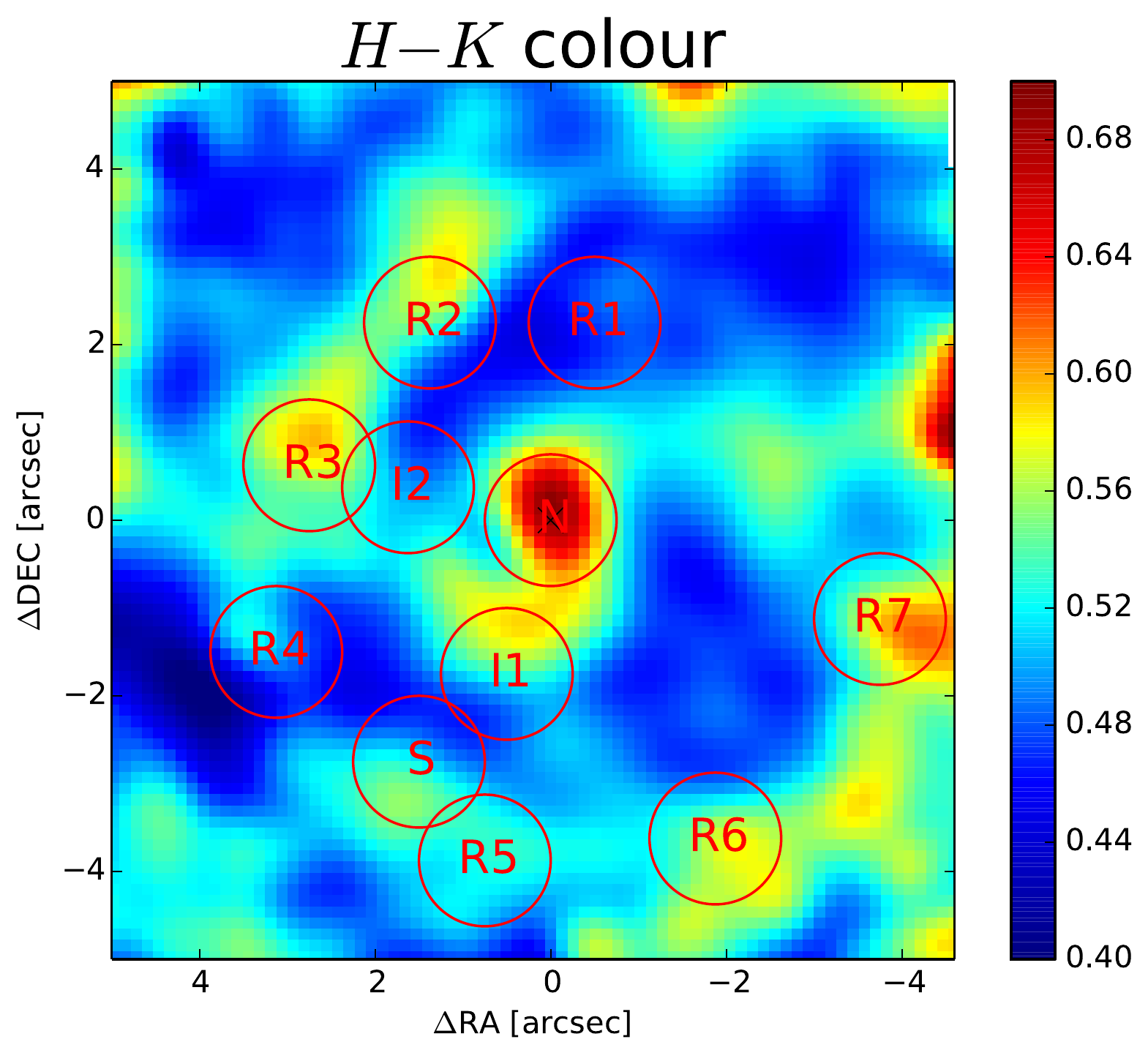}
\includegraphics[height=0.30\linewidth]{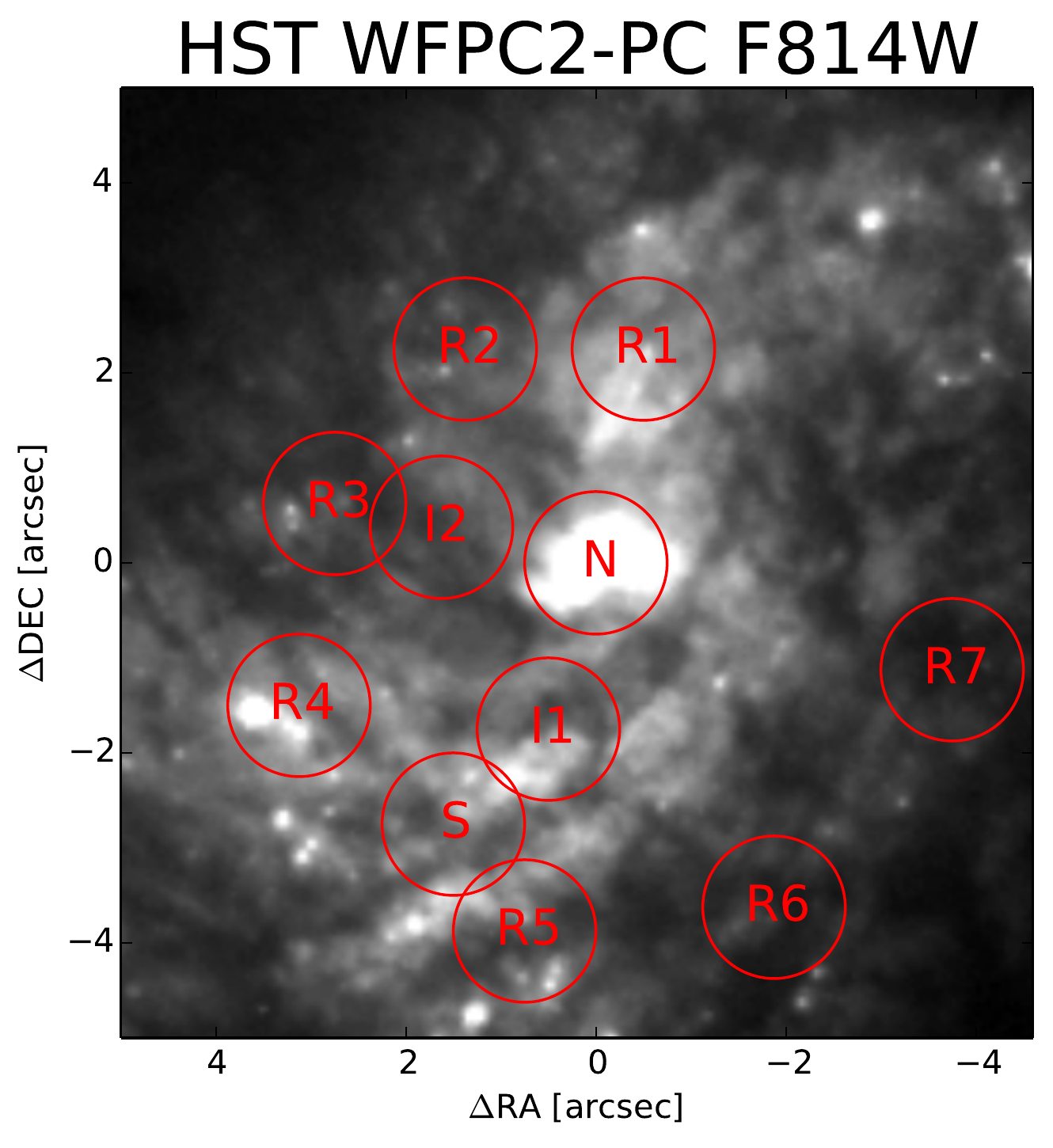}
\caption{\emph{From left to right:} Map of the visual extinction ($A_V$) in mag (derived from the line ratio Pa$\beta$/Br$\gamma$), the $H-K$ colour, and a HST F814W image. The nucleus is denoted by a cross. The apertures, introduced in Sect.~\ref{sec:fluxdistr} are marked with red circles for orientation.}
\label{fig:extinction}
\end{figure*}

In the left-hand panel of Fig.~\ref{fig:extinction}, we present the extinction map ($A_V$ in mag) derived from the Pa$\beta$/Br$\gamma$ ratio as explained above. We see that the extinction in the nucleus and in some of the spots along the ring is quite high, ranging from a few up to 5 mag, indicating that they might be dust-enshrouded starbursts. Furthermore, there are two regions (``I1'' south from the nucleus and the region north of the nucleus) that show enhanced extinction of about 3 mag. The extinction map is consistent with the $H-K$ colour map and the optical HST image (Fig.~\ref{fig:extinction}, middle and right panel), that is dusty regions (e.g. between R2 and R3, and from R5 to R6 and R7) show higher extinction.

\subsubsection{Diagnostic line ratios}

\begin{figure*}
\centering
\includegraphics[width=0.33\linewidth]{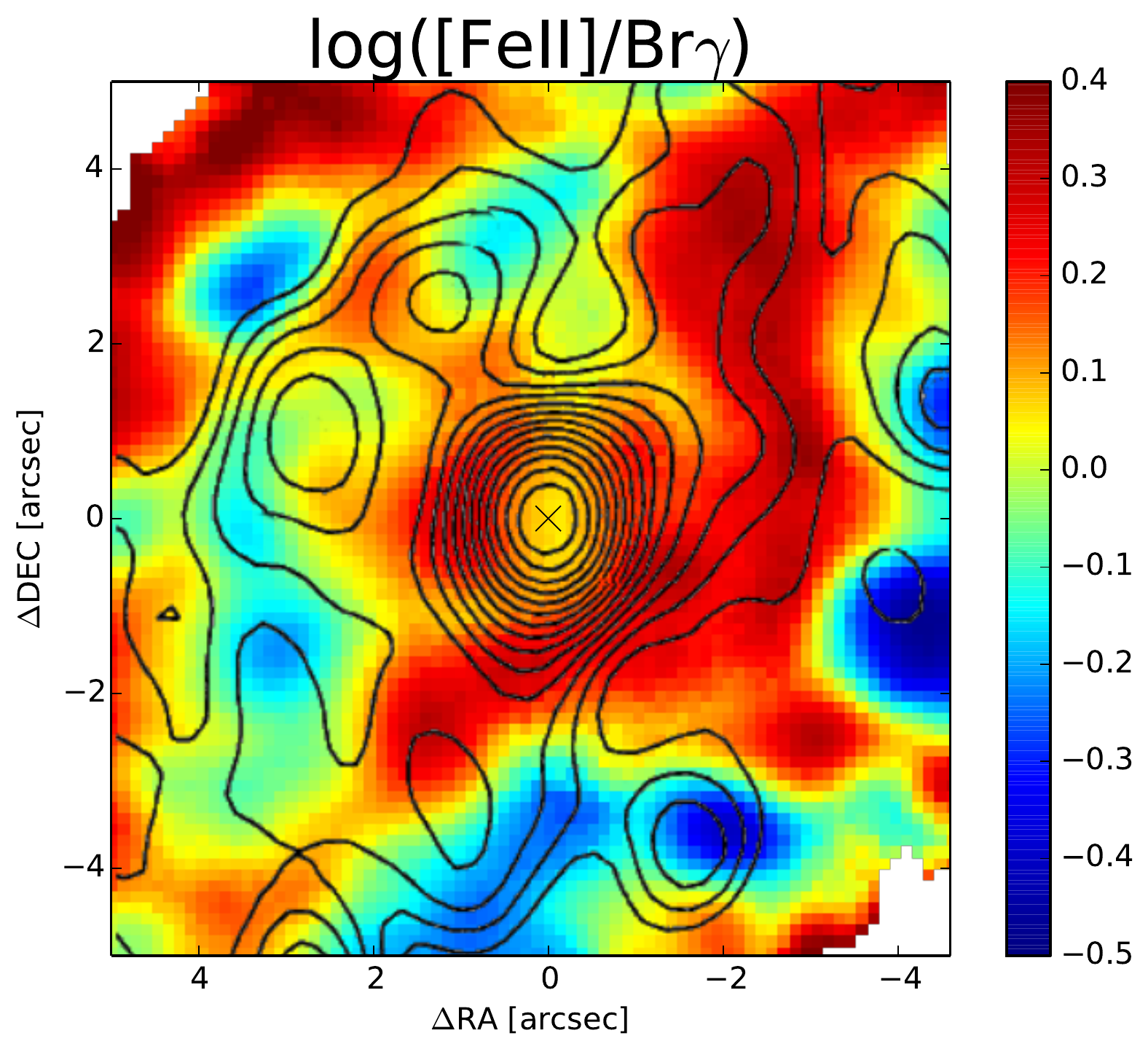}
\includegraphics[width=0.33\linewidth]{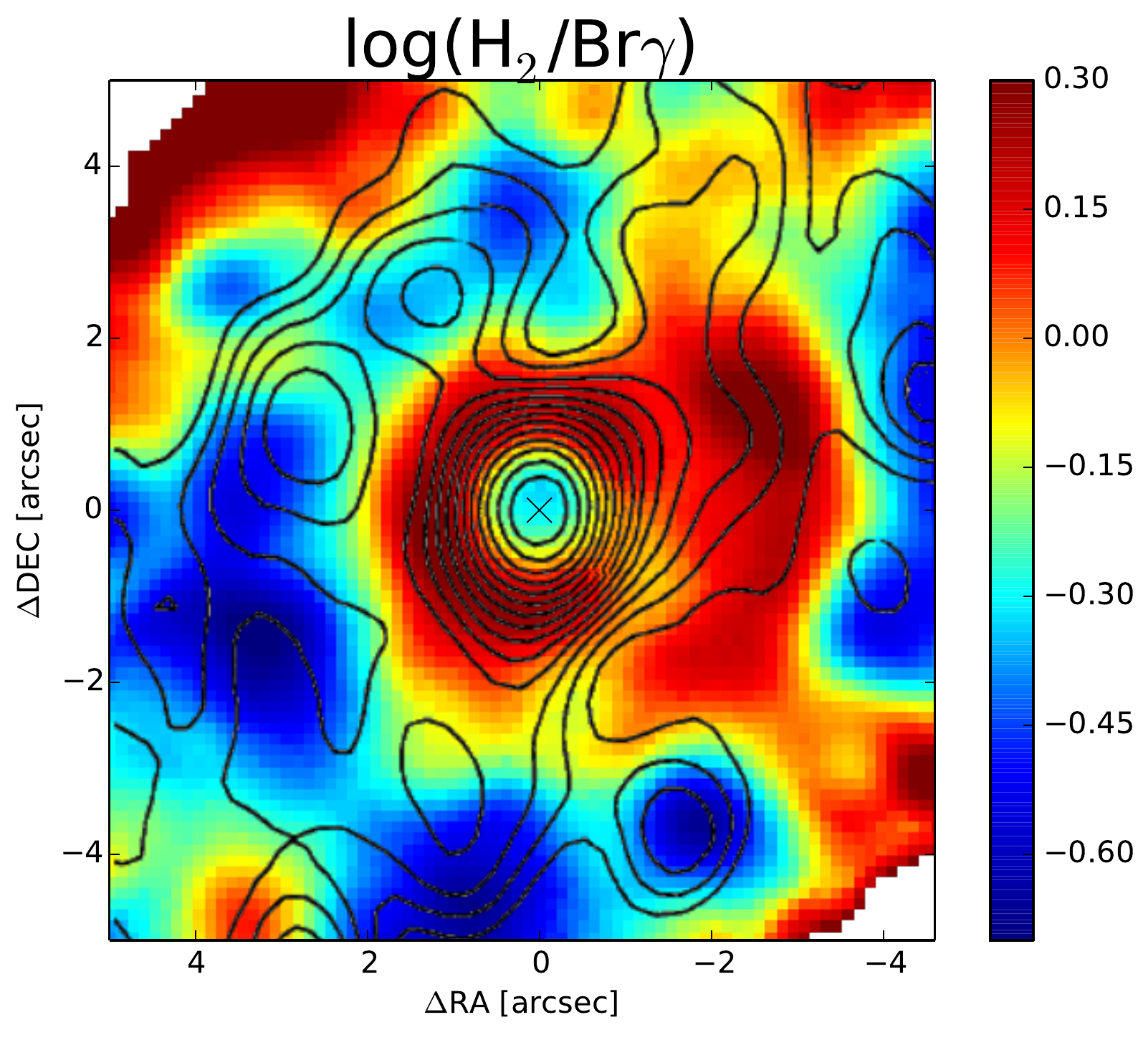}
\caption{\emph{From left to right:} The emission-line ratios $\log([\ion{Fe}{ii}]/\mathrm{Br}\gamma)$ and $\log(\mathrm{H}_2\lambda 2.12\mm/\mathrm{Br}\gamma)$. The nucleus is denoted by a cross. We overlay the contours of the 3.6cm radio-continuum image.}
\label{fig:lineratios}
\end{figure*}

In the left panel of Fig.~\ref{fig:lineratios}, we show a map of the line ratio $\log([\ion{Fe}{ii}]/\mathrm{Br}\gamma)$ that can be used to analyse the [\ion{Fe}{ii}] excitation mechanism. In this and the next plot, we also show the 3.6cm radio-continuum image of \citet{1994MNRAS.268..203C} \citep[taken from ][]{1996A&A...313..771K} as contours. Low values in the line ratio $\log([\ion{Fe}{ii}]/\mathrm{Br}\gamma)$ in the circumnuclear ring (spots R1-R7) are typical for starbursts. On the other hand, in the regions S and I1-2, as well as the region to the north-west of the FOV shocks could be present, since they show rather high values. Both, the suspected star formation regions with low line ratios as well as the suspectedly shocked regions with high line ratios, also show high radio-continuum fluxes.

In the right-hand panel, we show a map of the line ratio $\log(\mathrm{H}_2\lambda 2.12\mm/\mathrm{Br}\gamma)$. In the ring, we find very low ratios down to -0.6, that are indicative for starbursts. In the nuclear region however, we find values that are consistent with AGN excitation. On the western side, between nucleus and ring, high line ratios that are indicative for shock excitation, are found.

\subsection{Stellar kinematics}
\label{sec:res_stellkin}

In order to trace the stellar kinematics, that is the stellar line-of-sight velocity (LOSV) and the stellar velocity dispersion ($\sigma_*$), we fit the region containing the CO band heads at around $2.3\mm$. Since stellar absorption features generally show a lower signal-to-noise (S/N) compared to strong emission lines which we plot emission line flux maps of below, we need to bin the data. Only then, we reach sufficient S/N in all parts of the FOV for a reliable fit. The Voronoi binning method is an adaptive smoothing algorithm which bins the data to a constant S/N while preserving optimal spatial resolution.
We use the \textsc{DPUSER}\footnote{\textsc{DPUSER} was written by Thomas Ott (MPE Garching) as a software package for reducing astronomical speckle data. \url{http://www.mpe.mpg.de/~ott/dpuser/index.html}} implementation of the code originally provided by \citet{2003MNRAS.342..345C}.

For the fit, we use the penalized Pixel-Fitting method \citep[pPXF,][]{2004PASP..116..138C}. As stellar template spectra, we take the Gemini Spectral Library of Near-IR Late-Type Stellar Templates \citep{2009ApJS..185..186W} which consists of 29 giant and supergiant stars with spectral classes from F7III to M3III that have been observed with the IFU of the GEMINI integral-field spectrograph GNIRS.

\begin{figure}
\centering
\includegraphics[width=0.8\columnwidth]{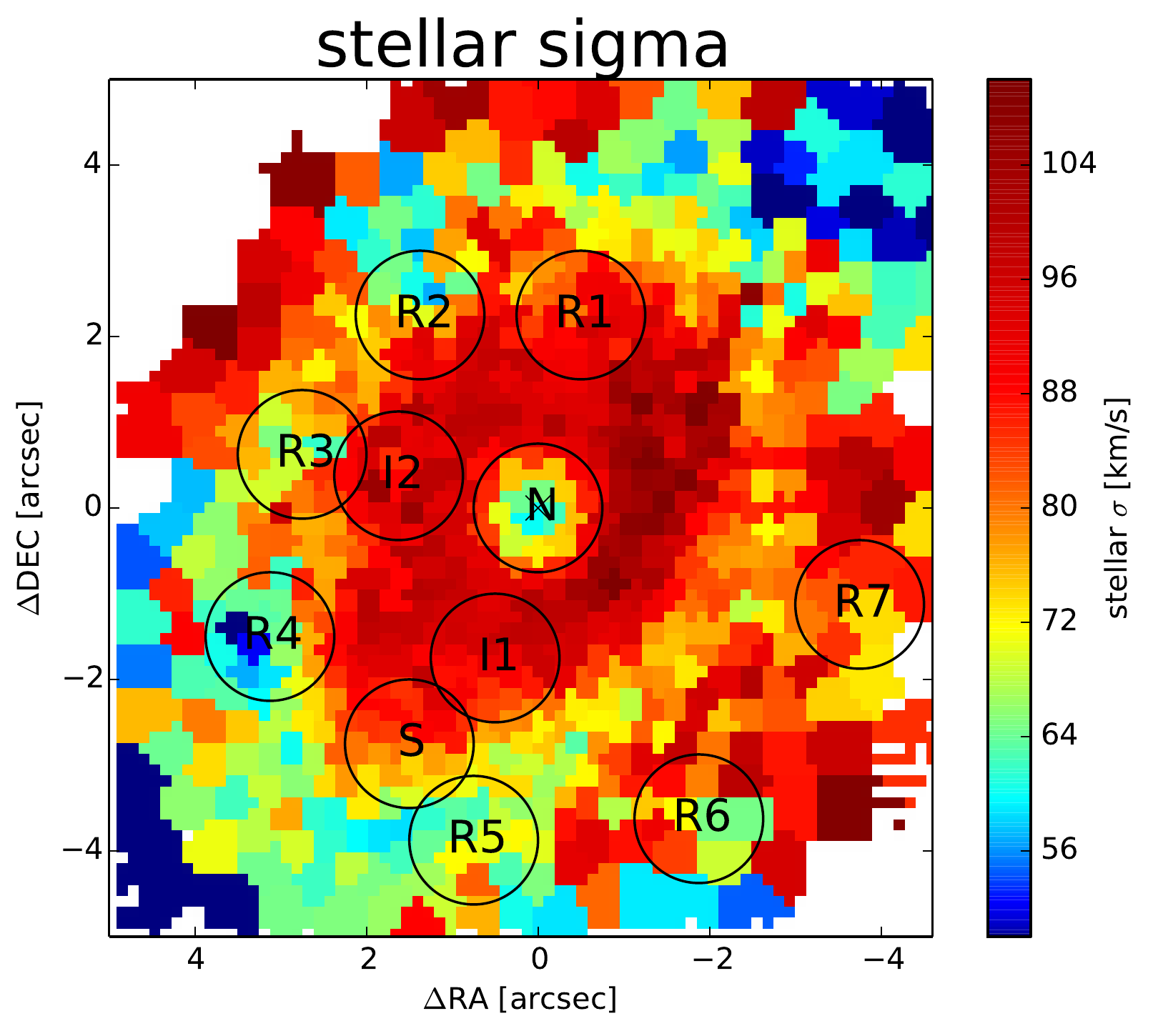}
\caption{Stellar velocity dispersion obtained by fitting the region around the CO band heads at $2.3\mm$.}
\label{fig:stellarvel}
\end{figure}

A map of the stellar velocity dispersion is presented in Fig.~\ref{fig:stellarvel}. It shows values from around $60\kms$ to $110\kms$. The centre shows a ``dip'' with a lower velocity dispersion of only around $60\kms$. Furthermore, the map shows a ring-like structure with lower velocity dispersion of around $70\kms$. This ring-structure has a radius of about $4\arcsec$ ($\approx 240\pc$). Assuming intrinsic circular shape, we get an inclination of $i=50^\circ - 60^\circ$ from the eccentricity.

\begin{figure*}
\includegraphics[width=0.3\linewidth]{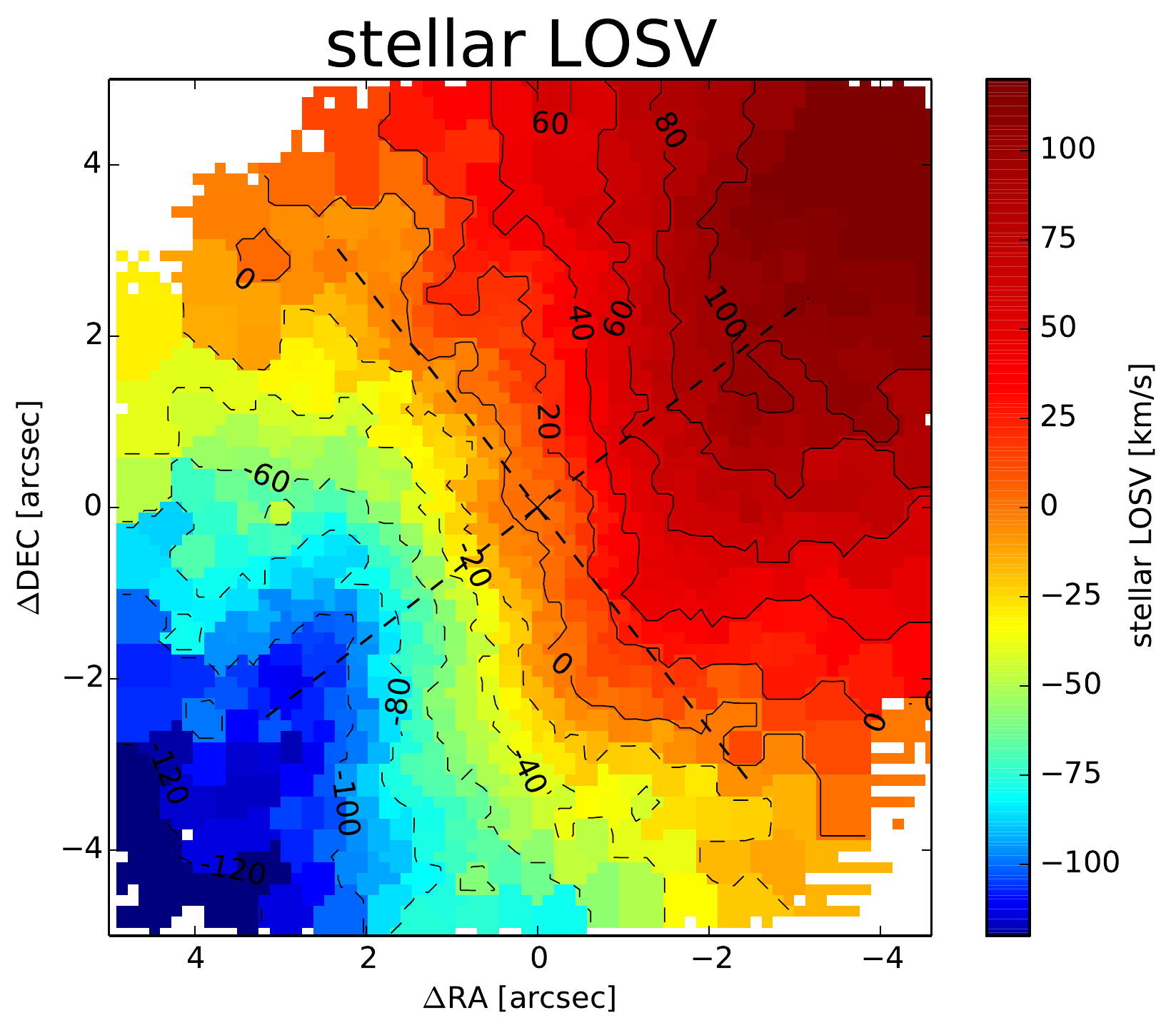}
\includegraphics[width=0.3\linewidth]{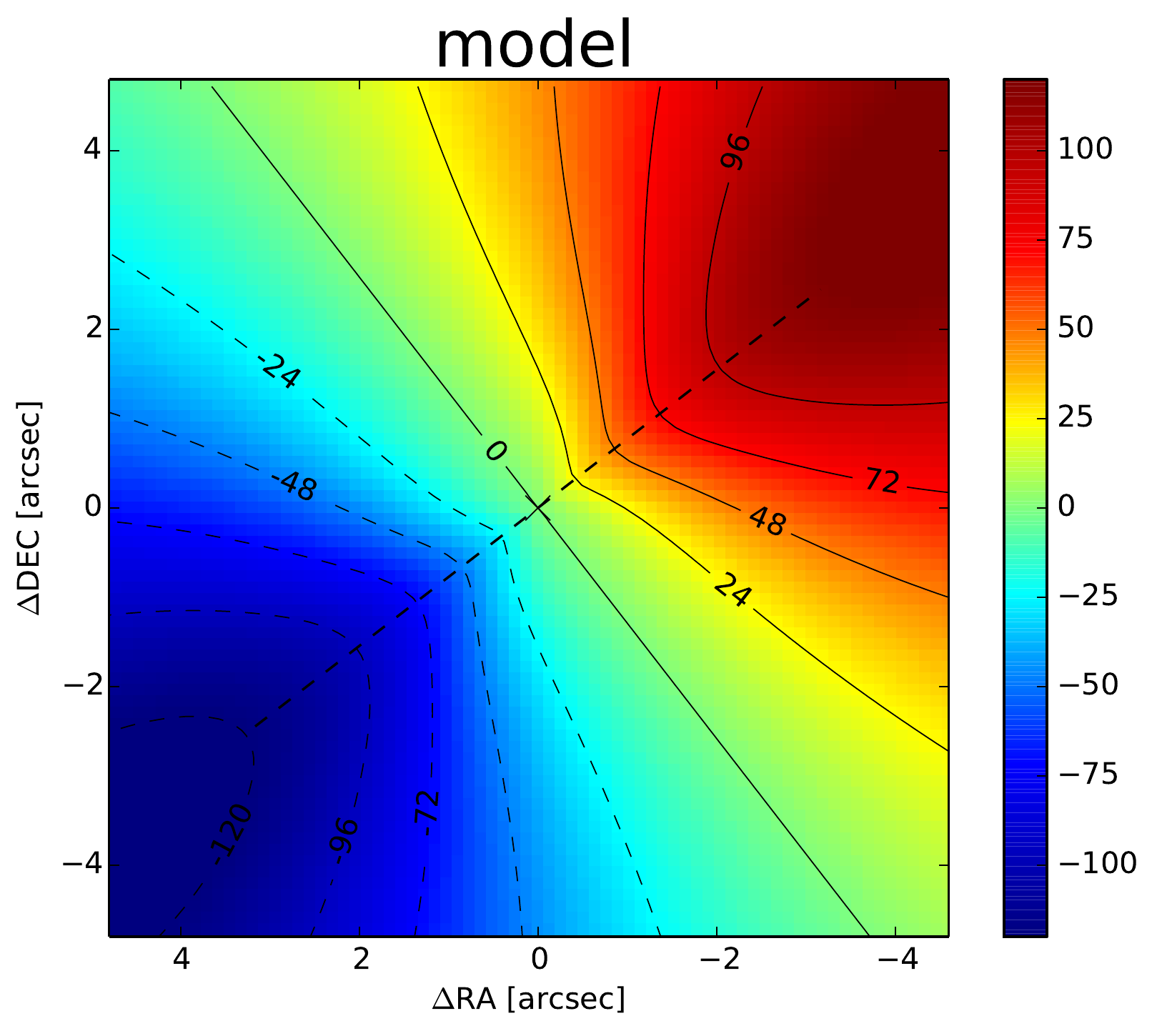}
\includegraphics[width=0.3\linewidth]{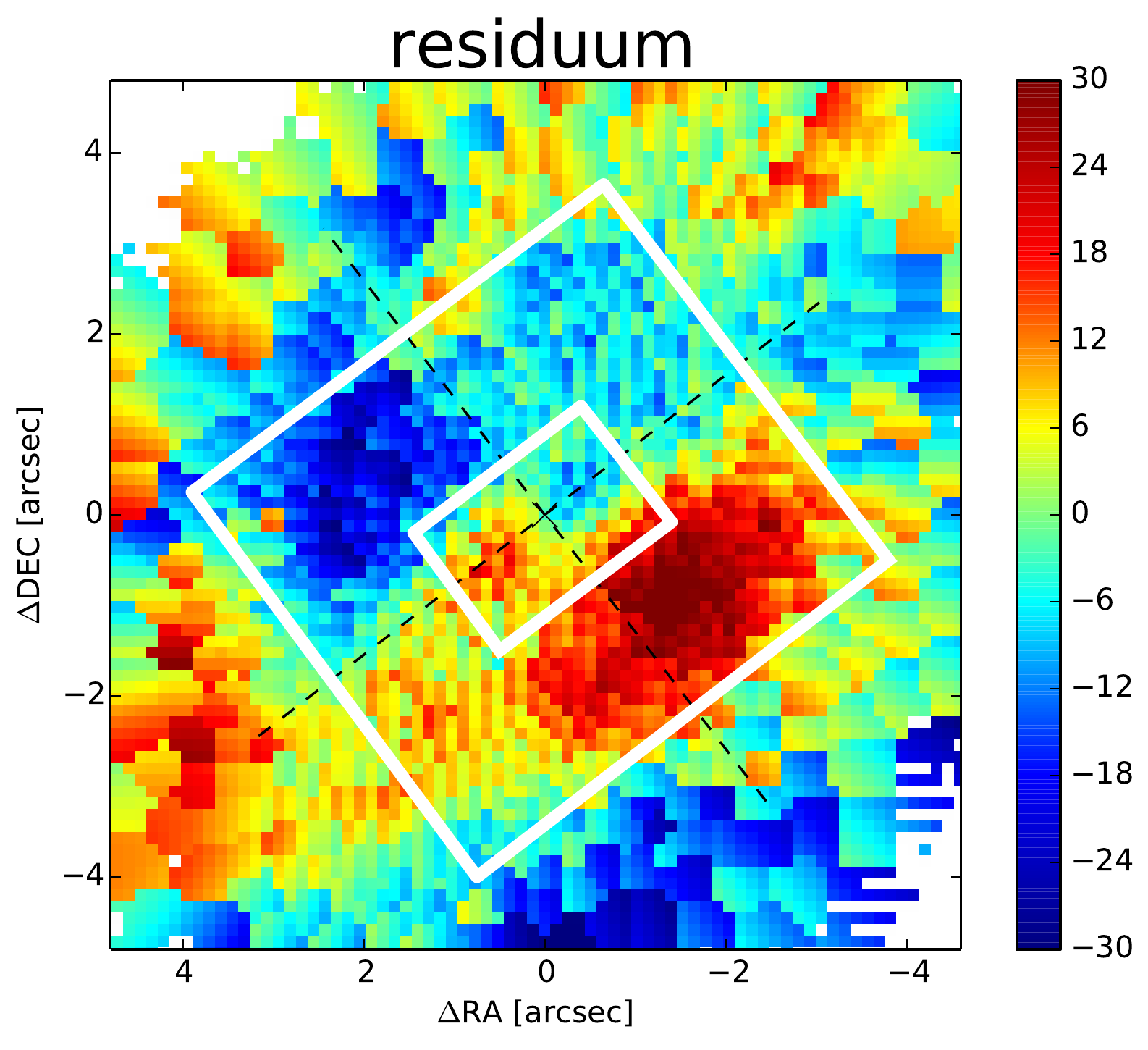}
\caption{\emph{Left:} Stellar line-of-sight velocity obtained by fitting the region around the CO band heads at $2.3\mm$. \emph{Middle:} Rotating disc model fitted to the stellar velocity field. \emph{Right:} Residual between LOSV and model. The residual shows sub-structures that are indicated with white boxes. The cross denotes the peak of the $K$-band continuum emission, the black dashed lines show the line-of-nodes and the zero-velocity line of the model respectively.}
\label{fig:stars_model}
\end{figure*}

The stellar velocity field is shown in the left panel of Fig.~\ref{fig:stars_model}. It shows a clear rotation pattern with redshift to the north-west and blueshift to the south-east. Maximum velocities are of the order of $100-120 \kms$. Assuming that the spiral arms seen in the optical images in Figs.~\ref{fig:ngc1808_cgs} and \ref{fig:hst} are trailing, the near side of the disc is in the south-west. The iso-velocity lines show an S-shape that is indicative for a disturbance of the velocity field, e.g. by a nuclear bar. This feature will be discussed below.

\subsection{Gas kinematics}

The gas kinematics gives insight into the overall dynamics of the host galaxy in comparison to the stellar kinematics. In addition the gas kinematics allows us to find and study zones of enhanced activity (due to the nucleus or due to star formation) via regions of enhanced velocity dispersion or excitation. 

The fit routine that we use in Sect.~\ref{sec:fluxdistr} to derive the emission-line flux distributions also delivers central positions and widths of the emission-line fits from which we can derive the line-of-sight velocity (LOSV) and velocity dispersion ($\sigma$).

\begin{figure*}
\includegraphics[width=0.33\linewidth]{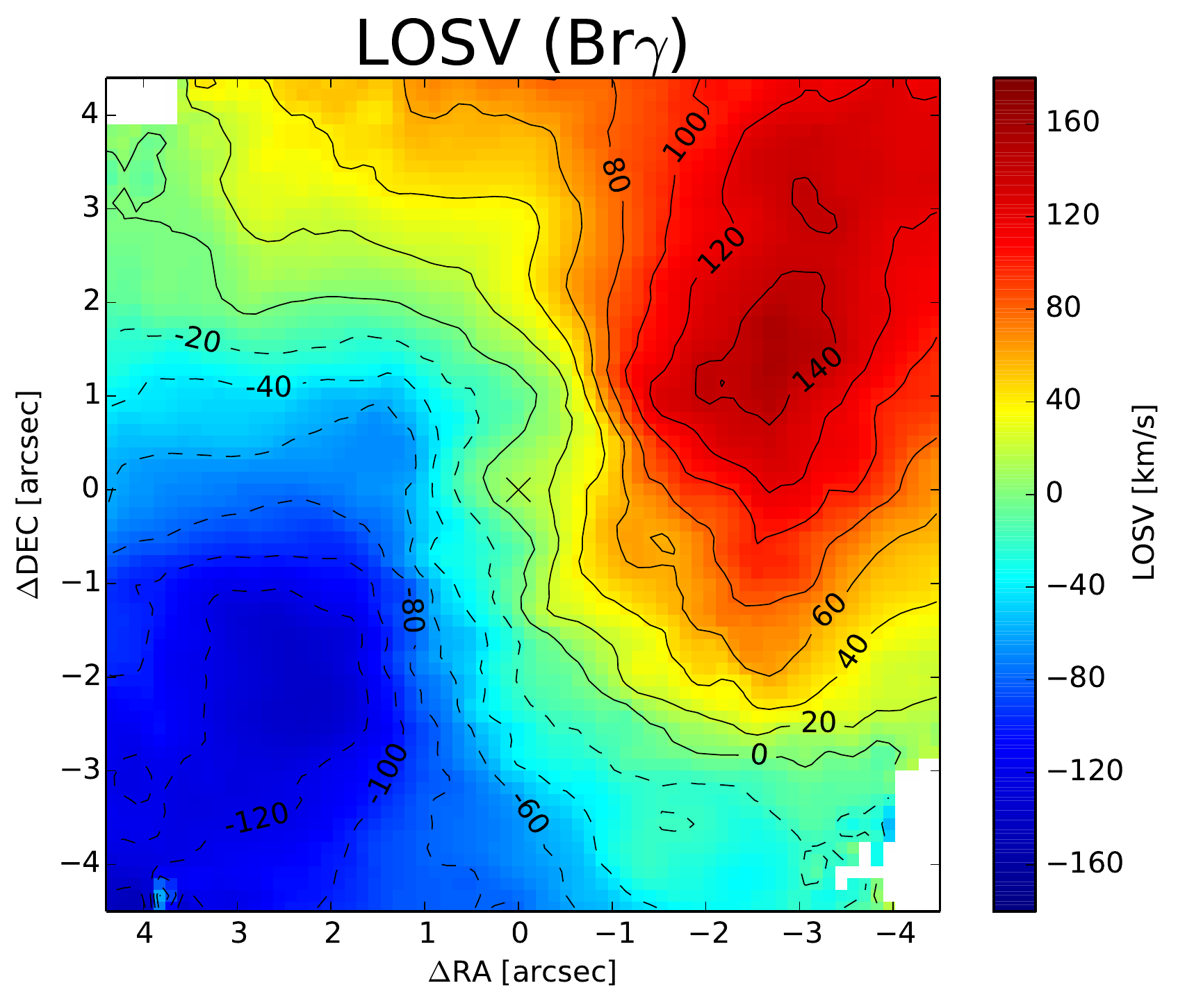}
\includegraphics[width=0.33\linewidth]{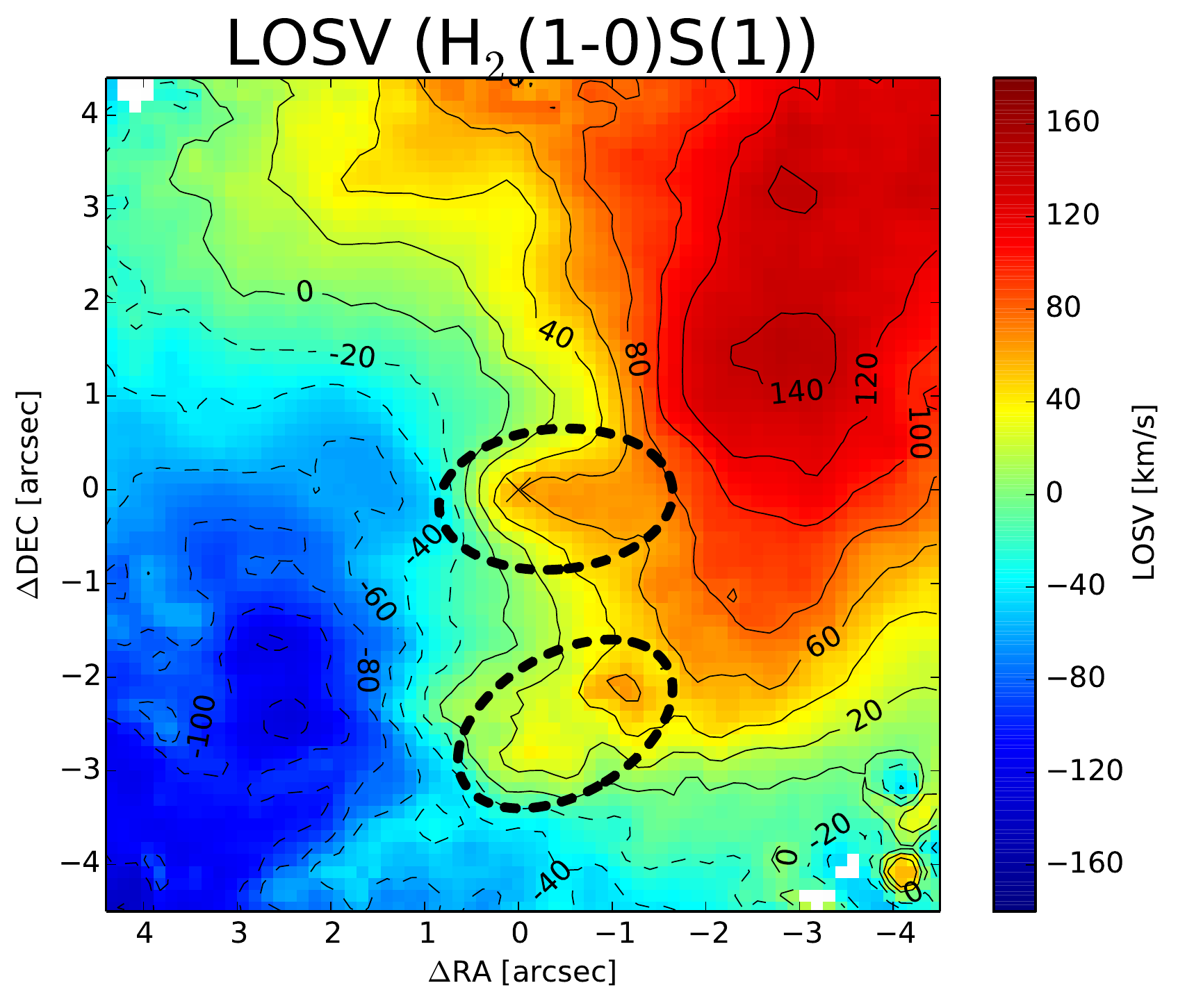}
\includegraphics[width=0.33\linewidth]{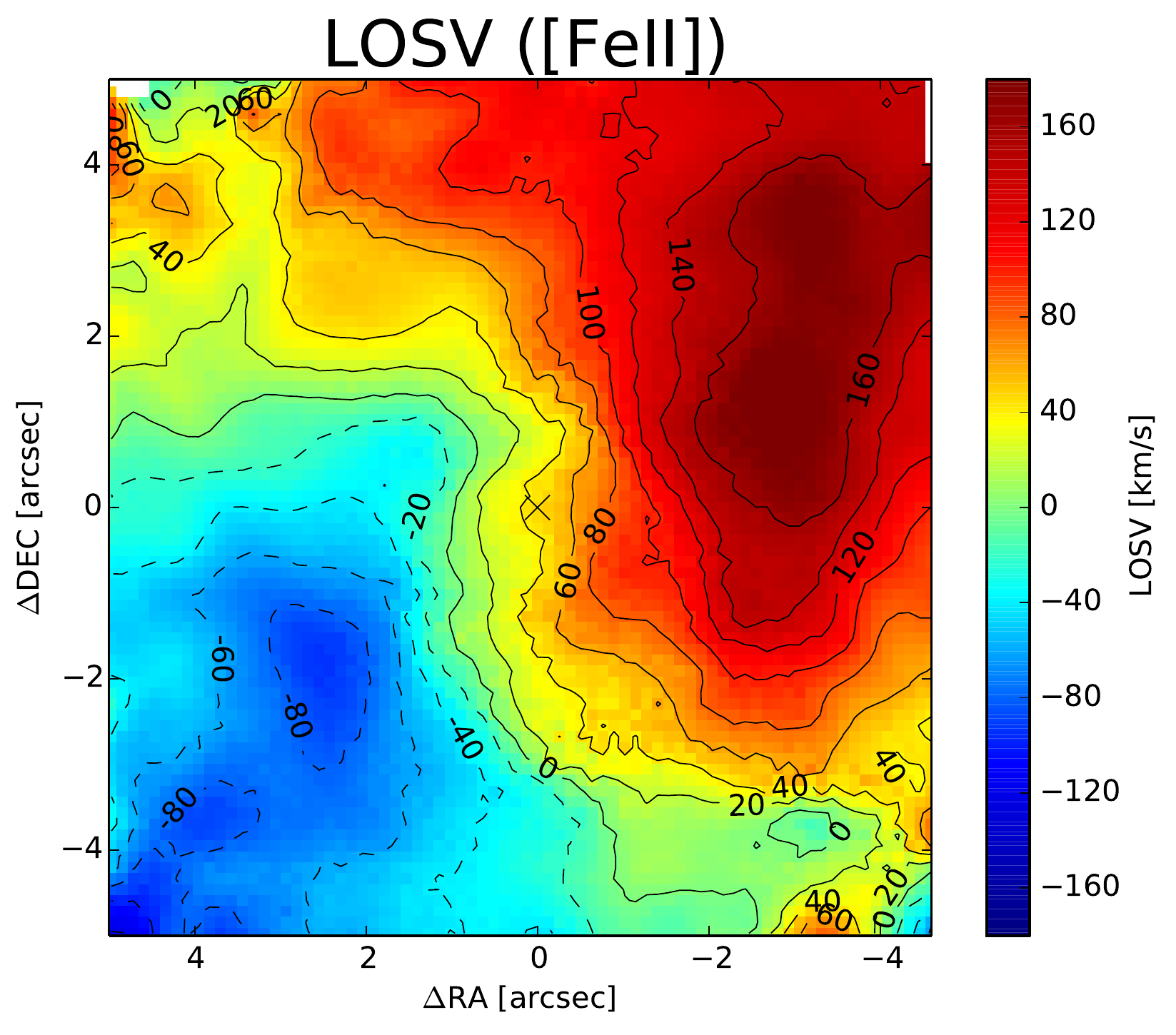}
\caption{Line-of-sight velocity maps of ionised gas (Br$\gamma$), molecular gas (H$_2\lambda 2.12\mm$), and partially ionised gas ([\ion{Fe}{ii}] $\lambda 1.644\mm$). The cross denotes the center, i.e. the peak of the continuum emission. In the $H_2$ LOSV-map, two elongated features are marked with dashed black ellipses.}
\label{fig:losv-maps}
\end{figure*}

In Fig.~\ref{fig:losv-maps}, we show the line-of-sight velocity fields of Br$\gamma$, H$_2\lambda 2.12\mm$, and [\ion{Fe}{ii}] $\lambda 1.644\mm$. The systemic velocity of $1056.8\kms$ (derived in Sect.~\ref{sec:stellkin_disc}) has been subtracted. The maps for Br$\gamma$ and H$_2\lambda2.12\mm$ have been derived from the $K$-band cube that has higher spectral resolution, whereas the [\ion{Fe}{ii}] map has been derived from the $H+K$-cube (but we use the $H+K$-data to cross-check the maps derived from $K$-band cubes). All velocity maps show redshifts to the north-west and blueshift to the south-east, in accordance with the stellar kinematics. However, the velocity fields of the gas show more pronounced S-shaped zero-velocity contours than the stellar field. Furthermore, the strong velocity gradient to the north-west from the nucleus is striking. While the Br$\gamma$ and the [\ion{Fe}{ii}] velocity fields are rather similar, the H$_2$ field differs from those: The centre shows a redshift of around $40\kms$ (after subtraction of the systemic velocity) and shows redshifted gas in elongated features reaching from the west to the nucleus as well as an elongated feature along the north-west to south-east axis, around $3\arcsec$ south from the nucleus. These features will be discussed in a later Section.

\begin{figure*}
\includegraphics[width=0.33\linewidth]{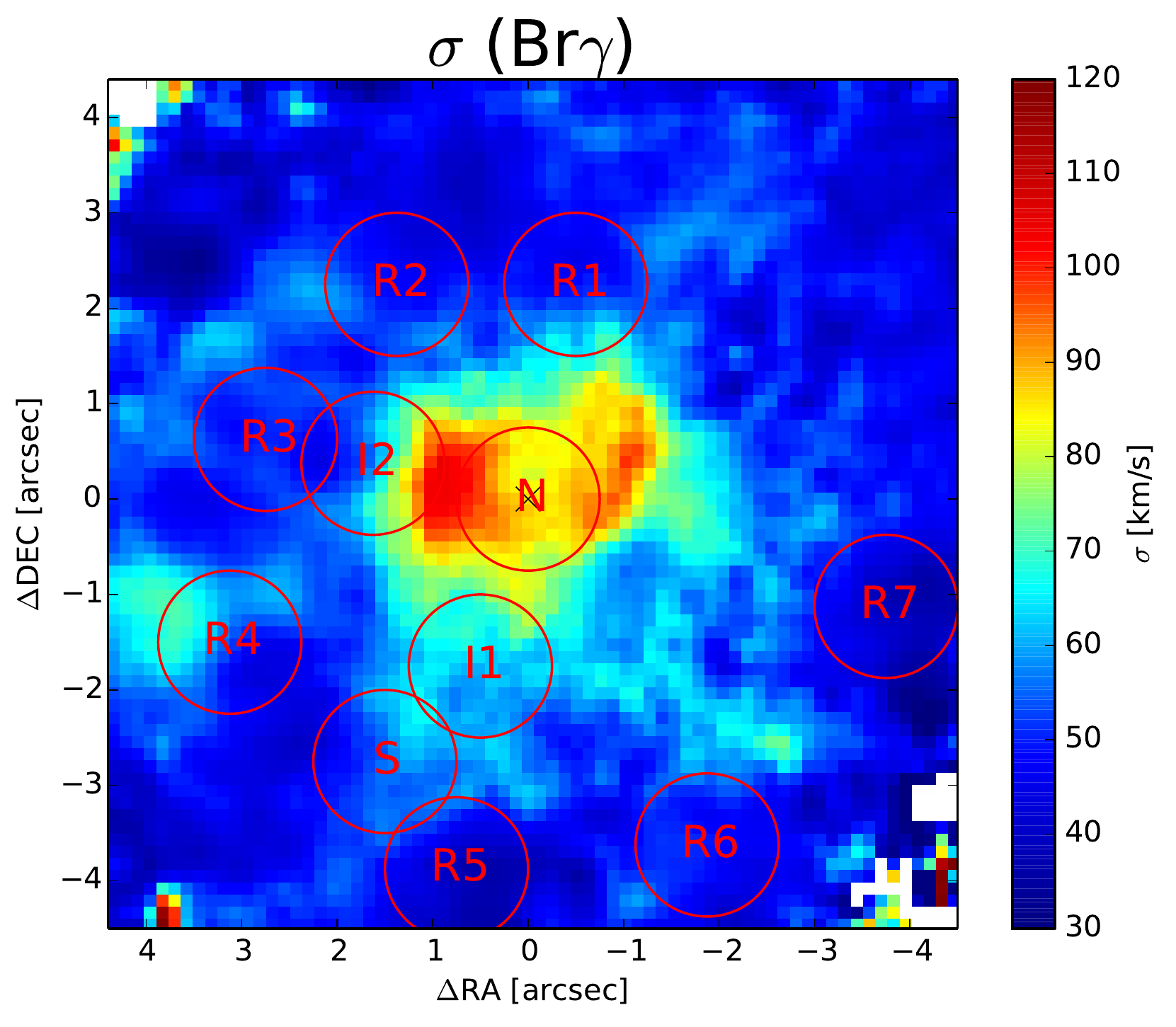}
\includegraphics[width=0.33\linewidth]{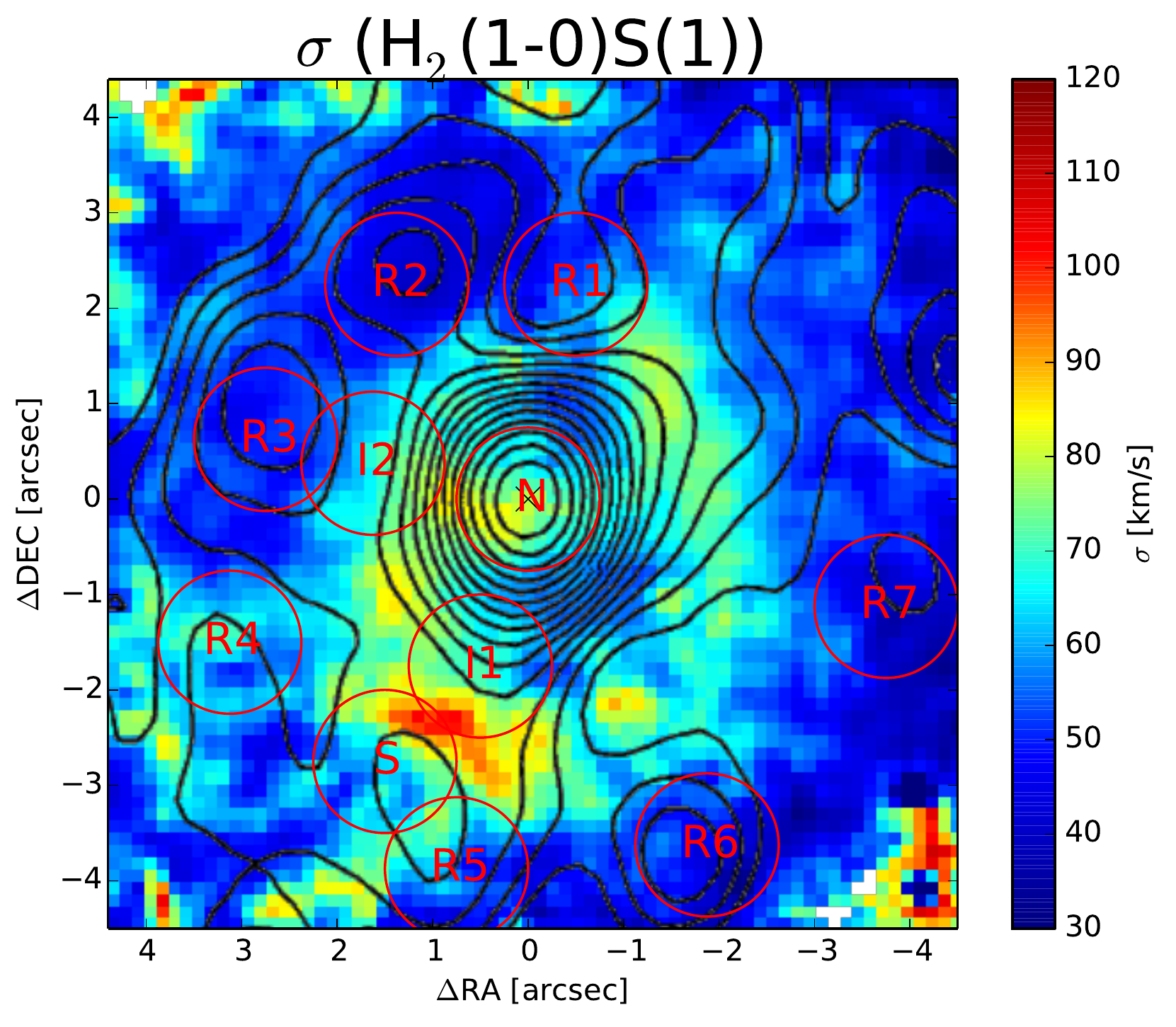}
\includegraphics[width=0.33\linewidth]{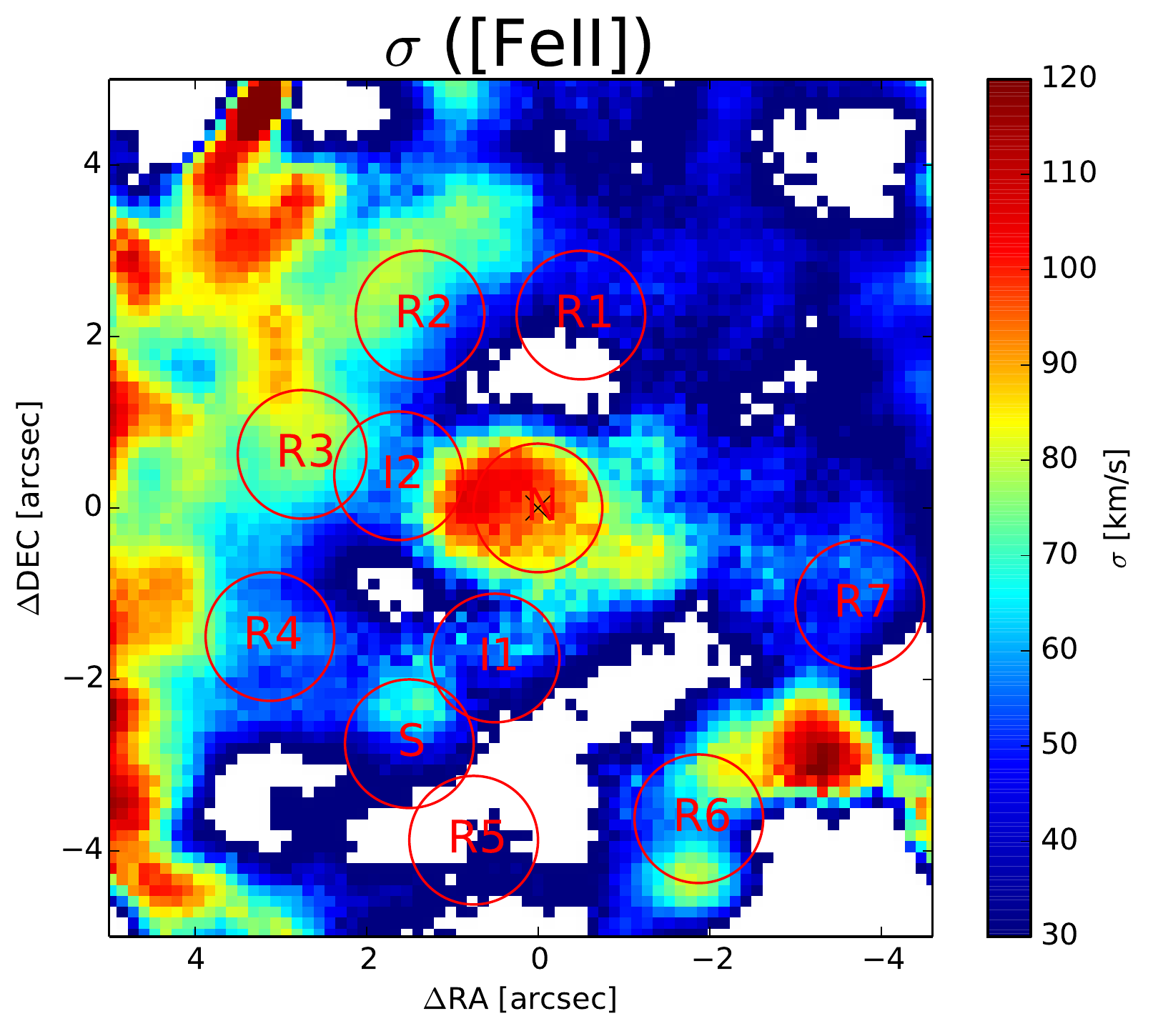}
\caption{Velocity dispersion maps of ionised gas (Br$\gamma$), molecular gas (H$_2\lambda 2.12\mm$), and partially ionised gas ([\ion{Fe}{ii}] $\lambda 1.644\mm$). The cross denotes the center, i.e. the peak of the continuum emission. In the map of molecular hydrogen, we overlay the radio contours.}
\label{fig:sigma-maps}
\end{figure*}

In Fig.~\ref{fig:sigma-maps}, we show the velocity dispersion fields for the same emission lines. The maps have been corrected for the instrumental broadening ($33\kms$ for the $K$-band maps Br$\gamma$ and H$_2\lambda 2.12\mm$, $97\kms$ for the $H+K$-band map [\ion{Fe}{ii}] $\lambda 1.644\mm$). In Br$\gamma$, the velocity dispersion shows maximum values of $\sim 100\kms$ at around $0\farcs9$ to the east and west of the nucleus. In the centre, the velocity dispersion drops down to $\sim 80\kms$. The molecular hydrogen line H$_2\lambda 2.12\mm$ shows the maximum velocity dispersion values of $\sim 80\kms$ in a ring-like shape, with a connection to the centre. This structure lies inside the circumnuclear star forming ring which is visible in Br$\gamma$ and Pa$\alpha$ emission. The highest values are seen in a region to the south-east of the centre, between spots I1 and S with values of $\sim 100\kms$. The velocity dispersion of [\ion{Fe}{ii}] $\lambda 1.644\mm$ peaks close to the centre ($\sim 0\farcs7$ to the east) with values of $\sim 110\kms$. Another peak is seen in the region where also the velocity dispersion of the H$_2\lambda 2.12\mm$ line peaks, between the spots I1 and S ($\sim 70\kms$). Further regions with enhanced velocity dispersion are between the spots R6 and R7, as well as outside the mentioned Br$\gamma$ ring, to the east side, particularly in the north-east. However, these regions are at the edges of the field-of-view and the signal-to-noise is therefore much lower.
Interestingly, we find that the velocity dispersion in the star formation ring and the radio hot spots is relatively low, indicating that the gas in these regions is rather undisturbed.

\section{Discussion}
\label{sec:discussion}

\subsection{Stellar kinematics}
\label{sec:stellkin_disc}

The stellar velocity field, as presented in the left panel of Fig.~\ref{fig:stars_model}, suggests rotation. Assuming circular orbits, we fit a Plummer gravitational potential, in which the velocity distribution is given by:
\begin{equation}
v_r=v_s + \sqrt{\frac{R^2GM}{(R^2+A^2)^{3/2}}} \frac{\sin(i)\cos(\Psi-\Psi_0)}{\left[\cos^2(\Psi-\Psi_0)+\frac{\sin^2(\Psi-\Psi_0)}{\cos^2(i)}\right]^{3/4}}.
\label{eq:plummer}
\end{equation}
where $v_s$ is the systemic velocity, $R$ and $\Psi$ are the cylindric coordinates of each pixel in the projected plane of the sky, $G$ is the gravitational constant, $M$ is the mass inside $R$, $A$ is the projected scale length of the disc, $i$ is the inclination of the disc, and $\Psi_0$ is the position angle of the line-of-nodes \citep[e.g.][]{2006MNRAS.371..170B,2015A&A...583A.104S}. 
The inclination based on the D25 diameter in the blue band is $i=53^\circ$ \citep[][from NED database]{1991rc3..book.....D}. Based on the HST optical data in the center, we estimate an inclination of $i\approx 52^\circ$. In the fit, we fix the inclination to the latter value to lower the number of free parameters. 

The position angle of the line-of-nodes $\Psi = (128\pm1)^\circ$ is lower than the photometric position angle $133^\circ$ \citep[][from NED database]{1991rc3..book.....D}. All parameters have typical values for nearby galaxies, e.g. compared to the galaxies analysed in \cite{2006MNRAS.371..170B}. The systemic velocity $v_s=(1056.8\pm 0.4)\kms$ is slightly higher than $995\kms$ derived from \ion{H}{i} measurements \citep{2004AJ....128...16K}. We subtract the systemic velocity from the line-of-sight velocity maps of gas and stars in Figs.~\ref{fig:stellarvel} and \ref{fig:losv-maps}.

The stellar velocity field in Fig.~\ref{fig:stars_model} shows a clear deviation from pure rotation. When subtracting the rotating disc model from the stellar velocity field, we find systematic residuals that are indicated with white boxes. On scales of $\sim 3 \arcsec$, we see peaks in redshift in the south-west and in blueshift in the north-east. These structures occur because the S-shaped zero-velocity line is not included in the simple rotational model. S-shaped zero-velocity lines are a well-known signature for non-circular velocity fields \citep[e.g.][]{2004A&A...414..857C,2006MNRAS.365..367E,2011MNRAS.411..469R}, that are disturbed e.g. by a bar potential (the position of the nuclear bar is indicated in the $K$-band continuum image in Fig.~\ref{fig:continuum}).

Central drops in the velocity dispersion can indicate decoupled dynamics in the nucleus, e.g. by a nuclear disc. \cite{2006MNRAS.369..529F} find that a significant fraction of spiral galaxies in the SAURON survey show this feature and kinematically decoupled components are actually a common phenomenon. A central $\sigma$-drop in NGC 1808 was found by \cite{2001A&A...368...52E} in ISAAC-longslit spectra and is the first time visible in its full two-dimensional extent in our maps (Fig.~\ref{fig:stellarvel}). The most likely scenario, supported by numerical simulations and model calculations, is that these velocity drops are produced by young stars that form from cold gas with low velocity dispersion compared to the underlying old stellar population. Since they are more massive and brighter than the old stars, they will dominate the observed kinematics at near-infrared wavelengths \citep{2001A&A...368...52E,2003A&A...409..469W,2008A&A...485..695C}. This scenario is indeed consistent with our observations of young stellar clusters in the centre (see in particular Sect.~\ref{sec:diagndiagram} and \ref{sec:sfrates}). In addition to the central $\sigma$-drop, the circumnuclear ring (particularly the apertures R2-R5) show a lower velocity dispersion than the surroundings. This is indicative for younger stellar population in these regions \citep[e.g.][]{2011MNRAS.416..493R,2014MNRAS.438.2036M}.

Tight relations between the stellar velocity dispersion of the bulge and the mass of the central black hole have been found and interpreted in the context of galaxy evolution \citep{2000ApJ...539L...9F,2000ApJ...539L..13G}. Here, we use the stellar velocity dispersion $\sigma_*$ to get a rough estimate of the mass of the central black hole. Running \textsc{Ppxf} on the spectrum integrated over an aperture of $r=4\arcsec$ (using the settings described in Sect.~\ref{sec:res_stellkin}), we obtain $\sigma_* = (105\pm5)\kms$. This value is lower than the value in the HyperLeda database\footnote{http://leda.univ-lyon1.fr/ \citep{2014A&A...570A..13M}} $\sigma_* = (141.3\pm 8.7)\kms$ and the values derived from fitting the CaT by \cite{2005MNRAS.359..765G}, $\sigma_* = (119\pm 6)\kms$ using direct fitting method and $\sigma_* = (129\pm 4)\kms$ using cross-correlation method. However, \cite{2015MNRAS.446.2823R} find a a systematic lower $\sigma_*$ for spiral galaxies when measured at the $K$-band CO band heads ($\sigma_\mathrm{CO}$), compared to optical measurements ($\sigma_\mathrm{opt}$). Using their best fit, we find that our measurement corresponds to $\sigma_\mathrm{opt} \approx 135\kms$ which is consistent with the optical measurements. Since most $M_\mathrm{BH}-\sigma_*$ relations in the literature are based on optical measurements, we use the latter value for the black hole mass estimates. Applying different relations, we get the following estimates: $\log(M_\mathrm{BH}/M_\odot) = 7.7$ \citep{2013ARA&A..51..511K}, $\log(M_\mathrm{BH}/M_\odot) = 7.4$ \citep{2009ApJ...698..198G}, $\log(M_\mathrm{BH}/M_\odot) = 7.3$ and $\log(M_\mathrm{BH}/M_\odot) = 7.0$ \citep[][using the relation for non-barred or barred galaxies respectively]{2013ApJ...764..151G}. The difference between the values seems large but is within the scatter of the BH mass - host scaling relations (which is between 0.29 and 0.44 dex for the cited relations). We conclude that the black hole mass is of the order of a few $10^7\,M_\odot$.

However, two caveats have to be kept in mind: First, due to the complicated $\sigma$ structure, in particular the central $\sigma$-drop, it is not completely clear which value is representative for the bulge's stellar velocity dispersion. One could argue that the central pixels have to be excluded due to their decoupled kinematics. However, the extent of the structure is relatively small and does not significantly affect the (luminosity weighted) average. Second, central drops in the stellar velocity dispersion as well as the dusty structure in the centre (see Fig.~\ref{fig:hst} and Fig.~\ref{fig:extinction}) are commonly seen as indicative of a pseudobulge structure that is the result of secular evolution \citep[e.g.][]{2016ASSL..418...41F}. It is under discussion whether pseudobulges follow $M_\mathrm{BH}$-host correlations \citep{2011Natur.469..374K} which would put into question our black hole mass estimates.

\subsection{Gaseous kinematics}

\begin{figure*}
\centering
\includegraphics[width=0.3\linewidth]{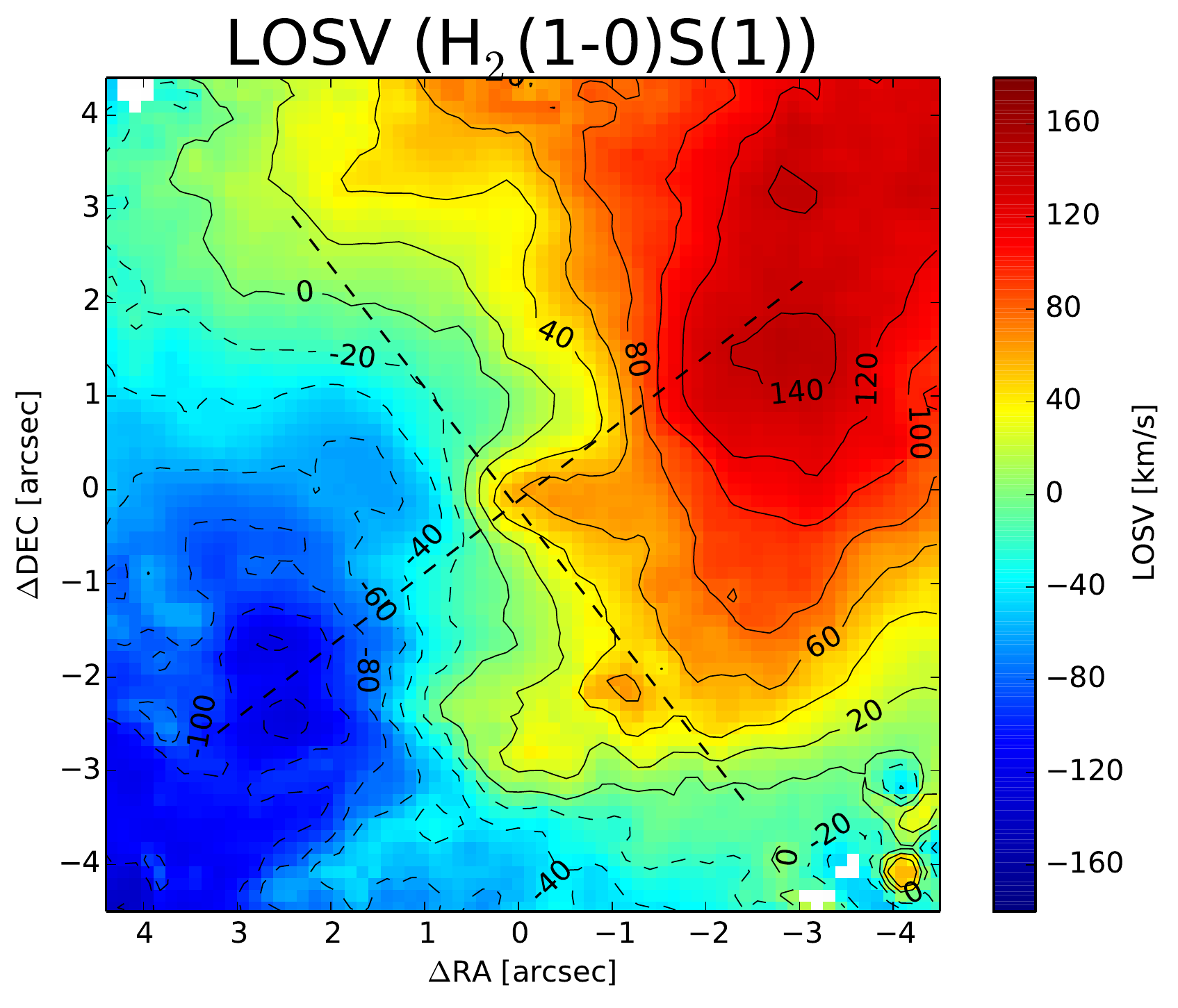}
\includegraphics[width=0.3\linewidth]{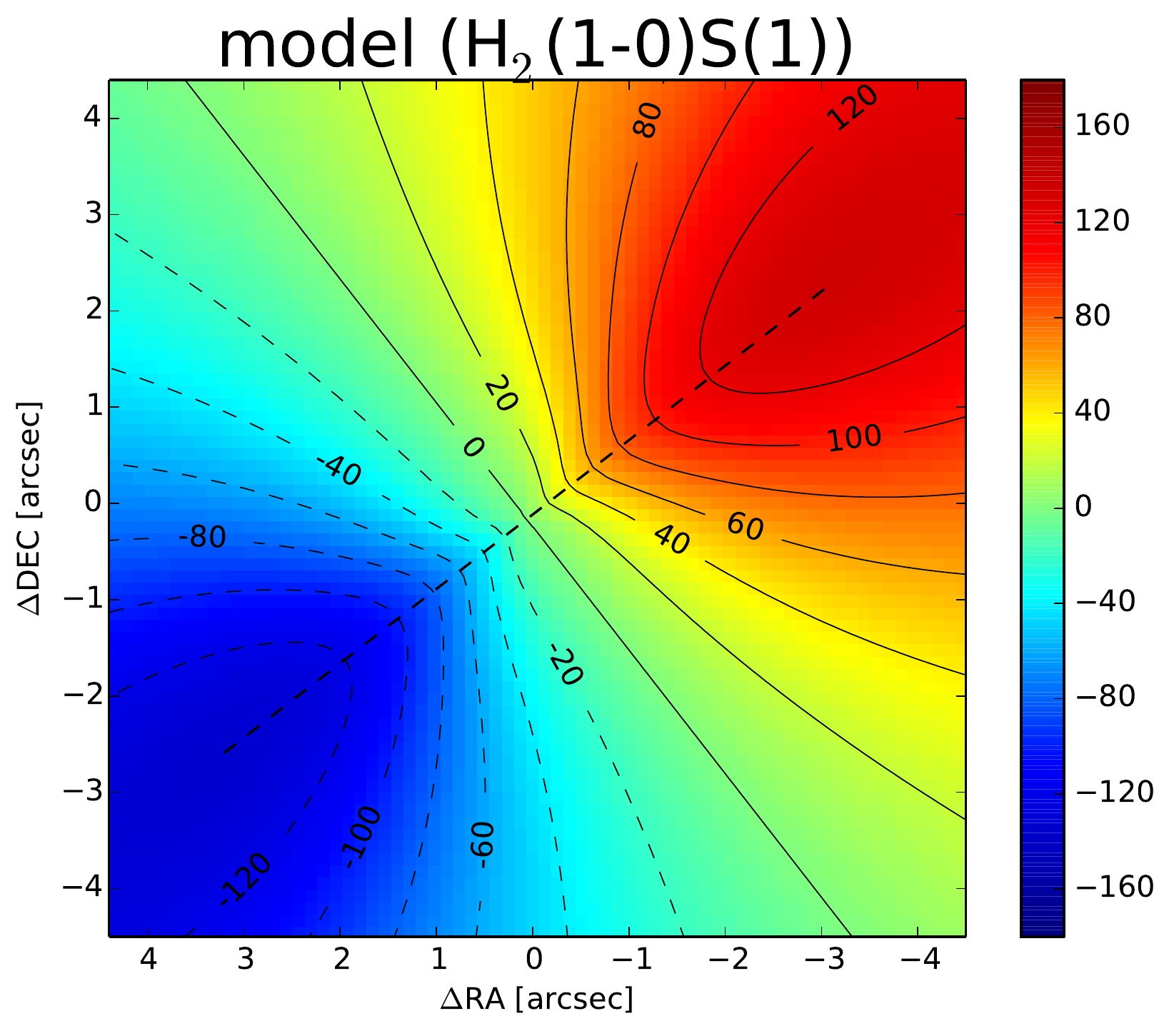}
\includegraphics[width=0.3\linewidth]{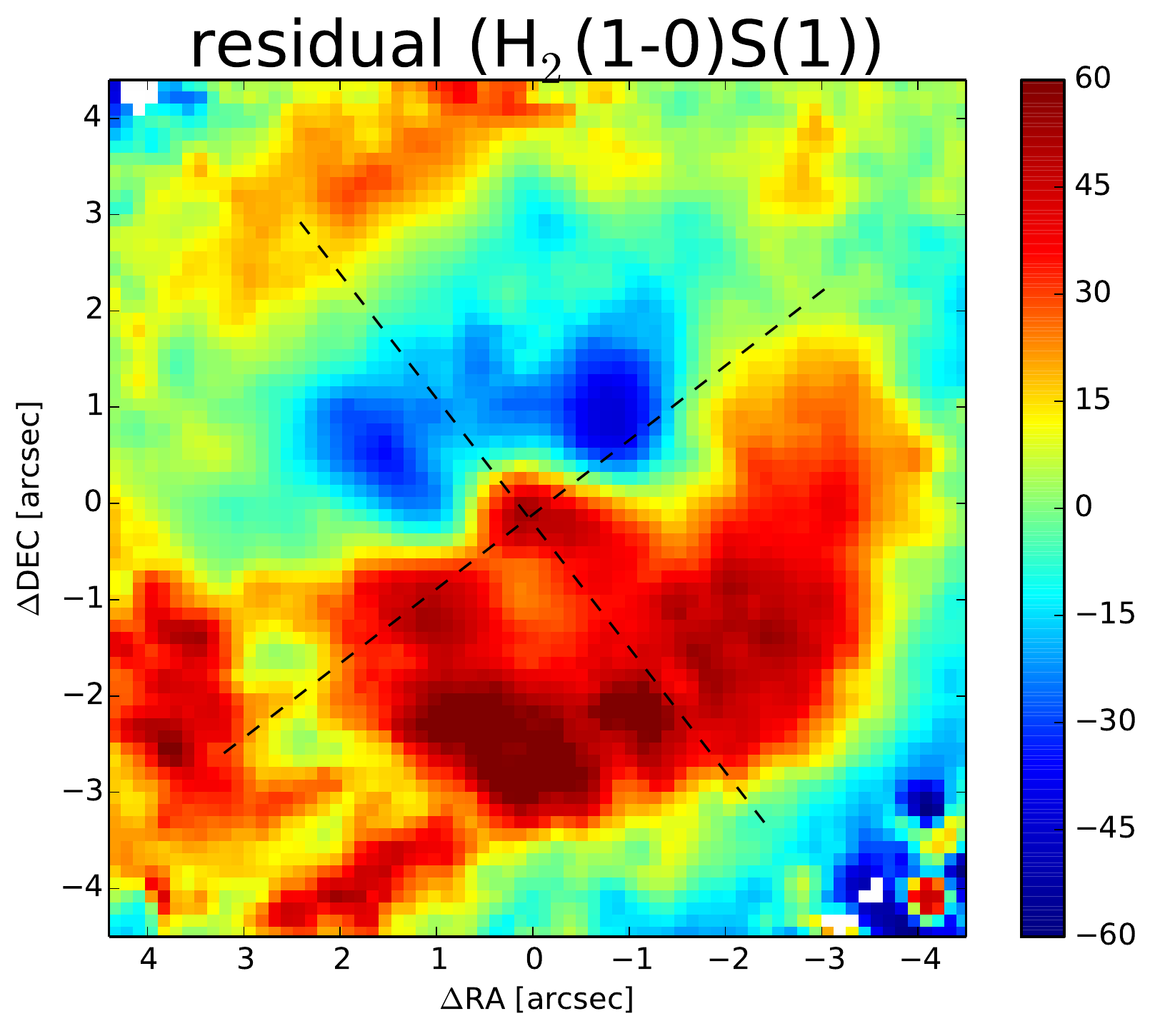}
\caption{\emph{From left to right:} Line-of-sight velocity map of H$_2$ emission line, pure rotational model, and residual.}
\label{fig:mgas-model}
\end{figure*}

The velocity fields of the gas in the central $200\pc$ show more obvious deviations from pure rotation than the stellar velocity field (see Fig.~\ref{fig:losv-maps}). In particular, a strong S-shaped zero-velocity line is striking. A likely explanation for this feature are non-circular motions, such as oval flows, secondary bars, or warps, with their axis not parallel to the symmetry axes of the large-scale velocity field \citep[e.g.][]{2002MNRAS.329..502M,2014MNRAS.438.2036M,2014ApJ...792..101D}. To subtract the regular velocity field, we performed the same fits as in Sect.~\ref{sec:stellkin_disc}. We notice that the gas velocity fields have a slightly higher redshift than the stars and furthermore, the scale length is smaller, which means that the gas is more centrally concentrated, particularly the [\ion{Fe}{ii}] emission. Despite the clear twist in the centres, the position angles of the line-of-nodes of gas and stellar velocity fields agree on larger scales (within a few degrees) which shows that the velocity fields are aligned. To better compare stellar and gaseous kinematics, we then fix most of the parameters (centre, inclination, position angle) to the results of the stellar fit. In Fig.~\ref{fig:mgas-model}, we show the resulting model and the residual based on the $H_2$ kinematics. The deviations from circular motions visible in the residua show similar structures and most important the same sign (though not the same amplitude) which indicates that they are produced by the same phenomenon (probably non-circular motions due to the bar which are amplified in the gas compared to the stars).

\subsubsection{Inflow or outflow?}

\begin{figure}
\centering
\includegraphics[width=\columnwidth]{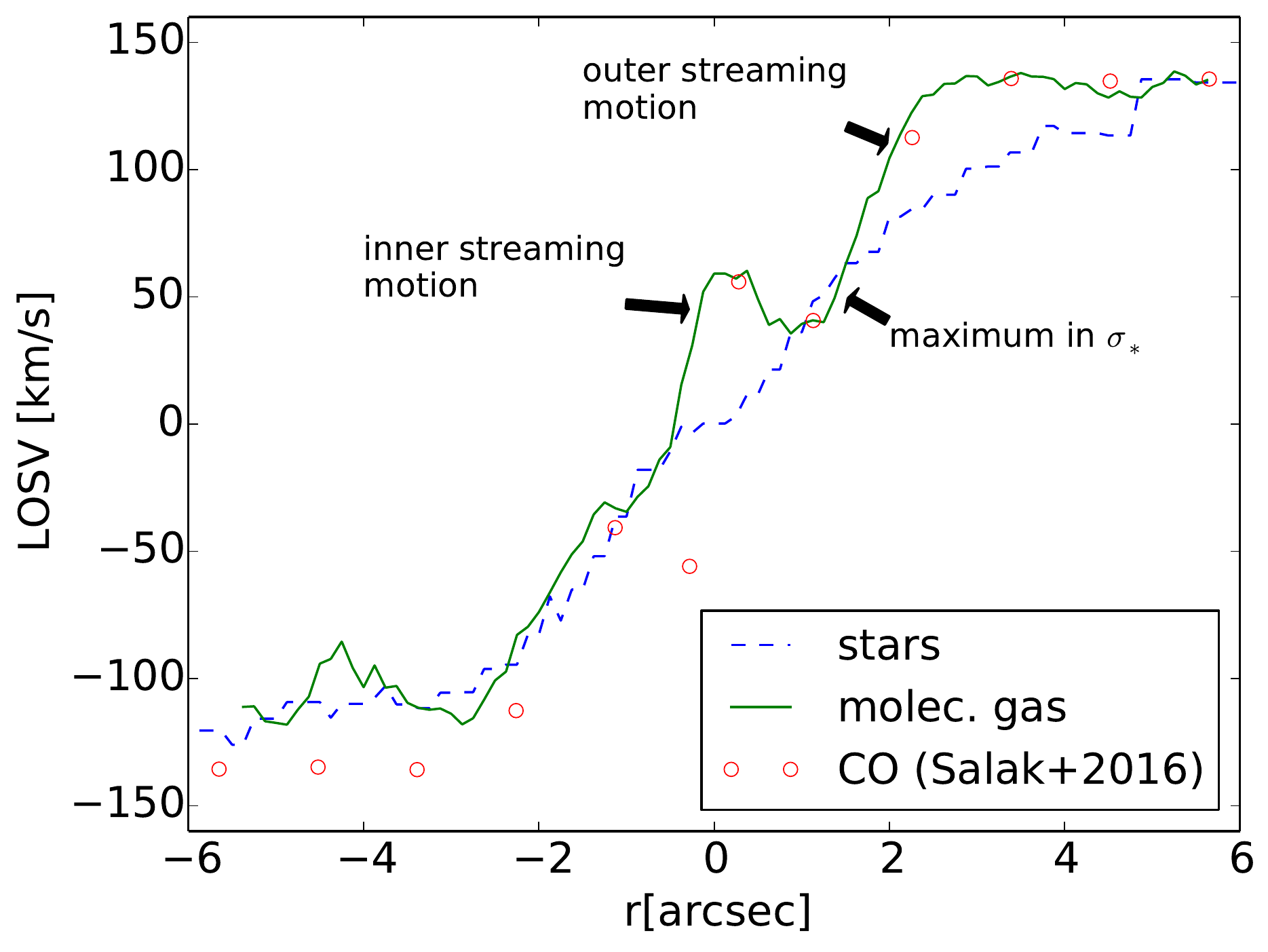}
\caption{LOSV cut from south-east to north-west along the line-of-node. The rotation curves of stars and hot molecular gas are derived from the SINFONI data. Data points for cold gas (CO) are added from \cite{2016ApJ...823...68S}.}
\label{fig:LOSV-cut}
\end{figure}

In the following we compare stellar and gaseous kinematics. It is known that in many galaxies the rotation velocity of gas has a higher amplitude than that of the stars. This comes from the fact that a significant fraction of the kinetic energy of the stars can go into random motions (velocity dispersion), while gas is more confined to the galaxy plane. We therefore compare the rotation curve (Fig.~\ref{fig:LOSV-cut}) of gas and stars. We find that in the south-east, the rotation curves do not show the aforementioned behaviour but are well aligned. In the north-west, two bumps are seen that coincide with features in the residuals. At the radius of the ring with high stellar velocity dispersion (see Fig.~\ref{fig:stellarvel}), the velocities are very similar. This is contrary to the expectation that the difference between gas and stellar velocity should be largest where the stellar velocity dispersion is largest. We suspect that the velocity amplitudes are not very different in this case due to the fact that the overall stellar velocity dispersion of the central component is quite low, probably due to the presence of a pseudobulge. In this case the kinematics of the stars in the centre would be dominated by rotation and not velocity dispersion (which would be the case in ``classical'' bulges).

We also add the rotation curve of the cold gas that \cite{2016ApJ...823...68S} derive from their CO measurements\footnote{The data points are derived from their rotational model, not the observed data, and listed in their Table 6. We multiply the inclination corrected values by $\sin(i)$ to compare them with our data.}. The rotation curves show a comparable degree of rotation. In the south-east, where the rotation curves deviate most, there are also residuals of the order of $+50\,\kms$ visible in their residual (their Fig.~29d).

\begin{figure*}
\centering
\includegraphics[height=0.3\linewidth]{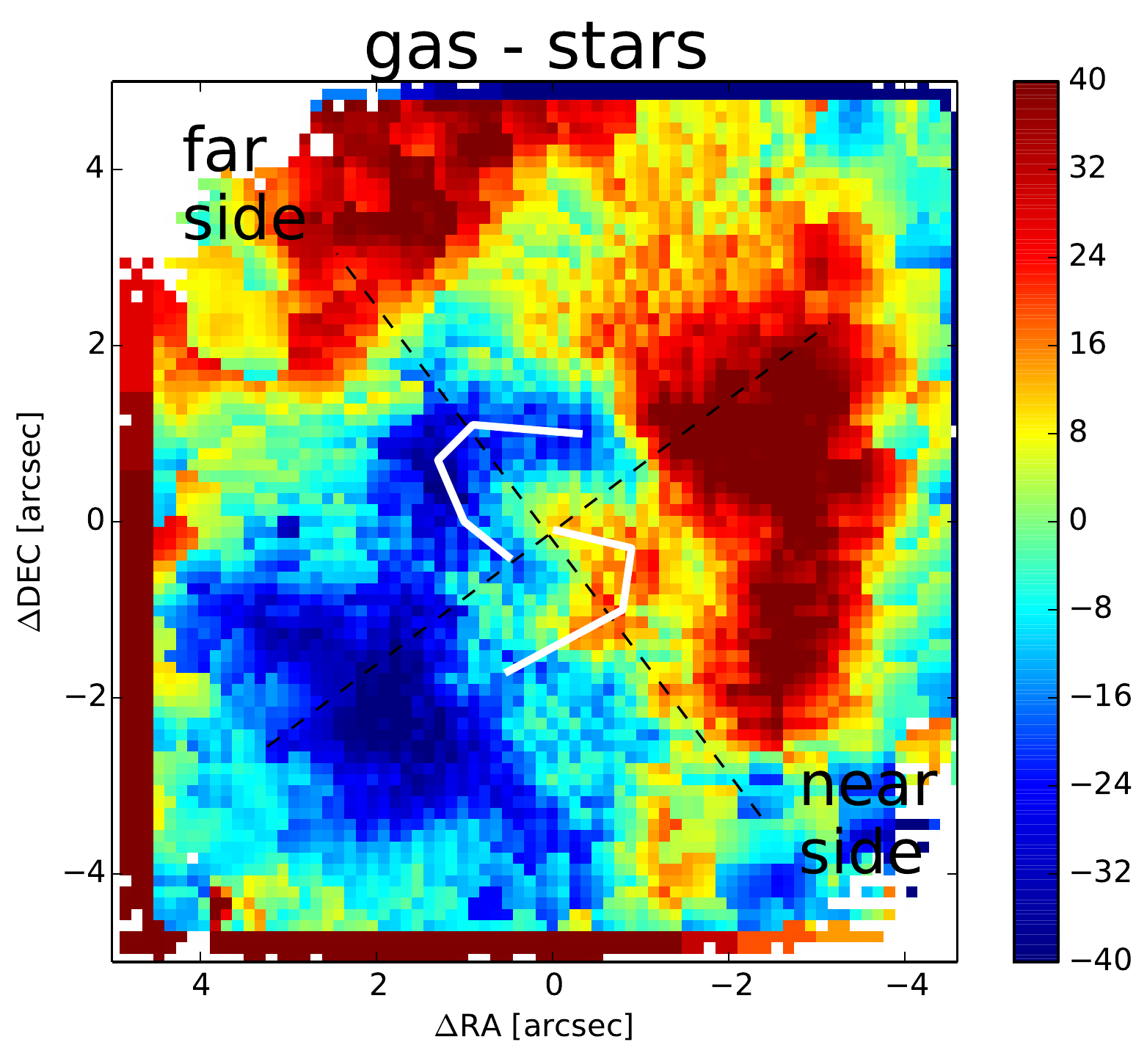}
\includegraphics[height=0.3\linewidth]{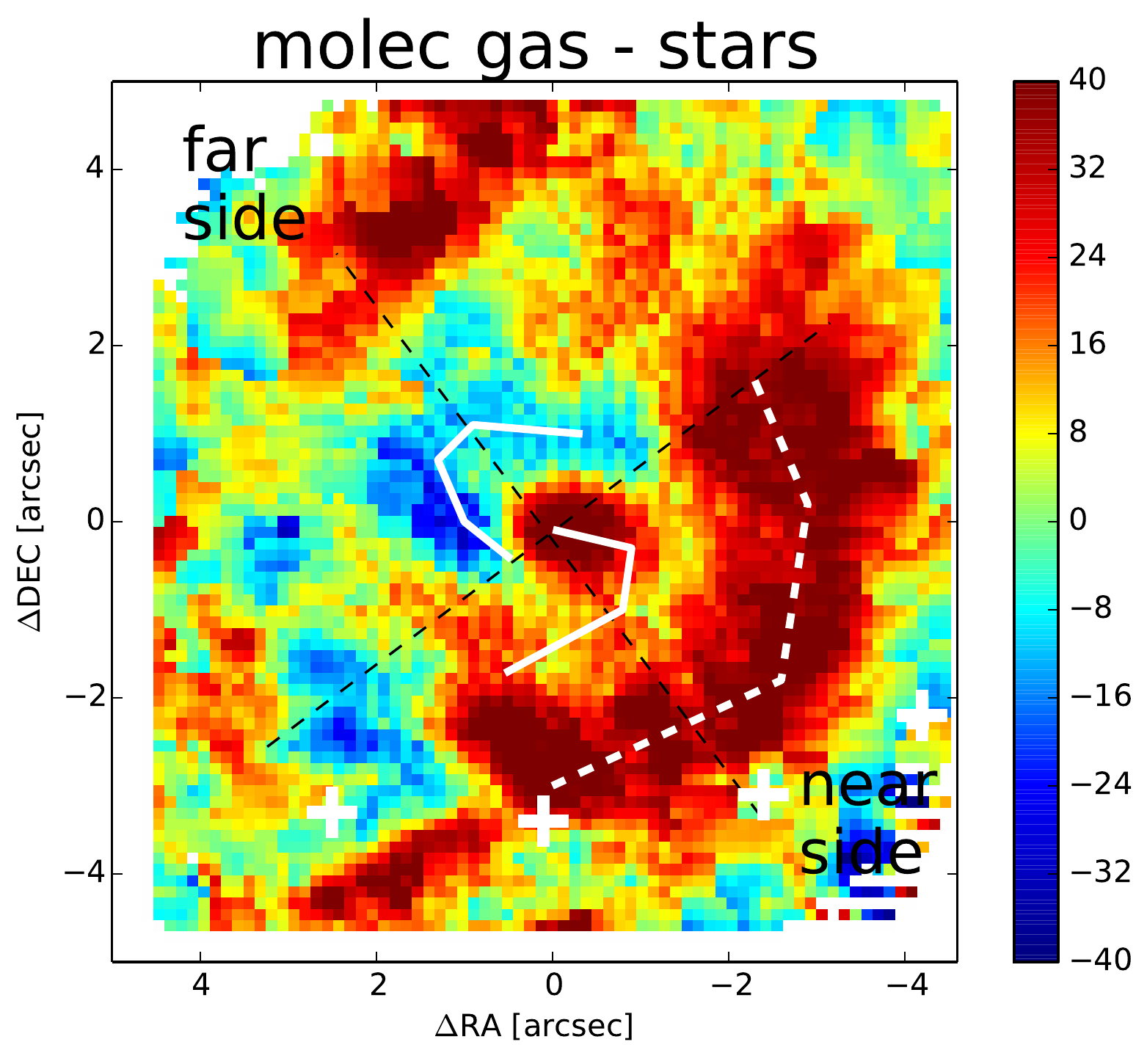}
\includegraphics[height=0.3\linewidth]{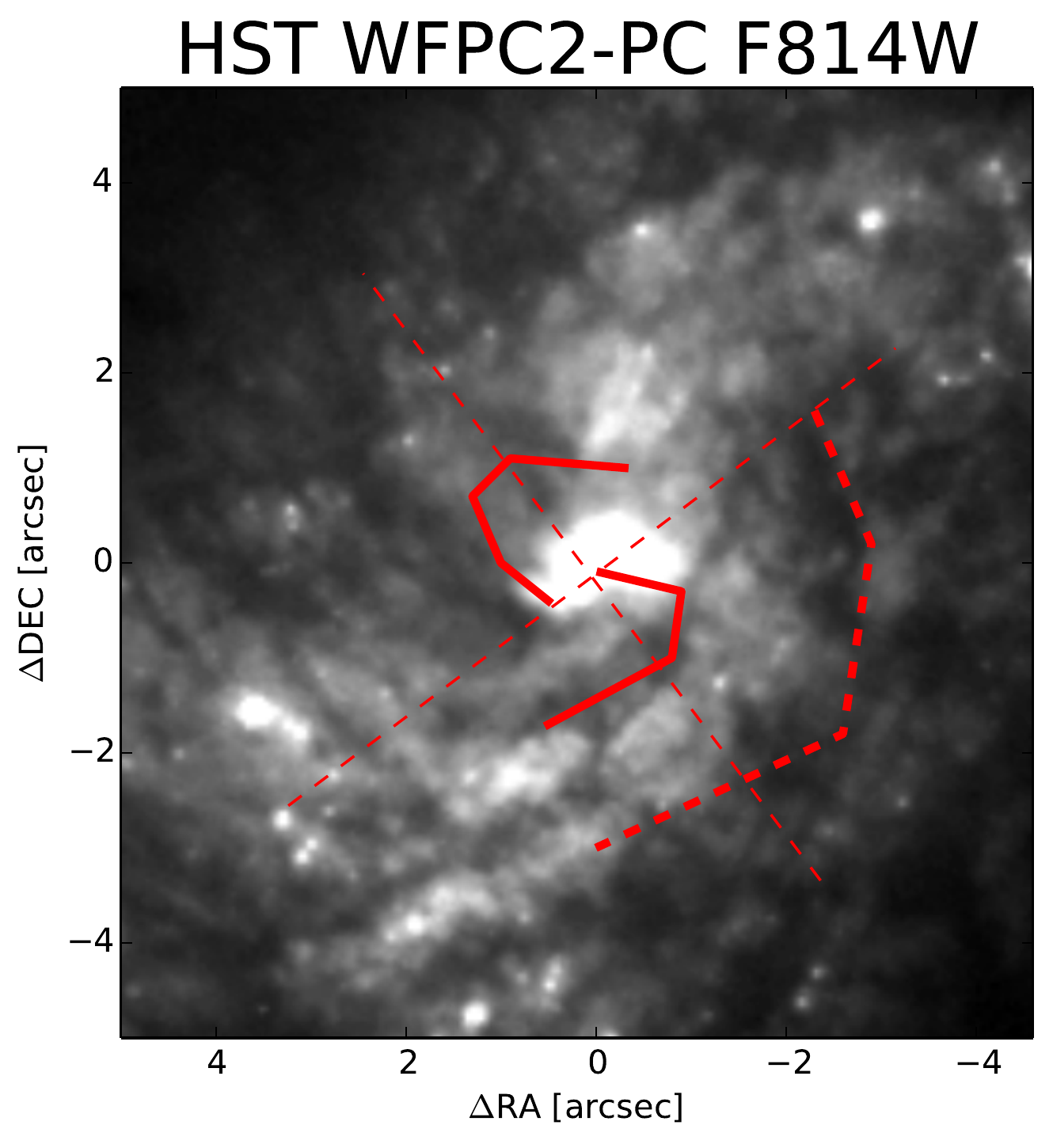}
\caption{Residual after subtraction of stellar velocity field from ionised gas (\emph{left}) and molecular gas (\emph{middle}). The streaming motion in the south-east coincides with the streaming motion that \cite{2016ApJ...823...68S} observe in CO. To guide the eye, we adopt the white crosses from their Fig.~29 a. \emph{Right panel:} HST F814W (I-band) map. Comparison of the residual velocities with the optical map shows that possible inflow motions coincide with dust lanes.}
\label{fig:gas-stars}
\end{figure*}

We conclude that hot and cold gas, as well as stars all show a very similar degree of rotation and therefore a simple residual map, which we obtain by subtracting the stellar velocity field from the velocity fields of Br$\gamma$ and H$_2$ (Fig.~\ref{fig:gas-stars}, left and middle panel), can give us an estimate of the non-rotational motions in the gas velocity fields.

The first prominent feature is a residual redshift around $3\arcsec$ to the south-west that shows a partial ring structure (marked with the dashed lines in Fig.~\ref{fig:gas-stars}). We note that this feature does not coincide with the star-formation ring but resembles the region inside this ring. In the line-ratio maps (Fig.~\ref{fig:lineratios}) this region shows high $[\ion{Fe}{ii}]/\mathrm{Br}\gamma$ and $\mathrm{H}_2/\mathrm{Br}\gamma$ line rations which could indicate shocks by inflowing gas. In their $^{12}\,\mathrm{CO}(J=1-0)$ map, tracing the cold molecular gas, \cite{2016ApJ...823...68S} find a spatially coincident streaming motion inside the gaseous nuclear spiral arm with a magnitude of $\sim\,50\,\mathrm{km}\,\mathrm{s}^{-1}$, which is fully consistent with our measurement, indicating that we trace the same motion in warm and cold molecular gas.

The second prominent feature is the nuclear two-arm spiral structure in the central $100\pc$ (marked with solid lines) that is only detectable in our near-infrared data due to the higher spatial resolution of $<1\arcsec$ ($\sim 50\,\mathrm{pc}$). Assuming that they are located within the disc plane and that the near side of the disc is in the south-west (see Sect.~\ref{sec:res_stellkin}), we conclude that the residual spiral arms could correspond to streaming motions towards the centre. Comparing to the optical HST F614W map (Fig.~\ref{fig:gas-stars}, right panel), we see that the possible inflow motions coincide with dust features. \cite{2014ApJ...792..101D} find that these chaotic circumnuclear dust structures are typically associated with external accretion in groups of galaxies.

To quantify the amount of cold gas that follows these streams and calculate the resulting gravity torques \cite[e.g.][]{2005A&A...441.1011G}, upcoming high-angular resolution sub-mm observations with ALMA will be helpful.

The velocity dispersion of [\ion{Fe}{ii}] shows enhanced values in the central region (Fig.~\ref{fig:sigma-maps}). This is likely caused by shocks due to supernova of an aging starburst. There is no indication for a jet that could be responsible for the shock. 
Furthermore, the velocity dispersion is enhanced in the region S and all the way to the centre, that is where the residual line-of-sight velocity of H$_2$ (Fig.~\ref{fig:gas-stars}) shows high redshifts. A possible explanation is that inflowing molecular gas shocks the interstellar medium, resulting in a higher flux level and velocity dispersion in the [\ion{Fe}{ii}] emission line.

In addition to the aforementioned features on 100 pc scales, a possible outflow or superwind in NGC 1808 on kpc scale has been discussed for long time. Dust lanes detected in optical images indicate the presence of a gas outflow \citep{1993AJ....105..486P}. This is supported by the detection of forbidden line ([\ion{O}{iii}], [\ion{N}{ii}], [\ion{S}{ii}]) emission and high [\ion{N}{ii}]/H$\alpha$ line ratios, indicative for shocked gas, at the base of the suspected outflow, $\gtrsim 1\,\mathrm{kpc}$ north-east of the centre \citep{2010ApJ...711..818S}. \cite{2016ApJ...823...68S} measure high velocity dispersion in this region and can even separate the CO emission lines in two components with a separation of $\gtrsim 100 \kms$. They further subtract a rotating disc model from the gas velocity field, resulting in blueshifted emission. Assuming an outflow perpendicular to the disc, the blueshifted velocity translates into a deprojected outflow velocity of $\sim 180 \kms$. On the other hand, assuming motions within the plane of the disc, the blueshifted emission would correspond to an inflow with $\sim 120 \kms$ which is similar to the rotational velocity. This is rather unlikely and therefore the authors favour the outflow scenario.

Since our field-of-view covers only the central $10\arcsec \times 10\arcsec$, we have only limited information about the suspected outflow region. Consistent with the mentioned previous observations, we find increased velocity dispersion of H$_2$ emission lines in the north-east corner (Fig.~\ref{fig:sigma-maps}, $\sim 200 \kms$ compared to $\sim 100 \kms$ in the star forming ring). However, we cannot separate two components with the available spectral resolution. Furthermore, the line ratios of H$_2$/Br$\gamma$ and [\ion{Fe}{ii}]/Br$\gamma$ have high values in this region, which is indicative for shocked gas. Unfortunately, this region is at the edge of our FOV and has low signal-to-noise. Further near-infrared integral-field spectroscopy focused on the suspected outflow region are desirable.

Besides this, based on our near-infrared data, we do not detect clear indications of this outflow in the central region ($|r|\lesssim 5\arcsec \approx 300\,\mathrm{pc}$), which raises the question where exactly the outflow starts and what the driver is.

\subsection{Gaseous excitation}

\subsubsection{NIR diagnostic diagram}
\label{sec:diagndiagram}
Diagnostic diagrams use line ratios of diagnostic lines to determine the dominating excitation mechanisms of the line-emitting gas. They are a very useful tool in the optical where they have first been established \citep[e.g.][]{1981PASP...93....5B,2001ApJ...556..121K,2003MNRAS.346.1055K,2007MNRAS.382.1415S,2013A&A...558A..34B,2014MNRAS.444.3961D,2015A&A...573A..93V}. 

The success of diagnostic diagrams in the optical motivated the search for similar tools in the near-infrared. Here, line ratios between star formation tracers (such as Pa$\alpha$, Pa$\beta$, Br$\gamma$) and shock tracers (such as different emission lines of H$_2$ or [\ion{Fe}{ii}]) are used to find the dominating source of excitation. 
One example is the line ratio H$_2 \lambda2.122\mm$/Br$\gamma$: The lowest values are found in starburst regions, where young OB stars produce strong emission in the hydrogen lines (e.g. Br$\gamma$). Higher values occur for older stellar populations, where the number of Br$\gamma$ emitting young OB stars has declined but on the other hand the rate of supernovae increased. Supernovae produce shocks that can be traced by shock tracing near-infrared lines such as H$_2$ or [\ion{Fe}{ii}].
The line ratio H$_2 \lambda2.122\mm$/Br$\gamma$ shows typical values of $\lesssim 0.4$ for star formation, low-ionisation nuclear emission-line regions (LINERs) show values of $\gtrsim 0.9$, while Seyferts typically have values in between \citep{1993ApJ...406...52M,1994ApJ...422..521G,1997ApJ...482..747A,2004A&A...425..457R,2005MNRAS.364.1041R,2013MNRAS.430.2002R}.

A two-dimensional diagnostic diagram in the near-infrared was introduced by \cite{1998ApJS..114...59L} and further developed by \cite{2004A&A...425..457R,2005MNRAS.364.1041R} and \cite{2013MNRAS.430.2002R}. They show that there is a transition from purely ionising radiation (starbursts) to pure supernova-driven shock excitation. AGNs are usually located in an intermediate region. Instead of single spectra per galaxy, \cite{2015A&A...578A..48C} use integral-field spectroscopy to spatially separate line emitting regions and place them in the diagram. By this method, they can find the typical locations of young star forming regions, older supernova dominated regions and the compact AGN dominated region, and show that they occupy different areas in the line-ratio space.

\begin{figure}
\includegraphics[width=\columnwidth]{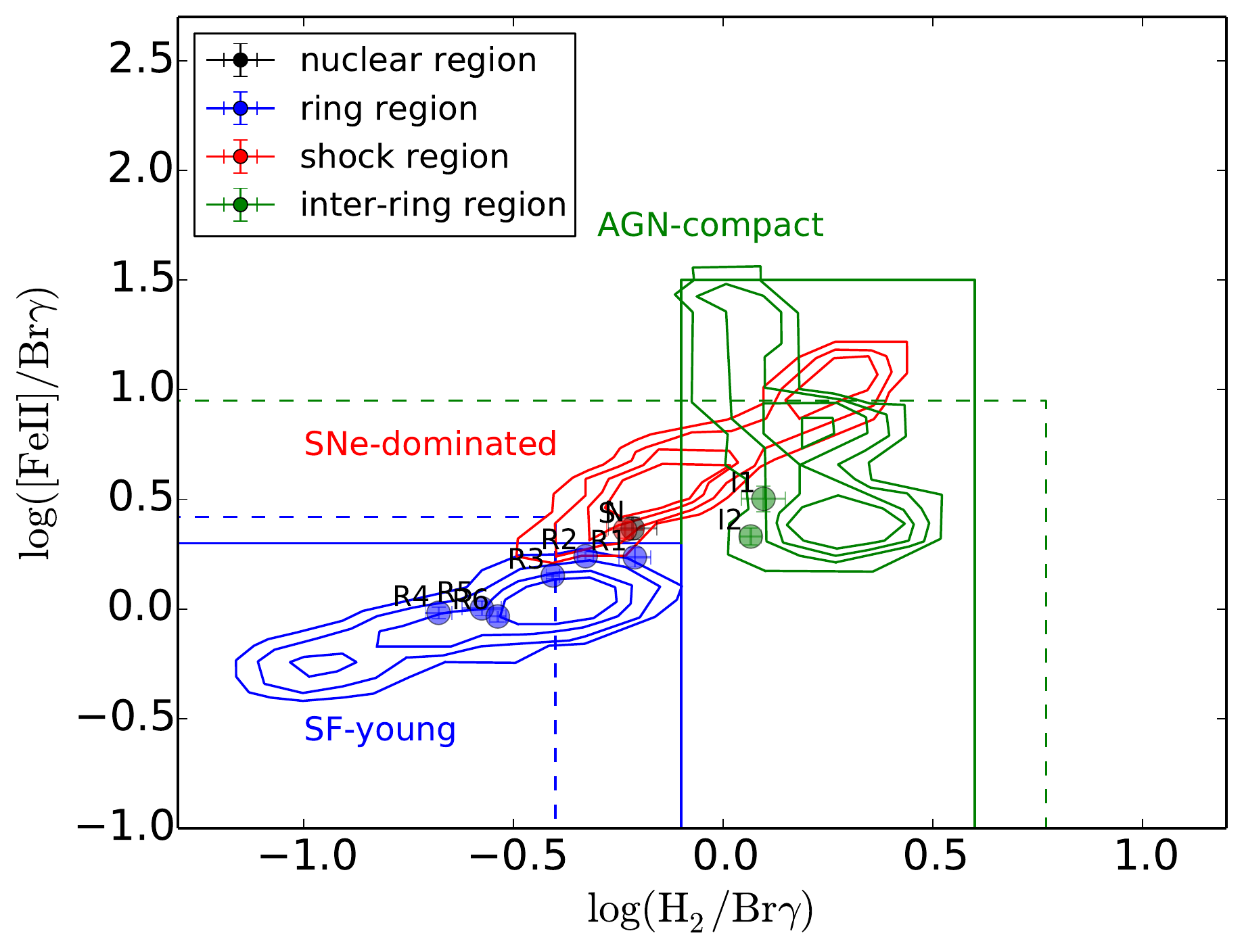}
\caption{Near-infrared diagnostic diagram with the line ratios $\log(\mathrm{H}_2\lambda2.12\mm/\mathrm{Br}\gamma)$ vs. $\log([\ion{Fe}{ii}]\lambda1.644\mm/\mathrm{Br}\gamma)$. The contours are from \cite{2015A&A...578A..48C} and denote the regions dominated by young star formation, supernovae, and the compact AGN. Solid lines are upper limits for young star-forming regions and AGN from the same study (based on 2D spectroscopy) while the dashed lines denote the upper limits from 1D spectroscopy studies done by \cite{2013MNRAS.430.2002R}.}
\label{fig:diagndiagram}
\end{figure}

We show the near-infrared diagnostic diagram in Fig.~\ref{fig:diagndiagram}. It uses the line ratios $\mathrm{H}_2\lambda2.12\mm/\mathrm{Br}\gamma$ and $[\ion{Fe}{ii}]\lambda1.644\mm/\mathrm{Br}\gamma$. The line ratio between the $H$-band [\ion{Fe}{ii}] and the $K$-band Br$\gamma$ line is corrected for extinction because at a typical extinction of $A_V=5$ the factor is already $\approx 1.6$. However, we did not correct the ratio between H$_2\lambda2.12\mm$ and Br$\gamma$ since they are close in wavelength and the correction factor would be only $1.03$. This means the introduced errors due to the uncertainty of the extinction value would be much larger than the actual correction. Therefore, we do not correct line ratios between lines that are both in the $K$-band. 

Placing the line ratios of our apertures in the diagram, we first note that all spots in the circumnuclear ring (``R'') are located in the region of young star formation. In particular, the spots R4-6 are at even lower ratios than the others, which is an indication for younger starbursts (which emit more Br$\gamma$ and show less supernovae). This is also visible in the line ratio maps in Fig.~\ref{fig:lineratios} where these spots clearly form a ring with particularly low values. The other spots have higher line ratios which shifts them in the region of (older) SNe-dominated stellar populations or the compact AGN region. This indicates that their stellar populations are significantly older than those in the circumnuclear ring. Furthermore, their line ratios are typical for AGN. This, however shows the limitation of the near-infrared diagnostic diagram: There is no confined location for AGN (in contrast to the optical BPT-diagram). Therefore, the position of the spots S, N, and I1-2 in the intermediate region just tells us that they show contributions from both shocks and photoionisation, but not necessarily induced by an AGN.

\subsubsection{Excitation of molecular hydrogen}
Rotational-vibrational transitions are an important cooling channel for molecular gas at temperatures of a few 1000 K. Molecular hydrogen emission lines in the near-infrared are excited by either thermal or non-thermal processes: Thermal processes include the heating of gas by shocks \citep{1989ApJ...342..306H}, e.g. due to the interaction of a radio jet with the interstellar medium, or heating by X-rays from the central AGN \citep{1996ApJ...466..561M}. A non-thermal processes is UV fluorescence \citep{1987ApJ...322..412B} where UV-photons with $912\AA<\lambda<1500\AA$ are absorbed by the H$_2$ molecule in the Lyman- and Werner bands, exciting the next two electronic levels ($B^1\Sigma-X^1\Sigma$ and $C^1\Pi-X^1\Sigma$). With a probability of 90\%, a decay into a bound but excited rovibrational level within the electronic ground level $X^1\Sigma$ will take place. By this mechanism, H$_2$ rovibrational levels will be populated, which could not be populated by collisions. Possible sources are OB stars or strong AGN continuum emission. 

In theory, all of these excitation processes would produce different H$_2$ spectra such that they could be distinguished by diagnostic line ratios and line population diagrams \citep{2003ApJ...597..907D,2005ApJ...633..105D,2004A&A...425..457R,2005MNRAS.364.1041R,2013MNRAS.430.2002R}. In practice, however, the different mechanisms usually occur together such that the H$_2$ spectra are mixed \citep[e.g.][]{2007A&A...466..451Z,2013MNRAS.428.2389M,2015A&A...575A.128B,2015A&A...583A.104S}. But still, H$_2$ line ratios can help estimating the dominating excitation mechanisms or constraining the contributing fractions of different mechanisms: While the H$_2$ 2-1 S(1)/H$_2$ 1-0 S(1) ratio can be used to distinguish between thermal and non-thermal excitation (such as UV fluorescence), the H$_2$ 1-0 S(2)/H$_2$ 1-0 S(0) (but also other line ratios between rotational levels in the same vibrational transition, such as H$_2$ 1-0 S(3)/H$_2$ 1-0S (1)) to separate thermal UV, shocks, and X-ray excitation.

Furthermore, the rotational excitation temperature can be determined from two ortho/para lines that belong to the same vibrational level, e.g. H$_2$ 1-0 S(0)/H$_2$ 1-0 S(2), whereas the vibrational excitation temperature can be determined by connecting two transitions with same $J$ but from consecutive $v$ levels, e.g. H$_2$ 2-1 S(1)/H$_2$ 1-0 S(1):
\begin{eqnarray}
T_{\mathrm{rot}(v=1)} = \frac{1113\,\mathrm{K}}{1.130+\ln \left( \frac{f_\mathrm{1-0\, S(0)}}{f_\mathrm{1-0\, S(2)}} \right)} 
\label{eq:temprot}\\
T_\mathrm{vib} = \frac{5594\,\mathrm{K}}{0.304+\ln \left( \frac{f_\mathrm{1-0\, S(1)}}{f_\mathrm{2-1\, S(1)}} \right)} .
\label{eq:tempvib}
\end{eqnarray}

\begin{figure}
\includegraphics[width=\columnwidth]{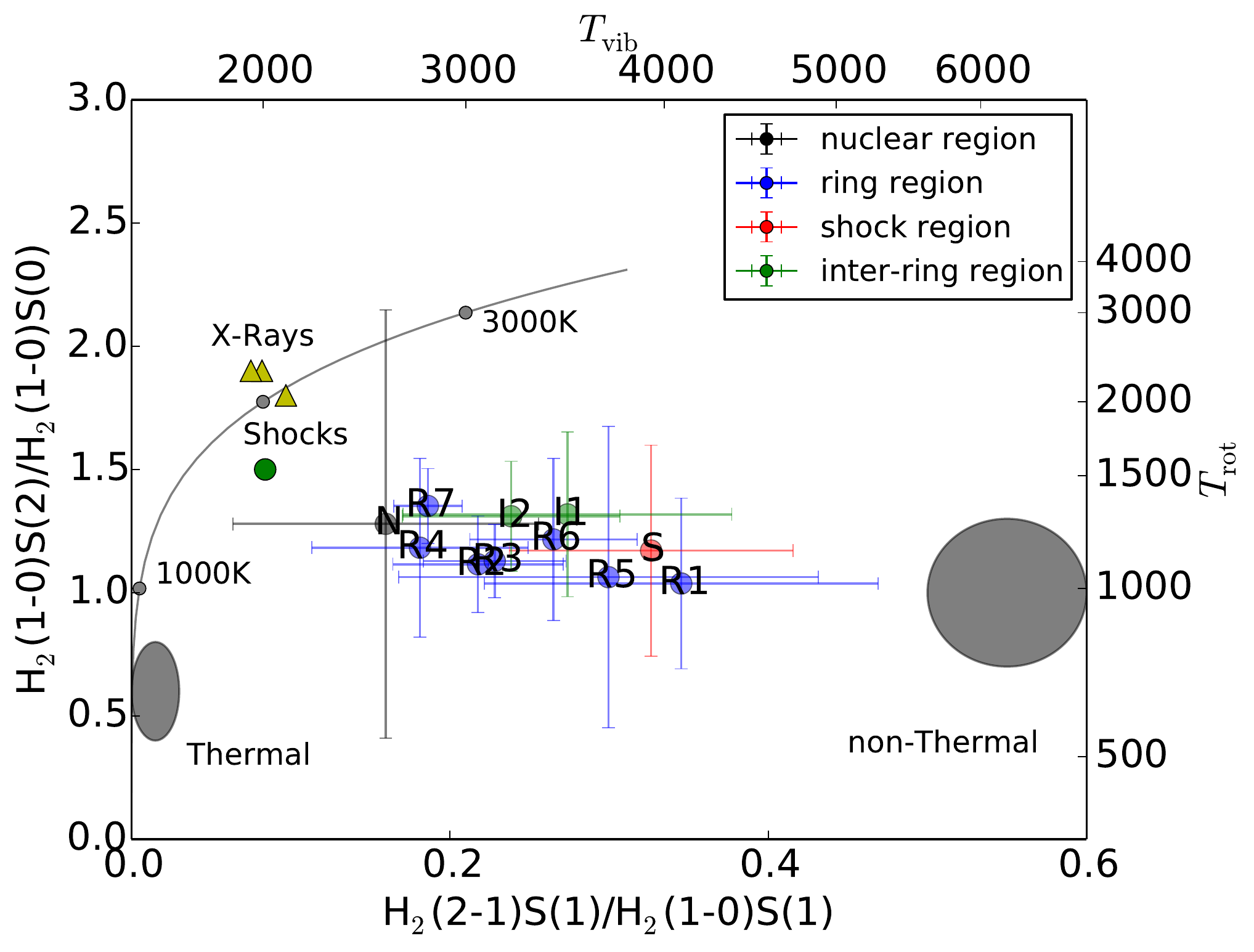}
\caption{Molecular hydrogen diagnostic diagram with 2-1S(1)/1-0S(1) vs. 1-0S(2)/1-0S(1) \citep{1994ApJ...427..777M}. The locations of the apertures are indicated. The location of thermal UV excitation \citep{1989ApJ...338..197S} and non-thermal models \citep{1987ApJ...322..412B}, as well as the thermal emission curve from 1000 K to 3000 K are plotted in gray. The X-ray heating models \citep{1990ApJ...363..464D} are marked by a yellow triangle, the shock-heating model \citep{1989MNRAS.236..929B} by a green circle.}
\label{fig:h2diagn}
\end{figure}

The line ratios of the analysed spots are displayed in the H$_2$ excitation diagram in Fig.~\ref{fig:h2diagn}. Furthermore, the vibrational and rotational excitation temperatures of the spots are denoted as upper x-axis and right-hand y-axis resp. We see that the rotational excitation temperature is in the range $1000\,\mathrm{K} < T_\mathrm{rot} < 1500\,\mathrm{K}$ for all spots. The vibrational temperature, however, reaches from values of $T_\mathrm{vib}\approx 2500\,\mathrm{K}$ up to values of $T_\mathrm{vib}\lesssim 4000\,\mathrm{K}$. Therefore, we conclude that the spots are not in local thermal equilibrium and a significant contribution of non-thermal fluorescent excitation is very probable. However, we stress that the line ratios do not lie in the region of purely non-thermal excitation which speaks for a mixture of thermal and non-thermal contribution. The low H$_2$ 1-0 S(2)/H$_2$ 1-0 S(0) ratios together with rather high values of H$_2$ 2-1 S(1)/H$_2$ 1-0 S(1) (fluorescent excitation) that we find for the spots in the circumnuclear ring are typical for starforming galaxies. In the near-infrared diagnostic diagram (Sect.~\ref{sec:diagndiagram}, Fig.~\ref{fig:diagndiagram}), all spots cannot be explained by pure shocks but show a contribution of photoionisation by young stars. This is in particular the case for the spots in the circumnuclear ring. This is consistent with the finding that all spots show significant contribution of non-thermal excitation.

\subsection{Mass of ionised and molecular hydrogen gas}
\label{sec:gasmass}

The mass of hot molecular gas can be estimated from the extinction corrected flux of the H$_2$ 1-0 S(1) $\lambda2.12 \mu$m emission line as \citep[e.g.,][]{1982ApJ...253..136S}
\begin{equation}
M_{\mathrm{H}_2} = \frac{2m_p F_{1-0\mathrm{S}(1)} 4\pi D_L^2}{f_{v=1,J=3} A_{1-0\mathrm{S}(1)} h c/\lambda}
\end{equation} 
where $D_L$ is the luminosity distance of the galaxy and $m_p$ the proton mass. Furthermore, we need the population fraction of the upper energy level $f_{v=1,J=3}$ and the Einstein coefficient $A_{1-0\mathrm{S}(1)}=3.47\times 10^{-7}\,\mathrm{s}^{-1}$ \citep{1977ApJS...35..281T} that is the probability that a particular transition from this level takes place. The population fraction depends strongly on the vibrational temperature of the system, with a dependency $f_{v,J}\propto g_J \times \exp\left(E_{v,J}/k_B T\right)$. Most authors assume (1) a local thermal equilibrium (LTE, from which follows that the level population is only determined by the system temperature) and (2) a typical vibrational temperature of $T_\mathrm{vib}=2000\,\mathrm{K}$. For this temperature the population fraction is $f_{v=1,J=3}=1.22\times 10^{-2}$. For better comparison, we will use this value and find the following relation between warm molecular gas mass and extinction corrected emission-line flux $F_{1-0\mathrm{S}(1)}$: 
\begin{equation}
M_{\mathrm{H}_2}=5.0776\times 10^{16} \left( \frac{D_L}{\mathrm{Mpc}} \right)^2 \left( \frac{F_{1-0\mathrm{S}(1)}}{\mathrm{W}\,\mathrm{m}^{-2}} \right) M_\odot,
\end{equation}
where $D_L$ is the distance to the galaxy in Mpc \citep{1982ApJ...253..136S,1998ApJS..115..293W,2010MNRAS.404..166R}.

As indicated before, the assumption of LTE might not always be justified. Furthermore, we see indications for a higher vibrational temperature. Therefore, we calculate the population fraction
\begin{equation}
f_{v=1,J=3} = \frac{g_{J=3} \times \exp \left( E_{v=1,J=3}/k_B T_\mathrm{vib}\right)}{\sum \limits_{v,J} g_J \times \exp \left( E_{v,J}/k_B T\right)}
\end{equation}
for higher temperatures as well. We find that $f(T=3000\,\mathrm{K})/f(T=2000\,\mathrm{K})\approx 1.8$ and $f(T=4000\,\mathrm{K})/f(T=2000\,\mathrm{K})\approx 2.0$. This means that for temperatures higher than $T=2000\,\mathrm{K}$, we would overestimate the gas mass.

In general, the hot molecular gas, as traced by the near-infrared H$_2$ emission lines, only represents the hot surface, and therefore a small fraction, of the total gas amount of the galaxy. A further difficulty is that the strength of the emission line does not only depend on the gas mass but also on the external energy source that is able to excite the ro-vibrational transitions (e.g. UV-photons from OB stars or shocks induced by supernovae or AGN outflows). Nevertheless, empirical relations between the near-infrared H$_2$ luminosities and CO luminosities (that are a common tracer of the total molecular gas mass) suggest that the luminosity of near-infrared H$_2$ lines can be used to estimate the cold gas mass. We will use the conversion factor $M_{\mathrm{H}_2\mathrm{(cold)}}/M_{\mathrm{H}_2\mathrm{(warm)}} = (0.3-1.6)\times 10^6$ from \cite{2013MNRAS.428.2389M}. Other consistent cold-to-warm H$_2$ gas mass conversion factors are $10^5-10^7$ \citep{2005AJ....129.2197D} and $2.5\times 10^6$ \citep[with a $1\sigma$ uncertainty of a factor of $2$;][]{2006A&A...454..481M}.

The mass of ionised hydrogen gas can be estimated from the measured and extinction corrected flux of hydrogen recombination lines such as Pa$\alpha$ or Br$\gamma$. Assuming an electron temperature of $T=10^4\,\mathrm{K}$ and a density in the range $10^2<n_e<10^4\,\mathrm{cm}^{-3}$, we can calculate the ionised gas mass from the extinction corrected Pa$\alpha$ and Br$\gamma$ fluxes, $f_{\mathrm{Pa}\alpha}$ and $f_{\mathrm{Br}\gamma}$ by 
\begin{eqnarray}
M_\ion{H}{ii} &\approx& 2.4\times 10^{21}\, \left(\frac{f_{\mathrm{Pa}\alpha}}{\mathrm{W}\,\mathrm{m}^{-2}} \right) \left(\frac{D}{\mathrm{Mpc}}\right)^2 \left(\frac{n_\mathrm{e}}{\mathrm{cm}^{-3}} \right)^{-1} \,M_\odot \\
&\approx& 2.9\times 10^{22}\, \left(\frac{f_{\mathrm{Br}\gamma}}{\mathrm{W}\,\mathrm{m}^{-2}} \right) \left(\frac{D}{\mathrm{Mpc}}\right)^2 \left(\frac{n_\mathrm{e}}{\mathrm{cm}^{-3}} \right)^{-1} \,M_\odot
\end{eqnarray}
where $D$ is the luminosity distance to the galaxy and $n_\mathrm{e}$ the electron density (see \citet{2015A&A...575A.128B} for details). Here, we use typical values of $T=10^4\,\mathrm{K}$ and $n_e=10^2\,\mathrm{cm}^{-2}$.

Given the mentioned systematic uncertainties and difficulties with the gas mass estimates that are much higher than those arising from (emission line flux) measurements, we refrain from stating formal errors when calculating gas masses but point out that these are only order-of-magnitude estimates.

We estimate the gas masses by summing over the field-of-view ($\sim 10\arcsec \times 10\arcsec$). We estimate the average extinction in this FOV to be $A_V \approx 2\,\mathrm{mag}$. Applying this correction and using the H$_2\,\lambda2.12\mm$ emission line, we estimate a hot molecular gas mass of $\sim 730\,M_\odot$ that corresponds to a cold molecular gas mass of $\sim (2-12)\times 10^8\,M_\odot$. This is in good agreement with the molecular gas mass of $4.2\times 10^8\,M_\odot$ that \cite{2014PASJ...66...96S} derive from CO-measurements with ASTE and $2\times 10^8\,M_\odot$ that \cite{2016ApJ...823...68S} derive from ALMA CO(1-0) measurements (in an aperture of $<250\,\mathrm{pc}$ corresponding to our field-of-view, however lower limit as it is not corrected for missing short baselines)\footnote{We translated their measurements of $3.0\times 10^8\,M_\odot$ and $1.4\times 10^8\,M_\odot$ at a luminosity distance of $10.8\,\mathrm{Mpc}$ to our assumed distance of $12.8\,\mathrm{Mpc}$}. Using the $K$-band emission-line Br$\gamma$, we estimate the ionised hydrogen mass $M_\ion{H}{ii}\sim 6.6\times 10^6\,M_\odot$. We see that the ionised hydrogen gas mass is $\sim 9000$ times higher than the hot molecular gas mass. This is in agreement with typical ratios of $10^3 - 10^4$ \citep[][and references therein]{2014MNRAS.442..656R}. For the nuclear regions of nearby Seyfert galaxies, the AGNIFS group obtained hot molecular gas masses with a range of $10^1 < M_{\mathrm{H}_2} < 10^3\,M_\odot$ and ionised gas masses with a range of $10^4 < M_\ion{H}{ii} < 10^7\,M_\odot$ \citep[][and references therein]{2015MNRAS.451.3587R}. For NUGA sources, cold molecular gas masses (derived from CO-emission) range from $10^8 - 10^{10}\,M_\odot$, with typical masses of the order of several $10^9\,M_\odot$ \citep[][and references therein]{2012nsgq.confE..69M}, whereas low-luminosity QSOs at higher redshift ($0.02 \leq z \leq 0.06$) have systematically higher gas reservoirs \citep{2015A&A...575A.128B,2016A&A...587A.137M}. The nuclear region of NGC1808 lies in the first range and shows therefore a typical behaviour for local Seyfert galaxies. However, we stress that the derived gas masses can only be used as order-of-magnitude estimates and only the combination of near-infrared observations together with high-resolution sub-mm interferometry (for example with ALMA) can provide a full view on the galactic centres and their warm and cold gas reservoirs.

We also measure the gas masses in the apertures. The masses are listed in Tables \ref{tab:emissionlines} and \ref{tab:h2lines}. In addition, we also list the gas mass surface densities of the cold molecular gas. These densities will be compared with the star formation surface densities in Sect.~\ref{sec:sfrates} to determine the efficiency of the formation of new stars out of the available gas amount.

\subsection{Star formation properties}
One of the most striking features in the central region of NGC 1808 is the patchy circumnuclear ring that is commonly associated with a star forming ring \citep[e.g.][]{1994ApJ...425...72K,1996A&A...313..771K,1996AJ....112..918T}. The positions in diagnostic diagrams strongly support this. In the near-infrared diagnostic diagram (Fig.~\ref{fig:diagndiagram}) the spots in the ring show very low line ratios, indicating pure photoionisation which is indicative for star formation. In the H$_2$ excitation diagram (Fig.~\ref{fig:h2diagn}), the spots show a significant contribution of non-thermal excitation, e.g. by UV fluorescence which is consistent with star forming clumps. Furthermore, we find a significantly lower stellar velocity dispersion in the region of the ring which is interpreted as due to the presence of younger stellar populations that have formed out of cooled molecular gas and thus not dispersed into the surrounding region (Sect.~\ref{sec:stellkin_disc}).

\subsubsection{Star formation rates and efficiency}
\label{sec:sfrates}
In starburst events, a large number of stars, some of them hot and luminous O and B stars, are formed. These immediately start to photoionise the surrounding interstellar medium. This produces large nebular regions which can show strong nebular emission like the hydrogen recombination lines H$\alpha$, H$\beta$ (in the optical), Pa$\alpha$, or Br$\gamma$ (in the near-infrared). Also UV photons are absorbed by dust which is heated to 30-60 K and then reradiates the energy at wavelengths around $60\mm$. 

The star formation rate is a good estimator of the power of the starburst. It can be calculated from the luminosity of hydrogen recombination lines since they are proportional to the Lyman continuum which is proportional to the star-formation rate (SFR). Independent of the star formation history, only stars with masses $\gtrsim 10\,M_\odot$ and ages $\lesssim 20\,\mathrm{Myr}$ contribute to the ionising flux. Therefore, hydrogen recombination lines trace the instantaneous SFR \citep{1998ARA&A..36..189K}. We derive the SFR with the calibration of \cite{2003A&A...409...99P}\footnote{While the uncertainty of the SFR introduced by line measurement errors can be estimated relatively straightforward, further uncertainties come from the uncertainty of the luminosity distance/cosmology, the extinction correction and most importantly the calibration of the SF-estimator. As a conservative estimate we have to assume an uncertainty of at least a factor of two. Therefore, we do not state particular uncertainties of the derived SFR values but advise the reader to consider them more as an order-of-magnitude estimate. The same is valid for the supernova rate that we estimate in the next section.}:
\begin{equation}
\mathrm{SFR}_{\mathrm{Br}\gamma} = \frac{L(\mathrm{Br}\gamma)}{1.585\times 10^{32}\,\mathrm{W}} \, M_\odot\,\mathrm{yr}^{-1}.
\end{equation}

In the apertures that are located in the star forming ring the star-formation rates range from 0.04 to 0.09 $M_\odot\,\mathrm{yr}^{-1}$. In the nuclear aperture, we find a star-formation rate of $\sim0.18\,M_\odot\,\mathrm{yr}^{-1}$. All measurements are listed in Table \ref{tab:sfr}. \cite{2014ApJ...780...86E} estimate the nuclear star-formation rate from the $11.3\mm$ PAH emission. In a slit with $0\farcs 35$ width, they measure $0.21\,M_\odot\,\mathrm{yr}^{-1}$. This is higher than the value that we measure in our larger $0\farcs 75$ nuclear aperture. However, the PAH feature traces star formation integrated over a few tens of Myr, while we use hydrogen recombination lines to estimate the instantaneous star-formation rate. With this in mind, the star-formation rate estimates are consistent.

To better compare the values with other measurements, we divide by the area of the apertures, which have a radius of $0\farcs75 \approx 50\,\mathrm{pc}$. The star formation surface density is then ranging from 5 to 13 $M_\odot\,\mathrm{yr}^{-1}\,\mathrm{kpc}^{-2}$ in the ring and $\sim 28\,M_\odot\,\mathrm{yr}^{-1}\,\mathrm{kpc}^{-2}$ in the nucleus. Typically, star formation surface density ranges from $(1-50)\,M_\odot\,\mathrm{yr}^{-1}\,\mathrm{kpc}^{-2}$ on hundreds of parsec scales, from $(50-500)\,M_\odot\,\mathrm{yr}^{-1}\,\mathrm{kpc}^{-2}$ on scales of tens of parsecs, and increases to up to some $1000\,M_\odot\,\mathrm{yr}^{-1}\,\mathrm{kpc}^{-2}$ on parsec scales \citep[][and references therein]{2012A&A...544A.129V}. We conclude that the star-formation rates in the apertures are in a typical order of magnitude.

\begin{figure}
\centering
\includegraphics[width=\columnwidth]{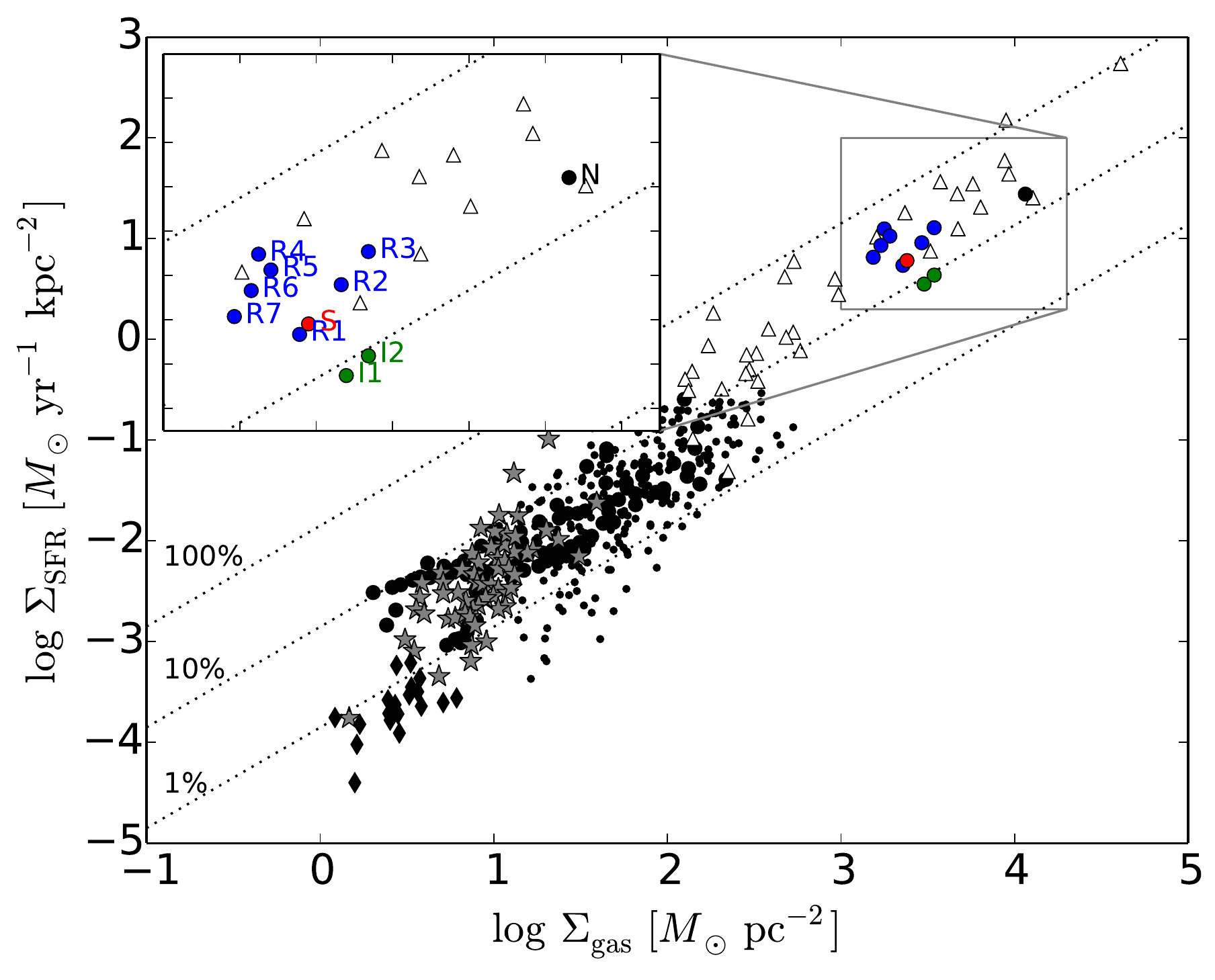}
\caption{Kennicutt-Schmidt relation of star formation: relation between the gas surface density, $\Sigma_\mathrm{gas}$, and the star-formation surface density, $\Sigma_\mathrm{SFR}$. The black/grey data points are collected by \cite{2008AJ....136.2846B}. Our measurements for NGC1808 are inserted with big black (nucleus), blue (star-formation ring), red (region S), and green (regions between ring and nucleus) circles.}
\label{fig:schmidtlaw}
\end{figure}

Another question is how efficient star formation, i.e. the transformation of gas into young stars, is. For this, we evaluate the location of our data in the Kennicutt-Schmidt law, which states that the star formation surface density scales with the local gas surface density as $\Sigma_\mathrm{SFR} \propto \left(\Sigma_\mathrm{gas}\right)^n$ \citep{1959ApJ...129..243S,1998ApJ...498..541K,2008AJ....136.2846B}. In Fig.~\ref{fig:schmidtlaw} we show the data points collected by \cite{2008AJ....136.2846B}, together with the locations of our apertures. We see that the spots in the ring that show high star-formation rates are all in the Kennicutt-Schmidt relation. Some of them (particularly the spots R3-R6 with high SFR) at the upper end, which indicates high star-formation efficiency. The apertures R1, S, and the nuclear aperture N are located at the lower part of the relation, which indicates less efficient star formation. The apertures I1 and I2, between ring and nucleus, show even lower star formation efficiencies which indicates that they do not host starbursts but the starbursts are confined in the star formation ring and the nuclear starburst.

\subsubsection{Star formation in the circumnuclear ring}

While it is clear that a circumnuclear star forming ring provides the necessary gas reservoir for star formation, a further question is the sequence in which star formation is taking place. \cite{2008AJ....135..479B} suggest two scenarios which they call ``popcorn'' and ``Pearls on a string''. In the first scenario, gravitational instabilities fragment the rings in the inner Lindblad resonance into several clouds. Starbursts will then occur whenever gas accretion leads to a critical density. The location of these starbursts is fully stochastic and therefore, no age gradient would be seen \citep{1994ApJ...425L..73E}. In the second scenario, star formation only occurs in (usually two, often located close to the point where the gas enters the ring) particular ``overdensity regions''. Then, the young clusters move along the ring, following the gas movement and meanwhile age, resulting in an age gradient along the ring \citep[e.g.][]{2014MNRAS.438..329F,2014ApJ...797L..16V}. 

\begin{figure}
\centering
\includegraphics[width=0.8\columnwidth]{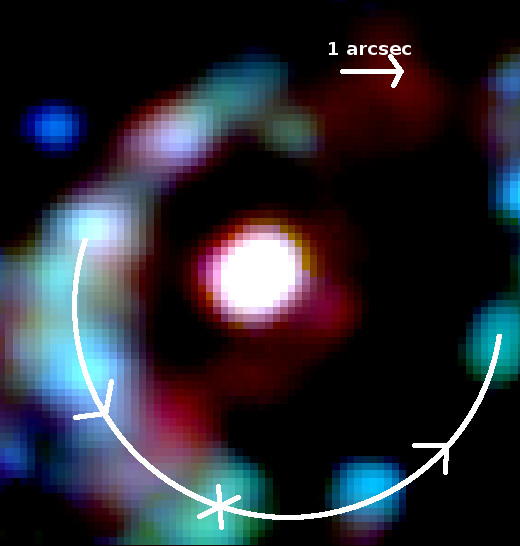}
\caption{False-colour map of the emission line fluxes in NGC1808 following the method described in \cite{2008AJ....135..479B}. The [\ion{Fe}{ii}] emission is red, Br$\gamma$ green, and \ion{He}{i} blue. The youngest regions appear in blue, the oldest in red. The direction of the rotation is indicated by arrows, the overdensity region where a possible gas spiral hits the ring by a cross. The centre is red, indicating that it is dominated by supernova remnants as traced by [\ion{Fe}{ii}] emission. In the ring, no clear age gradient is visible from the false-colour map.}
\label{fig:sf_rgb}
\end{figure}

While it is difficult to determine the absolute age of the young clusters, \cite{2008AJ....135..479B} suggest a method to at least determine the relative age of the clusters in order to identify possible age gradients. Following their method, we created a false-colour map, that shows the emission of \ion{He}{i} in blue, Br$\gamma$ in green, and [\ion{Fe}{ii}] in red. The idea is that since \ion{He}{i} has a higher ionisation potential than Br$\gamma$, it requires hotter and more massive stars than Br$\gamma$. \ion{He}{i} will therefore be only visible in the very early phase of an instantaneous starburst. [\ion{Fe}{ii}] on the other hand traces supernovae which become dominant after $\sim\,10\,\mathrm{Myr}$ when the most massive stars arrive at the end of their life time. Based on this argumentation, we can now identify an age sequence in the false-colour map, where blue traces the youngest clusters, green intermediate age, and red the oldest clusters. In Fig.~\ref{fig:sf_rgb}, we can indeed see that the clusters have different relative ages. In particular, the clusters in the south-east and south-west, which correspond to the apertures R4 and R6, show rather low ages, consistent with the NIR diagnostic diagrams (Figs.~\ref{fig:diagndiagram} and \ref{fig:h2diagn}). The central aperture, however, is dominated by supernova remnants. This is in agreement with previous age estimates of the nuclear starburst that hint to an age of $\gtrsim 10^7\,\mathrm{Myr}$ \citep{1994ApJ...425...72K,1996A&A...313..771K}. Based on our data, we do not find evidence for an age gradient as predicted by the ``Pearl on a string'' scenario. A more likely scenario is that the clusters form at random positions in the circumnuclear gas ring and then drift away following the more circular orbits of the stellar velocity field \citep{2008A&A...487..519G,2014mysc.conf...95G}.

\begin{figure*}
\centering
\includegraphics[width=0.45\linewidth]{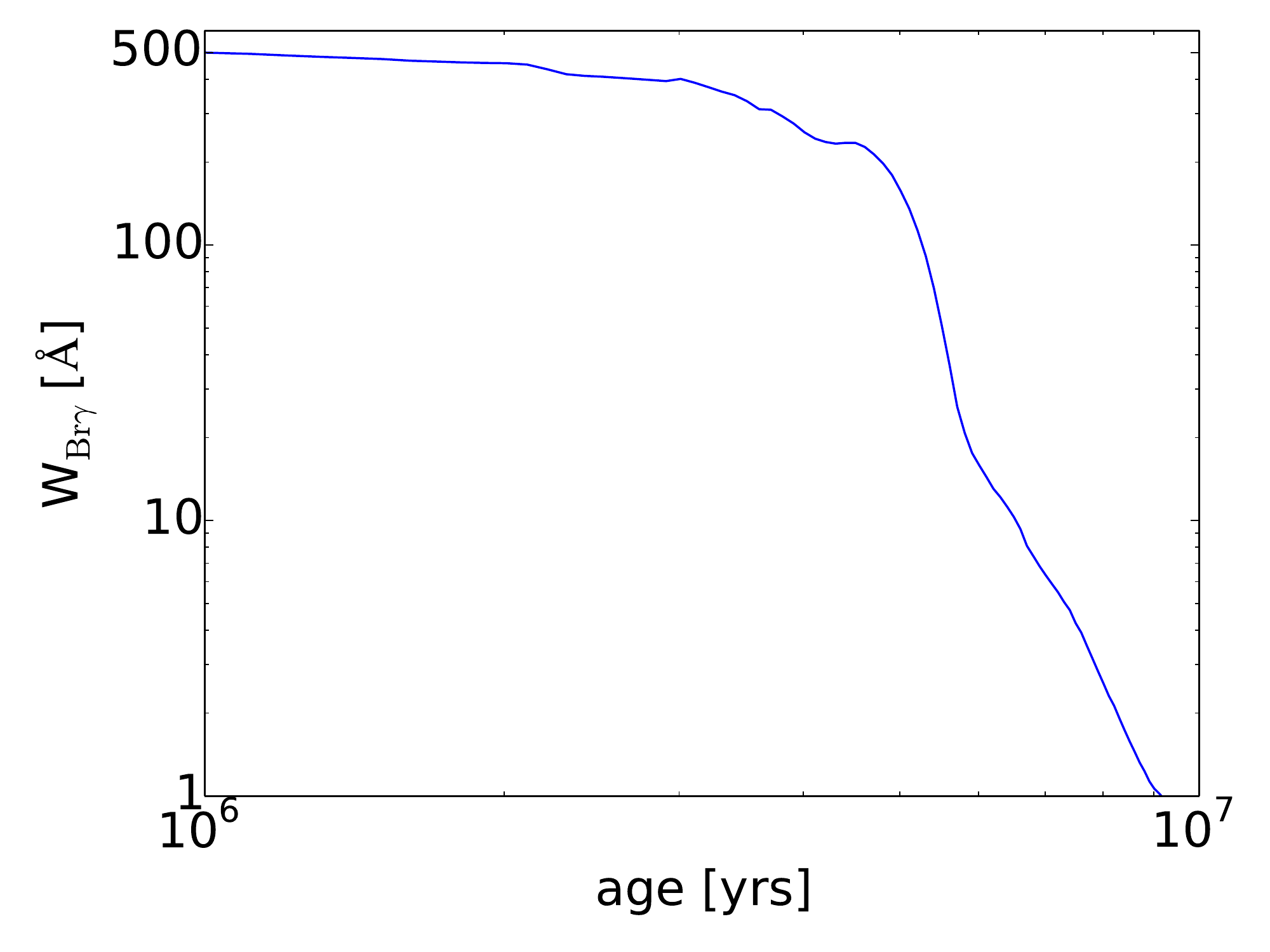}
\includegraphics[width=0.45\linewidth]{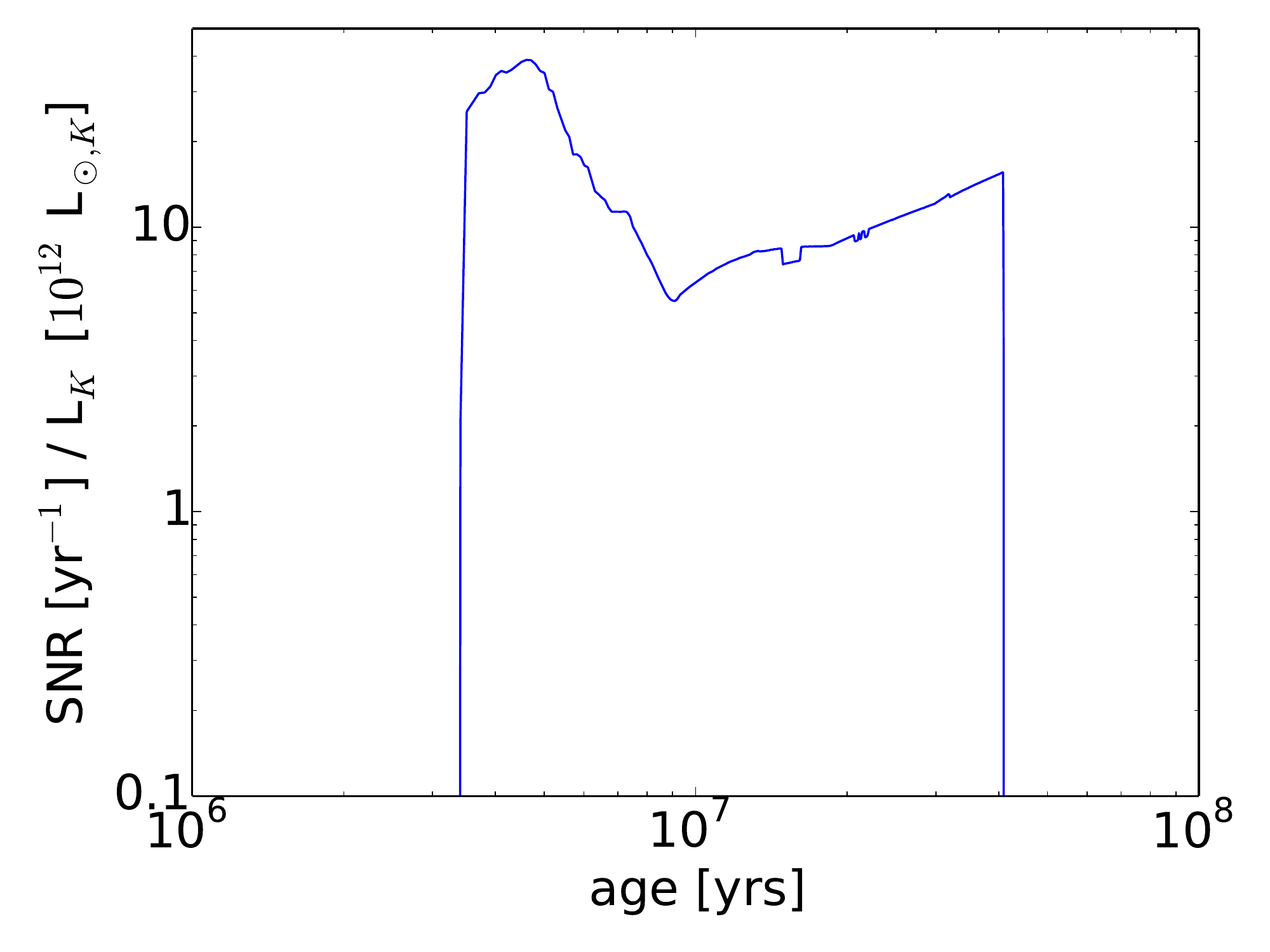}
\caption{Equivalent width of Br$\gamma$ (\emph{left}) and supernova rate (normalised by $K$-band luminosity, \emph{right}) as predicted by \textsc{Starburst99}.}
\label{fig:sb99}
\end{figure*}

In the following, we want to get an estimate of the absolute age of the starbursts. We use \textsc{Starburst99} \citep{1999ApJS..123....3L,2005ApJ...621..695V,2010ApJS..189..309L,2014ApJS..212...14L} to predict parameters such as the equivalent width of Br$\gamma$ for different star formation histories\footnote{We use an instantaneous starburst model with total stellar mass $10^6\,M_\odot$, a Kroupa IMF with two IMF intervals with $\alpha=1.3$ for the mass interval $0.1<m<0.5$ and $\alpha=2.3$ for $0.5<m<100$, and Geneva tracks, including AGB stars.}. These can then be compared to observations, to determine the age of the circumnuclear star formation regions and constrain the star formation history \citep[e.g.][]{2006ApJ...646..754D,2007ApJ...671.1388D,2008A&A...482...59D,2009MNRAS.393..783R,2013ApJ...779..170L,2015A&A...575A.128B}. In Fig.~\ref{fig:sb99}, we show simulations of the equivalent width of Br$\gamma$ and the supernova rate normalised by the $K$-band continuum luminosity. 

[\ion{Fe}{ii}] is a coolant in supernova remnants but absent in pure \ion{H}{ii} regions. Therefore, under the assumption that [\ion{Fe}{ii}] is primarily excited by shocks caused by supernova explosions, the supernova rate can be estimated from the [\ion{Fe}{ii}] luminosity. \cite{1992MNRAS.259..293F} argue that the radio hot spots in \object{NGC 1808} can be solely explained by supernova remnants. In Fig.~\ref{fig:deepmaps}, we show the [\ion{Fe}{ii}] emission together with the radio contours. They coincide well which supports that both, [\ion{Fe}{ii}] and radio emission, are indeed produced by the same mechanism, probably shocks due to supernovae. Furthermore, the radio spots coincide with mid-infrared hot spots which indicates that they are not isolated SNRs but associated with young star clusters \citep{2005A&A...438..803G}. We thus use the [\ion{Fe}{ii}] emission to estimate the supernova rate in the apertures, using the calibration of \cite{2003AJ....125.1210A}:
\begin{equation}
\mathrm{SNR} = 8.08 \, \frac{L_{[\ion{Fe}{ii}]}}{10^{35}\,\mathrm{W}}\,\mathrm{yr}^{-1}.
\end{equation}

Since both observables, Br$\gamma$ equivalent width and supernova rate from [\ion{Fe}{ii}], have to be normalised by the continuum luminosity, a proper estimation of the contribution due to the underlying (old) bulge population is crucial. \cite{2009MNRAS.393..783R} perform aperture photometry in the star formation regions and the neighbouring regions and find that the bulge contributes $\sim 50\%$ to the total flux of the star formation regions. We lay slits through the star formation spots and inspect the light profiles. In a first-order approximation, we then subtract a linear function from the profiles to distinguish between the smooth distribution of the underlying old population and the additional contribution of the young stellar clusters to the continuum flux in the apertures. In spots R2, R3, and R4, we find bulge contributions of 75\%, 80\%, and 70\% respectively. The observed equivalent widths have therefore to be increased by factors of 4, 5, and 3.3 respectively. In the other apertures, the contribution of the old population is $\gtrsim 90\%$. Thus, we have to keep in mind that the equivalent widths might be underestimated by a factor of at least ten. 

To estimate the age of the starbursts in R2, R3, and R4, we correct the Br$\gamma$ equivalent width and normalised SNR in Table \ref{tab:sfr} and compare them to the \textsc{Starburst99} predictions in Fig.~\ref{fig:sb99}. Assuming an instantaneous starburst, the favoured models give starburst ages of the order of 6 Myr (R2), 5 Myr (R3), and 6-8 Myr (R4). This is in good consistency with the estimates from near-infrared data which result in an age range between 5 and 20 Myr \citep{1994ApJ...425...72K,1996AJ....112..918T,1996A&A...313..771K} and from mid-infrared data which result in an age range between 3 and 6 Myr \citep{2005A&A...438..803G}.

The physical rotational velocity of the ring can be calculated from the observed velocity $v_\mathrm{obs}$ at an angle $\theta$ in the plane of the ring by
\begin{equation}
v_\mathrm{ring} = \frac{v_\mathrm{obs}}{\sin(i)\, \cos(\theta)}
\end{equation}
where $i$ is the inclination of $52^\circ$. From the Br$\gamma$ LOSV map (Fig.~\ref{fig:losv-maps}), we estimate the rotation velocity to be $v_\mathrm{ring}\approx170-180\kms$. With this velocity and the ring radius of $r\approx 190\pc$ at hand, we can calculate the travel time of a cloud in the ring to complete one orbit:
\begin{equation}
t_\mathrm{travel} = \frac{2\pi \times 190\pc}{170\kms} \approx 7\,\mathrm{Myr}
\end{equation}

Comparing the star cluster ages derived above with the travel time, we conclude that the clusters have travelled almost one complete orbit since formation. This is similar to the case of NGC 7552 \citep{2012A&A...543A..61B}.

\begin{table*}
\centering
\caption{Star formation rate (SFR) and SFR density, Br$\gamma$ equivalent width, supernova rate (SNR) and SNR normalised by $K$-band continuum luminosity for all apertures. All apertures have a radius of $0\farcs75$ which corresponds to a physical area of $6800\pc^2$.}
\label{tab:sfr}
\begin{tabular}{cccccc}
\hline \hline
 & SFR$_{\mathrm{Br}\gamma}$ & $\log(\Sigma_\mathrm{SFR})$ & $W_{\mathrm{Br}\gamma}$ & SNR & SNR$/L_K$ \\
aperture & [$10^{-3}\,M_\odot\,\mathrm{yr}^{-1}$] & [$M_\odot\,\mathrm{yr}^{-1}\,\mathrm{kpc}^{-1}$] & $\AA$ & [$10^{-5}\,\mathrm{yr}^{-1}$] & [$\mathrm{yr}^{-1}/10^{12}\,L_{\odot,K}$] \\
\hline

N & $188$ & $1.44$ & $4.5\pm0.3$ & $561$ & $2.5$ \\
R1 & $37$ & $0.73$ & $3.9\pm0.2$ & $81$ & $1.4$ \\
R2 & $62$ & $0.96$ & $9.9\pm0.2$ & $138$ & $3.9$ \\
R3 & $87$ & $1.11$ & $14.7\pm0.2$ & $159$ & $4.7$ \\
R4 & $85$ & $1.10$ & $13.3\pm0.2$ & $104$ & $2.6$ \\
R5 & $72$ & $1.03$ & $9.1\pm0.2$ & $93$ & $1.9$ \\
R6 & $58$ & $0.93$ & $14.7\pm0.2$ & $69$ & $3.1$ \\
R7 & $44$\tablefootmark{a} & $0.82$\tablefootmark{a} & $18.0\pm0.2$ & $29$\tablefootmark{a} & $1.6$ \\
S & $41$ & $0.78$ & $4.3\pm0.2$ & $121$ & $2.0$ \\
I1 & $24$ & $0.55$ & $1.9\pm0.2$ & $98$ & $1.4$ \\
I2 & $30$ & $0.64$ & $3.9\pm0.2$ & $81$ & $1.8$ \\

\hline
\end{tabular}
\tablefoot{
\tablefoottext{a}{Values are not corrected for extinction because extinction could not be determined in this aperture.}
}
\end{table*}

Simulations by \cite{2013ApJ...769..100S} show that the star formation rate could be decisive for the way star formation happens in circumnuclear rings: when the SFR is low, star formation mostly takes place in the contact points of the ring with the dust lanes, leading to an age gradient in the ring (``Pearls on a string''), while for high SFR, star formation is randomly distributed in the ring (``popcorn''). They find a critical SFR of $\sim 1\,M_\odot\,\mathrm{yr}^{-1}$. This is supported by a H$\alpha$ study of 22 circumnuclear rings \citep{2008ApJS..174..337M}. However, these authors find a higher critical SFR of $\sim 3\,M_\odot\,\mathrm{yr}^{-1}$ and a large spread in the SFR. Summing up the SFRs in the different apertures, we realise that the circumnuclear SFR in NGC 1808 is of the order of $~1\,M_\odot\,\mathrm{yr}^{-1}$, which is the suspected transition value between the two models.

\subsection{Nuclear activity: Does NGC1808 host an AGN?}
\label{sec:AGN}

Many authors have followed the old classification of NGC1808 as AGN by \cite{1985A&A...145..425V} which was based on the detection of two line systems with different widths. The one with the broader emission lines showed a flux ratio $\mathrm{H}\alpha/[\ion{N}{ii}]=0.94\pm0.05$ which is typical for Seyfert galaxies. More recent otpical spectroscopy is available from the survey S7 \citep{2015ApJS..217...12D} where the authors measure line ratios of $\log([\ion{O}{iii}]/\mathrm{H}\beta)=-0.58$, $\log([\ion{N}{ii}]/\mathrm{H}\alpha)=-0.08$, and $\log([\ion{S}{ii}]/\mathrm{H}\alpha)=-0.59$ which are typical line ratios of \ion{H}{ii} regions.
In accordance with this, several other authors do not find evidence to justify the old classifaction as Seyfert-galaxy and state that the central activity can be explained by intense star formation alone which comes along with strong supernova remnants \citep[e.g.][]{1992MNRAS.259..293F} and a superwind \citep[e.g.][]{1993AJ....105..486P}. \cite{2001ApJ...557..626K} establish a mid-infrared X-ray correlation and argue that the position of NGC 1808 in this diagram clearly indicates a pure starburst.

We analyse our emission-line measurements in the NIR diagnostic diagram in Fig.~\ref{fig:diagndiagram} and show that the nuclear aperture has line ratios typical for supernova dominated regions. The line ratios are close to the region of young star formation from which we conclude that there is no evidence for strong AGN activity and the nucleus shows characteristics of an aging starburst.

NGC 1808 was detected by XMM with a hard X-ray luminosity of $L_\mathrm{hard X-ray} = 10^{40.4}\,\mathrm{erg}\,\mathrm{s}^{-1}$. If we supposed that the X-ray flux was produced by an AGN, we could calculate the AGN luminosity, resulting in $L(\mathrm{AGN}) = 10^{41.2}\,\mathrm{erg}\,\mathrm{s}^{-1}$ \citep{2011MNRAS.413.1206B,2014ApJ...780...86E}. The Eddington ratio is defined by $\lambda \equiv L(\mathrm{AGN})/L_\mathrm{Edd}$ with Eddington luminosity $L_\mathrm{Edd} = 1.26\times 10^{38} \left(M_\mathrm{BH}/M_\odot \right)\,\mathrm{erg}\,\mathrm{s}^{-1}$. The black hole mass is of the order of $M_\mathrm{BH} \approx 5\times 10^7\,M_\odot$ (Sect.~\ref{sec:stellkin_disc}). With these values, the Eddington ratio is $\log(\lambda) = -4.4$ which is the Eddington ratio of a low-luminosity AGN. We conclude that an AGN cannot be fully ruled out. However, if present it would be very weak. The dominating excitation mechanism in the centre is star formation and associated supernovae.

\section{Conclusions and summary}
\label{sec:conclusions}

We have analysed NIR IFS of the central kiloparsec of the  galaxy NGC 1808 with the following results:

\begin{itemize}
\item We study the star formation history of the hotspots in the circumnuclear ring. We determine the ages to be $\lesssim 10\,\mathrm{Myr}$. We also study the relative ages. We cannot find evidence for an age gradient that would be predicted in a ``Pearls on a string'' model. However, the travel time around the ring is comparable to the cluster ages which means that the clusters have already travelled a significant portion of the ring.

\item In a field-of-view of $10\arcsec \times 10 \arcsec$ (which corresponds to $\sim 600\,\mathrm{pc}$), we measure the H$_2\lambda 2.12\mm$ line emission to estimate the molecular gas mass. For the hot molecular gas, we find a mass of $\sim 730\,M_\odot$ which corresponds to a cold molecular gas mass of $(2-12)\times 10^8\,M_\odot$. From the Br$\gamma$ line, we estimate an ionised gas mass of $\sim 6.6 \times 10^6\,M_\odot$. All gas masses are typical for NUGA sources.

\item From the $M_\mathrm{BH}-\sigma_*$ relation, we calculate a black hole mass of $\sim(1-5)\times 10^7\,M_\odot$. Since \object{NGC 1808} shows the signatures of a pseudobulge this should be interpreted as a rough estimate only.

\item We fit the CO absorption band heads in the $K$-band to obtain the stellar velocity field. It shows an overall rotation. However an S-shaped zero-velocity line, a drop in the central velocity dispersion, and residual structures after subtracting a pure rotation model indicate the presence of a bar and probably a nuclear disc.

\item The gaseous kinematics also shows an overall rotation. However, the deviations from pure rotation are much larger than in the stellar velocity field. By subtracting the stellar velocity field from the gas velocity field, we find a residual structure in the central $\sim 100\,\mathrm{pc}$ that has the shape of a two-arm nuclear spiral and could indicate an inflowing streaming motion.

\end{itemize}

To summarise, we find a large gas reservoir and a disturbed gas velocity field, that even shows signs of inflowing motion to the inner tens of parsecs. However, we do  not find indications for strong AGN activity (if AGN is present, its Eddington ratio is as low as $\log(\eta) \lesssim -4.4$). Instead the gas seems to be used up for star formation that is occuring in the circumnuclear ring and in the nuclear starburst. This shows that for the presence of an AGN, it is not enough to have gas and drive it to the centre. Furthermore it raises the question: What decides whether gas is used for AGN fuelling or star formation? Is there any sufficient condition for AGN activity?

Upcoming ALMA observations with higher angular resolution will trace the cold gas mass and structure and will show whether they show similar inflowing streaming motions like the warm gas.

\begin{acknowledgements}
The authors thank Florian Pei\ss ker and the team of the Paranal Observatory who conducted the SINFONI observations. Furthermore, we thank the anonymous referee for comments that helped to clarify the manuscript.

This work was supported by the Deutsche Forschungsgemeinschaft (DFG) via SFB 956, subproject A2. G.~Busch and N.~Fazeli are members of the Bonn-Cologne Graduate School of Physics and Astronomy (BCGS). M.~Valencia-S.~received funding from the European Union Seventh Framework Programme (FP7/2007-2013) under grant agreement No. 312789.

Based on observations made with the NASA/ESA Hubble Space Telescope, and obtained from the Hubble Legacy Archive, which is a collaboration between the Space Telescope Science Institute (STScI/NASA), the Space Telescope European Coordinating Facility (ST-ECF/ESA) and the Canadian Astronomy Data Centre (CADC/NRC/CSA).
\end{acknowledgements}

\bibliographystyle{aa}
\bibliography{ngc1808} 

\end{document}